\documentclass[prb,nofootinbib,twocolumn,superscriptaddress,showpacs,floatfix]{revtex4}
\usepackage{mathrsfs}
\usepackage{mathdots}
\usepackage{amsmath,amsfonts,amssymb}
\usepackage{epsfig}
\usepackage{graphicx}
\usepackage{wasysym}
\usepackage{bbm}
\usepackage{psfrag}
\usepackage{color}
\usepackage{pstool}
\usepackage{braket}
\usepackage{lmodern}
\usepackage{textcomp}
\usepackage[T1]{fontenc}
\usepackage[dvips, bookmarks, colorlinks=true, plainpages = false, citecolor = blue, linkcolor = blue, urlcolor = blue, filecolor = blue]{hyperref}
\def\be{\begin{equation}}
\def\ee{\end{equation}}
\def\bea{\begin{eqnarray}}
\def\eea{\end{eqnarray}}

\begin{document}
\title{Floquet topological phases of higher winding numbers in extended  Su-Schrieffer-Heeger model under quenched drive}
\author{Rittwik Chatterjee}\email{rittwikchatterjee@gmail.com}
\author{Asim Kumar Ghosh}
 \email{asimkumar96@yahoo.com}
\affiliation {Department of Physics, Jadavpur University, 
188 Raja Subodh Chandra Mallik Road, Kolkata 700032, India}
\begin{abstract} In this study topological properties of static and dynamic
  Su-Schrieffer-Heeger models with staggered further neighbor hopping terms
  of different extents are investigated. Topological characterization
  of the static chiral models is established in terms of
  conventional winding number while Floquet topological character is
  studied by a pair of winding numbers.
  With the increase of extent of further neighbor terms topological phases
  with higher winding numbers are found to emerge in both static and dynamic
  systems. Topological phase diagrams of static models for four different
  extents of further neighbor terms are presented, which has been generalized
  for arbitrary extent afterwards. Similarly, Floquet topological
  phase diagrams of four such dynamic models have been presented.
  For every model four different parametrizations of hopping terms
  are introduced which
  exhibits different patterns of topological phase diagrams.
  In each case emergence of `0' and `$\pi$' energy edge states
  is noted and they are found to consistent to the
  bulk-boundary correspondence rule applicable for chiral topological systems.

  \vskip 1 cm
  Corresponding author: Asim Kumar Ghosh,
  \footnote{Corresponding author: Asim Kumar Ghosh}

  Email: asimkumar96@yahoo.com 
\end{abstract}
\maketitle
\section{INTRODUCTION}
Investigation on topological quantum matters has become an intriguing part in condensed matter
studies for their exotic properties, such as symmetry-protected surface states
and anomalous transport phenomena, that are unaffected by  
deformations as well as presence of impurity \cite{Asboth,Qi}. 
In this process 
Su-Schrieffer-Heeger (SSH) model is the most successful
one-dimensional (1D) model for the
understanding of basic features of topological insulator\cite{SSH1,SSH2,SSH3}.
Topological properties based on single particle dynamics of this dimer
(two-sublattice) model
can be studied in terms of the ratio of intercell and intracell hopping strengths, 
$r$, between two different nearest-neighbor (NN) bonds, which 
is a measure of staggeredness between the pair of NN hopping amplitudes. 
Value of winding number ($\nu$) serves as one of the topological invariant
in order to distinguish the topological phases in the presence of
band gap. The staggered system noted by $r \ne 1$ develops a band gap 
and in this case non-trivial topological phase marked by 
$\nu=1$ emerges when $r>1$, while it becomes trivial ($\nu=0$)
for $r<1$.
Topological phases are associated with the symmetry protected edge states
for the chiral systems. Topological signature of
those models are generally demonstrated experimentally
by the presence of edge states. Such edge states
are found in a one-dimensional Stub lattice
of coupled waveguides with staggered hopping which
mimics the SSH model \cite{Vicencio}.
Evidence of topological phases has been 
detected in similar systems of
optical lattices composed of ultracold atoms 
and photonic systems of helical and cylindrical waveguides
\cite{Atala,Rechtsman,Li3}. Topological signature of
SSH model has been demonstrated in a vacancy lattice of a chlorine
monolayer on a Cu(100) surface \cite{Drost}. 

Topological phase diagrams of extended SSH (eSSH)
models with long range hopping terms are more exotic.  
For example, in order to induce nontriviality within the
region $r<1$, a third intercell hopping term between
further-neighbor (FN) sites of different sublattice 
has been introduced which essentially leads to
a new topological phase with $\nu=-1$ \cite{Li}. 
Moreover, this phase emerges both in the regions
defined by $r<1$ and $r\ge 1$, depending on the
strength of the third hopping. Topological phases with higher
winding numbers have been emerged with the increase of 
extent of the third hopping term \cite{Rakesh1}. 
In the same way, effect of a pair of inter-cell hopping term between
FN sites of different sublattice 
has been studied where an additional topological phase with $\nu=2$
is found to emerge \cite{Maffei}.
However in this case staggeredness of
either NN or FN hopping terms is necessary in order to induce
nontrivial topology.
Effect of disorder on the topological phases of higher winding numbers
in eSSH models has been investigated using Lindblad equation \cite{Giuliano}. 
In this study, effect of the extent of this pair of
FN hopping terms has been investigated where additional topological
phases with higher winding numbers are found. 

Understanding of periodically driven quantum lattice systems has been advanced
considerably during the past decades using the Floquet-Bloch formalism 
\cite{Floquet,Bloch,Shirley}, 
which has been employed in a variety of systems in due course
\cite{Dunlap,Sambe}. 
Nontrivial topology may be induced by the periodic drive which
is known as Floquet topological phases \cite{Kitagawa1,Platero1,Rahul}.
Topological state of Floquet systems has been demonstrated
in a number of tight-binding models \cite{Kitagawa2,Linder,Torres},
in various dimensions  \cite{Fruchart,Arghya}. 
Character of these topological
phase could be controlled by the amplitude, frequency and phase of the
periodic drive. Periodic drives may be categorised in terms of
either continuous or quenching protocols where the system experiences a
sudden change for the later case. It means before and after the sudden
change the system remains static with two different configurations. 
Interestingly, SSH model has played crucial role in order to
exhibit Floquet topological states under both types of periodic drives.
Both Zak phase and winding number are used for the characterization of
Floquet topology \cite{Zak,Ghuneim}. For example, SSH model with continuously driven
NN hopping amplitudes of Gaussian pulses hosts several
Floquet topological phases
those are characterized by a pair of winding numbers \cite{Asboth1}.
Similarly, SSH model under sinusoidal
periodic modulation of NN hopping terms exhibits Floquet topological
phase which is identified by pair of Zak
phases  and winding numbers \cite{Lago,Balabanov}.

On the other hand, topological phases in a spin-1/2
and non-Hermitian double kicked rotor model
have been found under quenched driving protocol\cite{Gong1,Zhou1}.
Floquet topological phase has been characterized by a 
pair of winding numbers in another chiral symmetric 1D
non-Hermitian system in the presence of quenched drive\cite{Zhou2}.
Besides to the topological phase transition, quenching protocol have been used
in order to study the Floquet dynamical phase transitions \cite{Zhou3}.
Floquet topological phases of a non-Hermitian SSH model
have been investigated under quenched drive\cite{Wu1}.
Exotic topological phases are found in non-Hermitian
Weyl semimetal on a cubic lattice under periodically
quenched drive \cite{Wu2}.
Topological phases with higher Chern numbers have been found in
periodically quenched Haldane model \cite{Xiong}.
Floquet topological phases have been noted in periodically quenched
SSH model where two different SSH models around the
quenching point have opposite dimerization \cite{Jangjan}.
In this study, topological phases with winding number not more than one is
found to appear. Floquet topological properties
of SSH trimer model has been studied under quenched drive 
 \cite{Ghuneim}.

In this work emergence of Floquet topological phases
in the periodically driven eSSH model with pair of extended FN
hopping terms has been noted besides the topological phases of
the same static eSSH models. Dynamic eSSH model is constituted by the
pair of static eSSH models with FN terms of different extents 
around a particular quenching point within a period.
However the extent of the FN terms for a particular set is kept fixed.
Here properties of Floquet topological phases have been characterized
for four different extents of FN terms and in every case
topological phases with winding number higher values are 
found to appear. Moreover, phases with higher values of
winding numbers emerge with the increase of extent of FN terms.

The primary motivation behind this study is to propose
theoretical models 
which are capable to exhibit topological phases of
higher winding numbers in the static and Floquet systems.
Notably, topological phases of higher winding number have
been induced by introducing FN hopping terms of larger extent 
within the standard SSH model, both for the static and dynamic cases,
as investigated here. 
In this context, the basic concern is whether this long-range
hopping is experimentally realizable. The realization of
long-range hopping paths is possible in the systems
synthesized by various routes those are found successful 
for the SSH dimer\cite{Thomale,Bitan} 
and trimer models\cite{Alvarez,Kwapinski}, 
as well as for the Kitaev's 1D $p$-wave
topological superconducting model earlier\cite{Kitaev,Thomale}.
As stated before topological signature of
these models have been demonstrated experimentally
by finding the edge states in coupled waveguides\cite{Vicencio},
optical lattices composed of ultracold atoms \cite{Atala},
and photonic systems of helical and cylindrical waveguides
\cite{Rechtsman,Li3}.  It has been demonstrated in
topolectrical circuits\cite{Thomale,Bitan}, a vacancy lattice of a chlorine
monolayer on a Cu(100) surface \cite{Drost}, and very recently in 
water wave channel\cite{Anglart}. In all these hybrid systems
topological character was established by detecting the pair of  
edge modes. Remarkably, in this effort the organic polymer,
$trans$-polyacetylene is of no help, even though the
SSH model has been formulated on this polymer.
So, it is expected that long-range hopping path could be
realized in these hybrid material platforms, and the
experimental realization of topological phases in terms of
topological edge modes will be made possible through these potential routes. 
Upon comparing various hybrid platforms, it turns out that 
long-range hopping amplitudes are more easily realizable by
topological integrated circuits as they are found successful
both for the SSH model\cite{Thomale,Bitan}, 
and the Kitaev's model in early attempts\cite{Kitaev,Ezawa}. 

Hamiltonians with the extended FN terms have been formulated in
Sec \ref{model-static} as well as the topological
phases in terms of winding number
have been presented for the static models.
Periodically driven dynamic models have been presented 
in Sec \ref{Floquet} using the Floquet formalism. 
Dynamic system is composed of two different static 
Hamiltonians, say $H_{n1}$ and $H_{n2}$ within a single period,
where $n$ marks the extent of the FN terms. 
System switches from $H_{n1}$ to $H_{n2}$ at a time $t_p$
within a complete period. The corresponding static model
can be generated when $H_{n1}=H_{n2}$.
Variety of Floquet topological phases are
presented in Sec \ref{Floquet-numerical} for four different
sets of Hamiltonians defined by four values of extents
$n=1,2,3,4$. 
A discussion based on those results is available 
in Sec \ref{Discussion}.
\section{Extended SSH model with staggered further neighbor hopping terms}
\label{model-static}
The eSSH model with staggered FN hopping terms
is defined on a 1D bipartite lattice in which primitive
cell comprises of two different sites, A and B. The corresponding Hamiltonian
is expressed as
\be
H_{n}\!=\!\!\sum_{j=1}^N\!\!\left(t_1\,a^\dag_{j}b_{j}\!+\!t_2\,a^\dag_{j+1}b_{j}
\!+\!t_3\,a^\dag_{j}b_{j+n}\!+\!t_4\,a^\dag_{j+n+1}b_{j}\right)+\, {\rm h.c.},
 \label{essh}
\ee
where $a_{j}$ and $b_{j}$ are the annihilation operators of
electron on sublattices A and B, respectively, for the $j$-th primitive cell. 
$N$ is the total number of primitive cells where $t_1$ and $t_2$ are the
NN intracell and intercell hopping amplitudes, respectively.
$t_3$ and $t_4$ are the FN intercell
hopping strengths where $n=1,2,3,\cdots$, represents the separation between
the relevant cells.
The staggeredness of this model can be understood
in terms of the ratios of NN and FN terms, say, $r=t_2/t_1$,
and $r'=t_4/t_3$. 
The standard SSH model \cite{SSH1} will be recovered when
both $t_3=0$ and $t_4=0$.
In this case, 
energy spectrum is gapless when $r=1$, while there is
a band gap when $r \ne 1$. Among the two different limits 
one is topologically trivial ($\nu=0$) when $r<1$,
while the system exhibits nontrivial topological phase with $\nu=1$ 
on the other limit when $r>1$. In addition, winding number is
undefined for the isotropic chain when $r=1$.

 In this model,
$t_3$ is the strength for hopping over the $n$ intermediate
unit cells while $t_4$ is that for hopping over the $n+1$ intermediate
cells. The system exhibits nontrivial topological phases
in the presence of FN staggered hopping terms
for either $r'\ne 1$, when  $r=1$, or $r'= 1$, when  $r\ne 1$, 
for any value of $n$, as well as when both $r \ne 1$ and $r'\ne 1$.
Again winding number is 
undefined when $r=1$ and $r'=1$.
Under the Fourier transformations, 
\bea a_{j}&=&\frac{1}{\sqrt N}\sum_{ k\in {\rm BZ}}a_{ k}\,e^{i kj},\nonumber\\ [0.4em]
b_{j}&=&\frac{1}{\sqrt N}\sum_{k\in {\rm BZ}}b_{ k}\,e^{i kj},\nonumber
\eea 
where the summation extends over BZ,
and periodic boundary condition (PBC) is assumed, 
the Hamiltonian in the $k$-space becomes
\[H_n=\sum_{ k\in {\rm BZ}}\Psi^\dag_{k}H_n( k)\Psi_{ k},\]
where 
\be H_n( k)=\left(
\begin{array}{cc}0&g_n( k)\\[0.4em]
 g^*_n( k) &0\end{array}
 \right),\label{Hnk}\ee
$\Psi^\dag_{ k}=[a^\dag_{k}\;b^\dag_{k}],$  and 
 $g_n(k)=t_1+t_2e^{-ik}+t_3e^{ink}+t_4e^{-i(n+1)k}$,
 as shown in the  Appendix-A. 
 Here, lattice parameter is assumed unity. 
 The energy dispersion relation can be
  given by $E_n(k)=|g_n(k)|$. 
It can be shown that $H({\rm k})$ satisfies the
following transformation relations under the three
different operators:
\[\left\{\begin{array}{l}
\mathcal T H_n(k) \mathcal T^{-1}=H_n(- k),\\ [0.4em]
\mathcal P H_n(k) \mathcal P^{-1}=-H_n(-k),\\ [0.4em]
 \sigma_z H_n(k) \sigma_z=-H_n(k),\end{array}\right. \]
where $\mathcal T = \mathcal K$, $\mathcal P = \mathcal K \sigma_z$ and
$\mathcal K$ is the complex conjugation operator. 
These relations correspond to the conservation of
time-reversal, particle-hole and chiral symmetries.
As a consequence, inversion symmetry is preserved as 
$\sigma_x H_n(k) \sigma_x=H_n(- k)$.
 
Topological phase has been determined by means of
winding number which is defined in terms of $g_n(k)$ as 
\[\nu_n=\frac{1}{2\pi i}\int_{-\pi}^\pi \,d{\rm k} \frac{d}{dk}\log{g_n(k)}.\]
Now introducing the complex variable
$z=e^{ik}$, winding number can be expressed as
\bea \nu_n&=&\frac{i}{2\pi}\oint_{\mathcal C: |z|=1} \,d \log{g_n(z)},
\nonumber\\ [0.4em]
&=& \mathcal N_{\rm p}-\mathcal N_{\rm z},
\label{Np-Nz}
\eea
in which
\be g_n(z)=t_1+\frac{t_2}{z}+t_3\,z^{n}+\frac{t_4}{z^{n+1}},
\label{gn}
\ee
where $\mathcal N_{\rm p}$, and $\mathcal N_{\rm z}$
are the number of poles and zeros of $g_n(z)$ within the unit circle
$|z|=1$.  Proof of this theorem is given in the Appendix-B. 

It is shown that topological phases of higher winding number
emerge in the eSSH model in the presence of FN hopping
terms\cite{Maffei,Rakesh1}.
For example, eSSH model with a single FN hopping term
could exhibit topological phase of any winding numbers,
where value of winding number increases with the
value of hopping extent \cite{Rakesh1}.
Topological phases with higher winding numbers emerge in eSSH
model with a pair of FN staggered hopping
terms with $r=1$, and $r'\ne 1$, when $n=1$ \cite{Maffei}.
Topological phase diagram in the $t_3$-$t_4$ plane has been
given by Eq. \ref{H-1},  
\bea 
\nu_1 &\!\!=\!\!&
\begin{cases}
2, & \!\!\!\text{if} \; 
\begin{cases} 
t_3 + t_4 < -2t_1, \; \text{and} \; r'<1, \\
t_3 + t_4 > -2t_1\cos{\theta_1}, \; \text{and} \; r'>1,
\end{cases} \\
1, & \!\!\!\text{if} \; -2t_1 < t_3 + t_4 < -2t_1\cos{\theta_1}, \; \text{and} \; r'<1, \\
0, & \!\!\!\text{if} \; -2t_1 < t_3 + t_4 < -2t_1\cos{\theta_1}, \; \text{and} \; r'>1, \\
-1, & \!\!\!\text{if} \; 
\begin{cases} 
t_3 + t_4 < -2t_1, \; \text{and} \; r'>1, \\
t_3 + t_4 > -2t_1\cos{\theta_1}, \; \text{and} \; r'<1,
\end{cases}
\end{cases}\nonumber
\\ &&{\rm when}\, r=1, \,{\rm and}\, \theta_1=2\pi/3.
\label{H-1}
\eea

Three different topological phases with
$\nu=2,1,-1$ emerge.  
The isotropic point defined by $r'= 1 $ does not
lead to new topological phase since winding number is
undefined over this line. Obviously staggeredness of the model
is lost at this point. Since the energy gap vanishes on
the topological phase transition point, or $E_n(k)=0$, location of the
phase boundary can be identified by the equation $|g_n(z)|=0$,
when $|z|=1$. 
For $k=\pi$, the equation $g_n =0$, leads to the condition $t_1+t_4=t_2+t_3$ 
($t_1-t_4=t_2-t_3$) for odd (even) values of $n$. Thus it represents  
a phase boundary. 
Similarly, equation $g_n=0$, for $k=0$ leads to the
condition, $t_1+t_2+t_3+t_4=0$. 
As a result, lines satisfying these equations on
the $t_3$-$t_4$ plane always represent 
phase boundaries for any values of $n$. However
other phase boundaries are found to depend on the value of $n$.
For example, when $n=1$, the third phase boundary
can be identified by $z=e^{\pm 2\pi i/3}$, which leads
to the equation $t_3+t_4=t_1$, by
solving the cubic equation, Eq. \ref{gn}, for $n=1$. 
In this study, eSSHs models with
arbitrary value of $n$ have been constructed in order to 
exhibit the emergence of topological
phase of higher winding numbers. For this purpose topological
phases for $n=1,2,3,4$ have been discussed here. 
\begin{figure}[h]
\psfrag{a}{(a)}
\psfrag{b}{(b)}
\psfrag{c}{(c)}
\psfrag{d}{(d)}
\psfrag{e}{(e)}
\psfrag{A}{\large A}
\psfrag{B}{\large B}
\psfrag{t1}{ $t_1$}
\psfrag{t2}{ $t_2$}
\psfrag{t3}{ $t_3$}
\psfrag{t4}{ $t_4$}
\psfrag{w}{ $t_3/t_1$}
\psfrag{z}{$t_4/t_1$}
\psfrag{nu}{$\nu_2$}
\psfrag{0}{$0$}
\psfrag{1}{$1$}
\psfrag{2}{$2$}
\psfrag{-3}{$-3$}
\psfrag{-2}{\hskip -0.2cm $-2$}
\psfrag{-1}{\hskip -0.2cm $-1$}
\psfrag{-4}{$-4$}
\psfrag{3}{$3$}
\psfrag{4}{$4$}
\psfrag{1.0}{$1.0$}
\psfrag{-1.0}{\hskip -0.2cm $-1.0$}
\psfrag{-0.5}{$-0.5$}
\psfrag{-1.5}{$-1.5$}
\psfrag{1.5}{$1.5$}
\psfrag{0.0}{$0.0$}
\psfrag{0.5}{$0.5$}
\psfrag{2.0}{$2.0$}
\psfrag{-2.0}{\hskip -0.2cm $-2.0$}
\includegraphics[width=240pt]{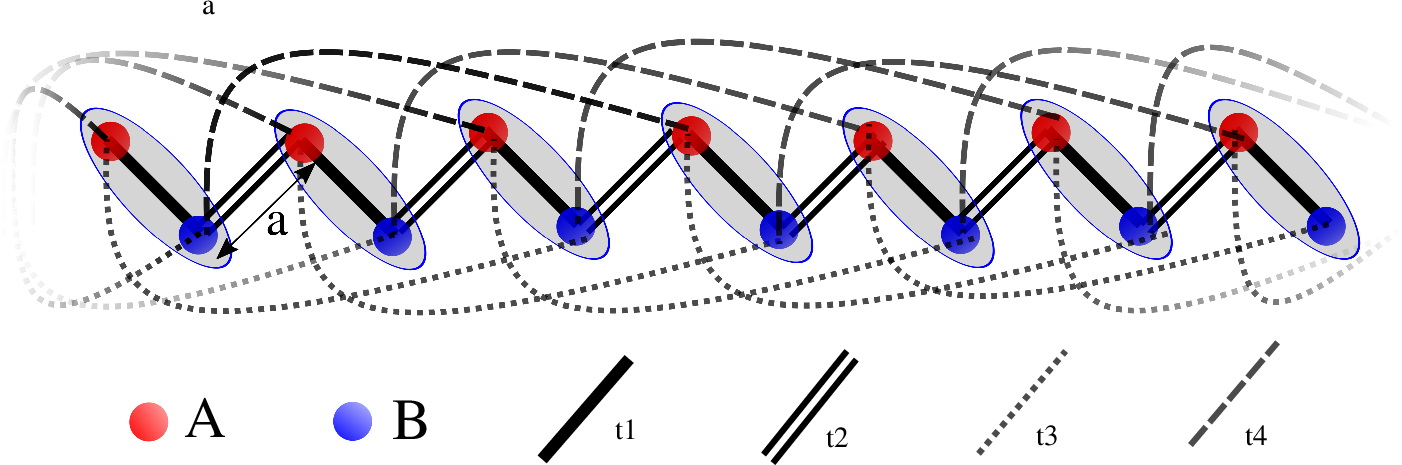}
\vskip 0.2 cm
\includegraphics[width=245 pt,angle=0]{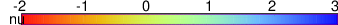}
\vskip -0.45cm
\hskip -.5 cm
\begin{minipage}{0.25\textwidth}
  \includegraphics[width=110pt,angle=-90]{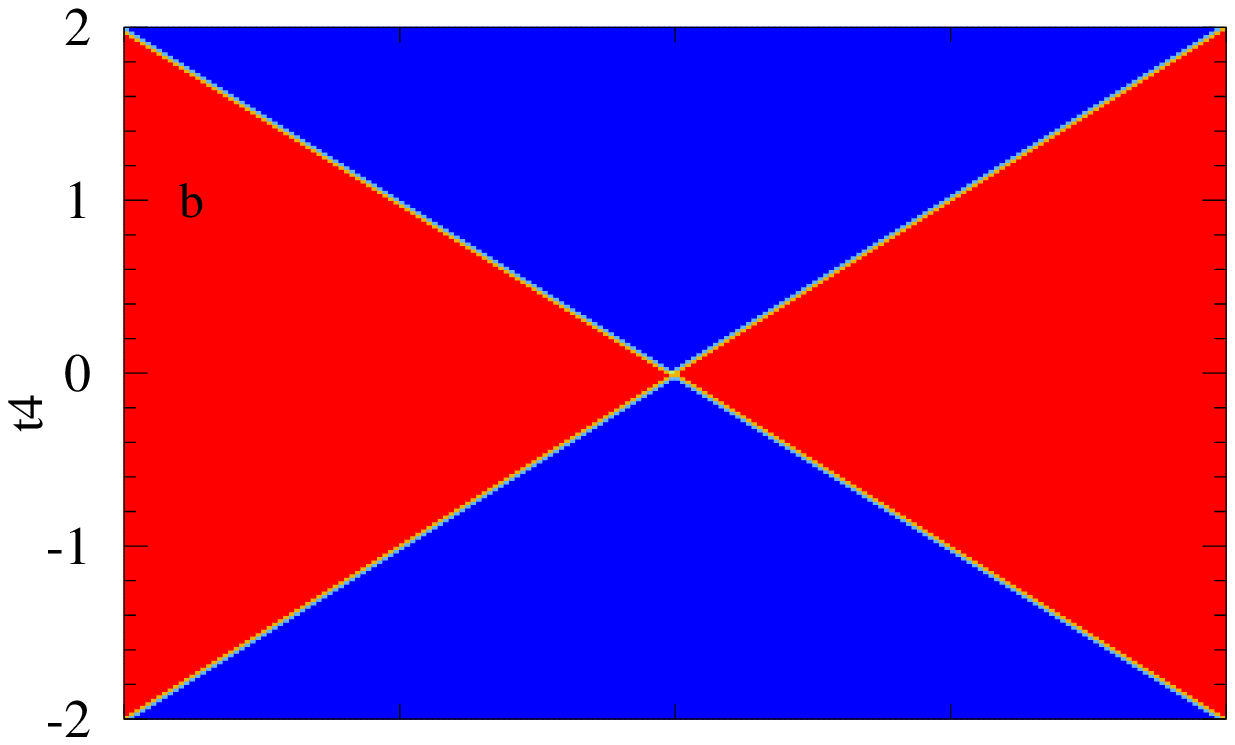}
  \end{minipage}\hskip -0.3cm
  \begin{minipage}{0.25\textwidth}
  \includegraphics[width=110pt,angle=-90]{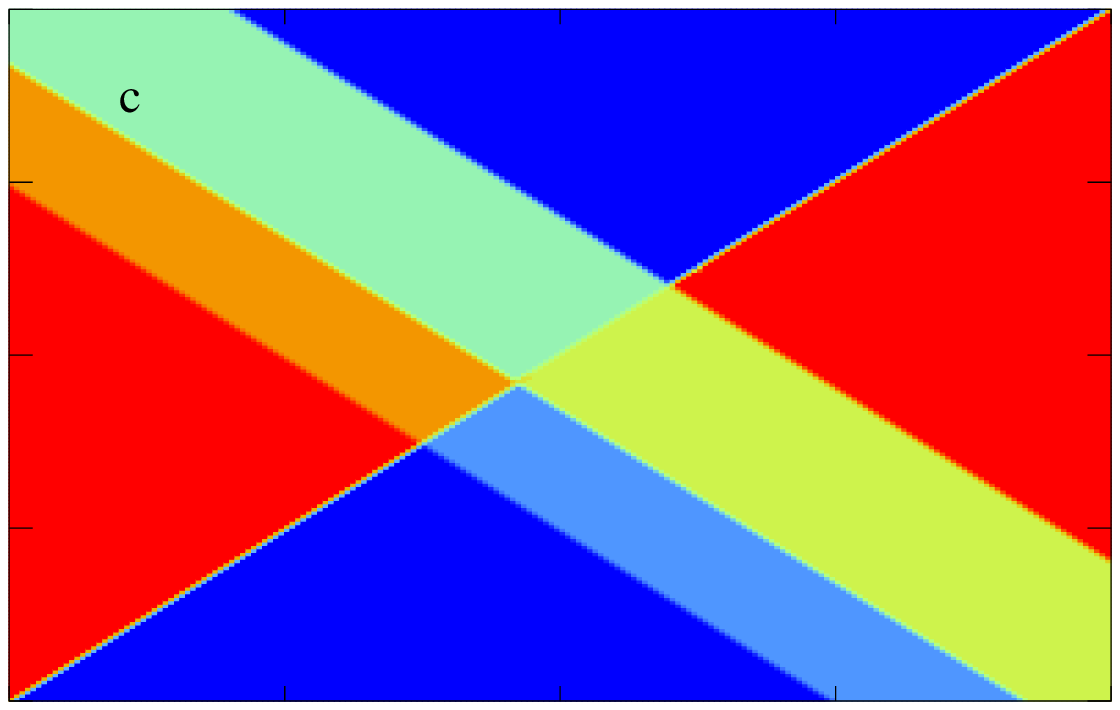}
  \end{minipage}
  \vskip -1.2 cm
  \hskip -0.5 cm
  \begin{minipage}{0.25\textwidth}
  \includegraphics[width=110pt,angle=-90]{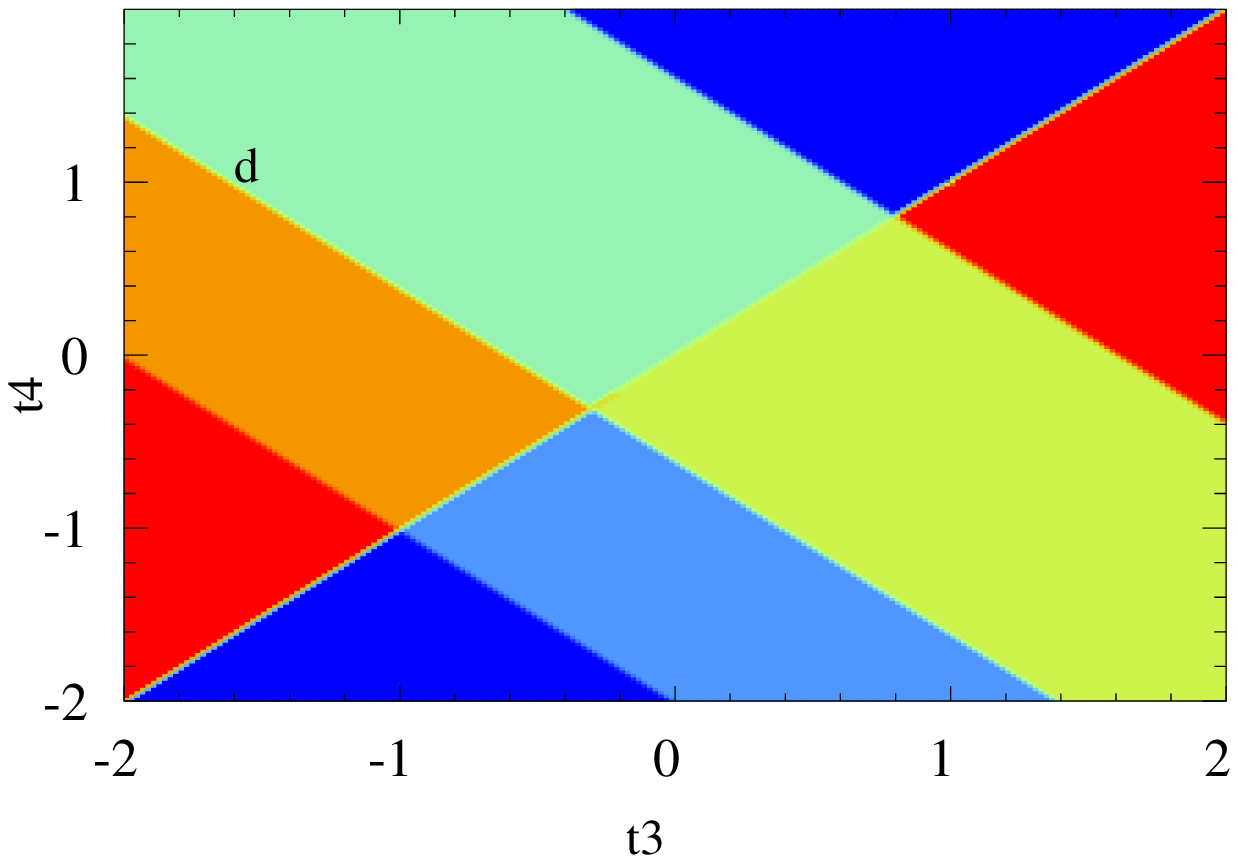}
  \end{minipage}\hskip -0.3cm
  \begin{minipage}{0.25\textwidth}
  \includegraphics[width=110pt,angle=-90]{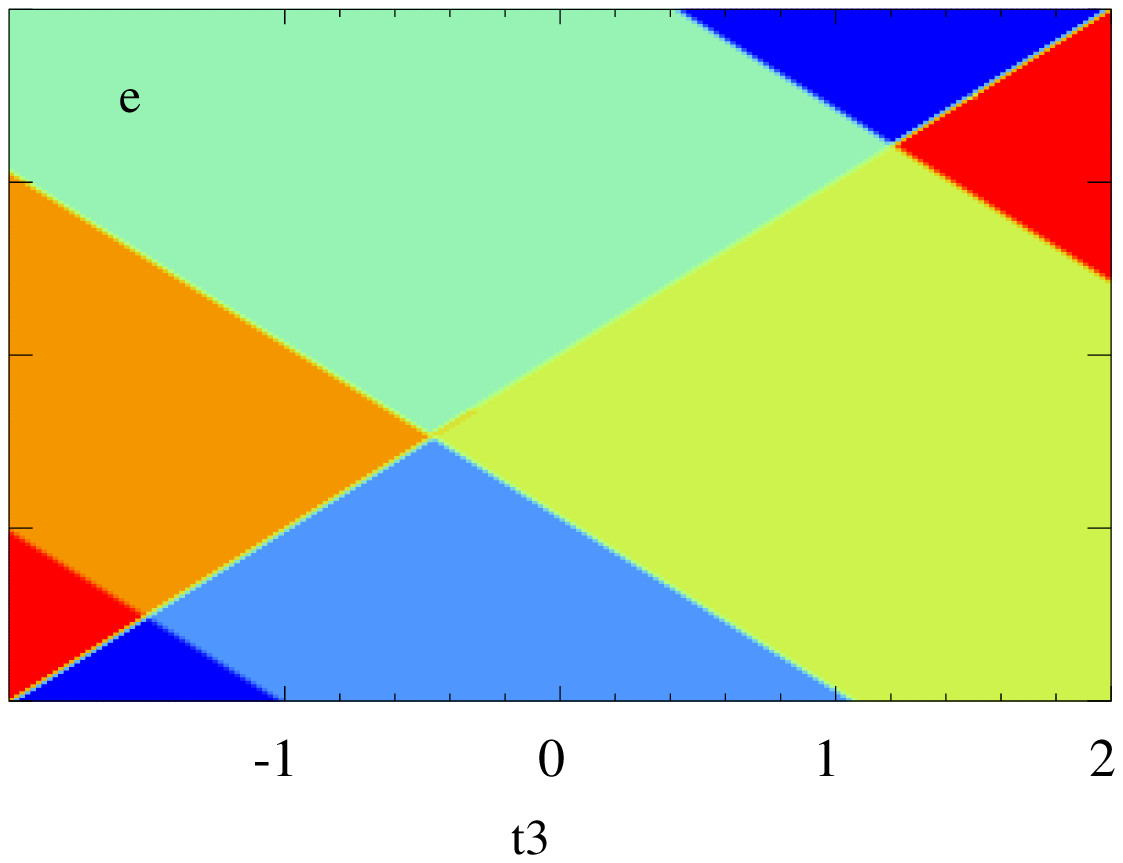}
  \end{minipage}
\caption{(a) Extended SSH model describing the hopping for the
  Hamiltonian, $H_2$. Lattice parameter is `a'. 
  Topological phase diagram of $H_2$ when $t_1=t_2$, for 
  (b) $t_1=0$, (c) $t_1=1/2$, (d) $t_1=1$, and (e) $t_1=3/2$. 
Horizontal colorbar indicates the variation of $\nu_2$.}
\label{Extended-SSH-n-2}
\end{figure}

In order to study the effect of competition between the NN
($t_1$, $t_2$) and FN ($t_3$, $t_4$) hopping amplitudes, topological phases for
two extreme cases have been explained in the beginning. 
In one extreme point, $t_3=t_4=0$, system becomes the standard SSH model,
whose topology is governed by the relation, 
\[\nu=\left\{\begin{array}{cc}1,&\;\left|\frac{t_2}{t_1}\right|>1,
\\[0.2em]
0,&\;\left|\frac{t_2}{t_1}\right|<1.\;
\end{array}\right. \;\]
for any value of $n$. 
For the other extreme point, $t_1=t_2=0$, topological properties
will depend on the value  of $n$, for 
obvious reason. Now, for the arbitrary value of $n$,
values of ${\mathcal N_{{\rm p}}}$ and ${\mathcal N_{{\rm z}}}$ of the
complex function, $g_n(z)$, Eq. \ref{gn}, 
can be determined in the following way. In this case,
 \[g_n(z)=t_3\,z^{n}+\frac{t_4}{z^{n+1}},\]
so, ${\mathcal N_{{\rm p}}}=n+1$, since $g_n(z)$ always has a pole of
order $(n+1)$ at the point $z=0$, for any values of $t_3$ and $t_4$. 
While ${\mathcal N_{{\rm z}}}$ has been determined by finding the 
roots of the equation, $t_3 z^{(2n+1)}+t_4=0$, within the circle, $|z|=1$. 
Since the roots are given by  $z_0^{(2n+1)}=-t_4/t_3$,
it leads to zero of order $(2n+1)$, and as a result, 
\[{\mathcal N_{{\rm z}}}=\left\{\begin{array}{cc}(2n+1),&\;\left|\frac{t_4}{t_3}\right|<1,
\\[0.2em]
0,&\;\left|\frac{t_4}{t_3}\right|>1.\;
\end{array}\right. \]

\begin{table}[htbp]
  \centering
\begin{tabular}{|c|c|c|c|c|} 
\hline  &&&&\\[-1.2em]\;$t_3\ne 0$\;&$\;\mathcal N_{\rm p}\;$&$\;|z_0|\;$&$\;\mathcal N_{\rm z}\;$ &$\;\nu_n\!=\!\mathcal N_{\rm p}\!-\!\mathcal N_{\rm z}\;$
\\[0.2em] \hline
&&&&\\[-1.1em] $\left|\frac{t_4}{t_3}\right|>1$ & $(n+1)$  &$|z_0|>1$ &  0 & $(n+1)$  \\[0.1em] \hline
&&&&\\[-1.1em] $\left|\frac{t_4}{t_3}\right|=1$  &  $(n+1)$  &$|z_0|=1$ &  ? &undefined   \\[0.1em] \hline
&&&&\\[-1.1em]  $\left|\frac{t_4}{t_3}\right|<1$  &  $(n+1)$  &$|z_0|<1$ &  $(2n+1$) &$-n$   \\[0.1em] \hline
\end{tabular}
\end{table}
Hence, topological phase is determined by the condition, 
\be \nu_n=\left\{\begin{array}{cc}(n+1),&\;\left|\frac{t_4}{t_3}\right|>1,
\\[0.2em]
-n,&\;\left|\frac{t_4}{t_3}\right|<1.\;
\end{array}\right.
\label{nun}
\ee
It means, system exhibits topological phases of extreme
values of winding numbers, which is
either $\nu_n= (n+1)$, or $\nu_n=-n$, when $t_1=t_2=0$.
No other phases with intermediate values, $-n<\nu_n<(n+1)$,
is found to appear in this case, which means not even the topologically 
trivial phase ($\nu_n=0$) is found to exist. In contrast for the
previous extreme case (or SSH model), system hosts the 
trivial phase, $\nu=0$. As a result, the competition between NN
and FN hopping amplitudes leads to the existence of
all intermediate nontrivial phases along with the trivial phase, which is
discussed in following subsections. 
\subsection{Topological phases for $n=2$:}
Topological phase diagram for the eSSH model $H_2$ is shown in
Fig. \ref{Extended-SSH-n-2} (b), (c), (d) and (e),
respectively for $t_1=0$, $t_1=1/2$, $t_1=1$, and $t_1=3/2$
when $t_1/t_2=1$, whereas the FN hoppings of the corresponding 
eSSH model is shown in Fig. \ref{Extended-SSH-n-2} (a).
Existence of two phases with $\nu_2=3$, and $\nu_2=-2$,
corresponds to Eq. \ref{nun} for $n=2$,  when $t_1=0$, as shown in
Fig. \ref{Extended-SSH-n-2} (b).  
Diagrams are drawn in the parameter space spanned by $\left|t_3\right|\le 2$, 
$\left|t_4\right|\le 2$. As a result, each of the phases with
$\nu_2=3$, and $\nu_2=-2$
appears twice in two different locations in this parameter space
as shown in every diagram of Fig. \ref{Extended-SSH-n-2}.

Apart from the phases with $\nu_2=3$, and $\nu_2=-2$, three
intermediate topological phases
with $\nu=-1,1,2$, appear in this case along with
the trivial phase, $\nu=0$,  when $t_1\ne 0$, as shown in
Fig. \ref{Extended-SSH-n-2} (c), (d) and (e).  
In order to study the evolution of the intermediate  phases,
additional phase diagrams for $t_1=1/2,\,1,$ and 3/2 have been drawn. 
Now comparing all these phase diagrams 
in Fig. \ref{Extended-SSH-n-2}, it is clear that 
with the increase of $t_1$, area of the region
for the intermediate phases increases, at the cost of the
area for phases with extreme values $\nu=3$, and $\nu=-2$.
 Equations of
several phase boundaries have been noted in Eq. \ref{H-2}, for the case
$t_1/t_2=1$. 
Winding number is undefined over the diagonal line $t_3=t_4$.
Phases with the largest ($\nu=3$) and the lowest ($\nu=-2$) winding numbers
appear both sides of the diagonal line, $t_3=t_4$.
Besides that, phases with odd
winding numbers ($\nu=1$ and $-1$) appear above the
line and phases with even winding number ($\nu =2$) and
the trivial phase ($\nu=0$) appear below the line.
Above the diagonal line phases appear with increasing value
of winding numbers with the increase of $t_3$, while
below that line phases appear with decreasing value
of winding numbers with $t_3$. 
\bea
\nu_2 \!\!&=&\!\!\!
\begin{cases}
3, & \!\!\!\text{if} \;
\begin{cases} 
t_3 + t_4 < -2t_1, \; \text{and} \; r'<1, \\
t_3 + t_4 > -2t_1\cos{\theta_1}, \; \text{and} \; r'>1,
\end{cases} \\
2, & \!\!\!\text{if} \; -2t_1 < t_3\!+\!t_4 < -2t_1\cos{\theta_2}, \; \text{and} \; r'<1, \\
1, & \!\!\!\text{if} \; -2t_1\cos{\theta_2} < t_3\!+\!t_4 < -2t_1\cos{\theta_1}, \; \text{and} \; r'>1, \\
0, & \!\!\!\text{if} \; -2t_1\cos{\theta_2} < t_3\!+\!t_4 < -2t_1\cos{\theta_1}, \; \text{and} \; r'<1, \\
-1, & \!\!\!\text{if} \; -2t_1 < t_3\!+\!t_4 < -2t_1\cos{\theta_2}, \; \text{and} \; r'>1, \\
-2, & \!\!\!\text{if} \;
\begin{cases} 
t_3 + t_4 < -2t_1,\; \text{and} \; r'>1, \\
t_3 + t_4 > -2t_1\cos{\theta_1}, \; \text{and} \; r'<1,
\end{cases}
\end{cases}\nonumber
\label{H-2}\\
&&{\rm when}\, r=1,\, \theta_1=4\pi/5,\, {\rm and}\, \theta_2=8\pi/5.
\eea

Topological phase boundaries on the $t_3$-$t_4$ plane
for $n=2$ can be given by the equations, $t_3=t_4$ and
$t_3+t_4=-2t_1$, when $t_1=t_2$, which is
identical to the previous case $n=1$,
as they corresponds to $z=\pm 1$.
Other two boundary lines can be given by the equations
$t_3+t_4=t_1(z^2+z^3)$, when $z=e^{2\pi i /5}$ and
$z=e^{4\pi i /5}$. These phase boundaries can be obtained by
the equations $E_2(k)=0$. 
Winding number is undefined over those lines
which separate different phases as shown in Fig. 
\ref{Extended-SSH-n-2} (b).

\begin{figure}[h]
\psfrag{a}{(a)}
\psfrag{b}{(b)}
\psfrag{c}{(c)}
\psfrag{d}{(d)}
\psfrag{e}{(e)}
\psfrag{A}{\large A}
\psfrag{B}{\large B}
\psfrag{t1}{ $t_1$}
\psfrag{t2}{ $t_2$}
\psfrag{t3}{ $t_3$}
\psfrag{t4}{ $t_4$}
\psfrag{nu}{$\nu_3$}  
\psfrag{0}{$0$}
\psfrag{1}{$1$}
\psfrag{2}{$2$}
\psfrag{-2}{\hskip -0.2cm $-2$}
\psfrag{-1}{\hskip -0.2cm $-1$}
\psfrag{-3}{\hskip -0.0cm $-3$}
\psfrag{-4}{$-4$}
\psfrag{3}{$3$}
\psfrag{4}{$4$}
\psfrag{1.0}{$1.0$}
\psfrag{-1.0}{$-1.0$}
\psfrag{-0.5}{$-0.5$}
\psfrag{-1.5}{$-1.5$}
\psfrag{1.5}{$1.5$}
\psfrag{0.0}{$0.0$}
\psfrag{0.5}{$0.5$}
\psfrag{2.0}{$2.0$}
\psfrag{-2.0}{$-2.0$}
\includegraphics[width=240pt]{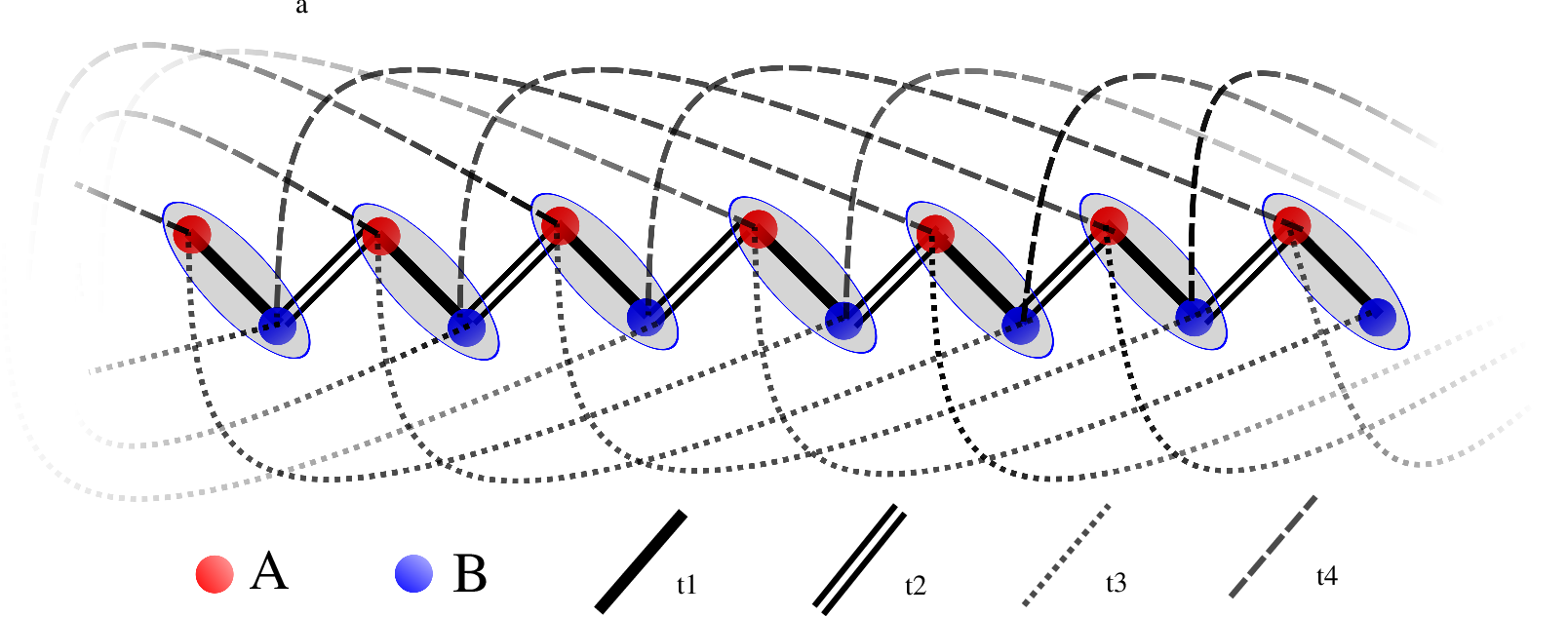}
\vskip 0.2 cm
\includegraphics[width=245 pt,angle=0]{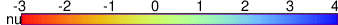}
\vskip -0.45cm
\hskip -.5 cm
\begin{minipage}{0.25\textwidth}
  \includegraphics[width=110pt,angle=-90]{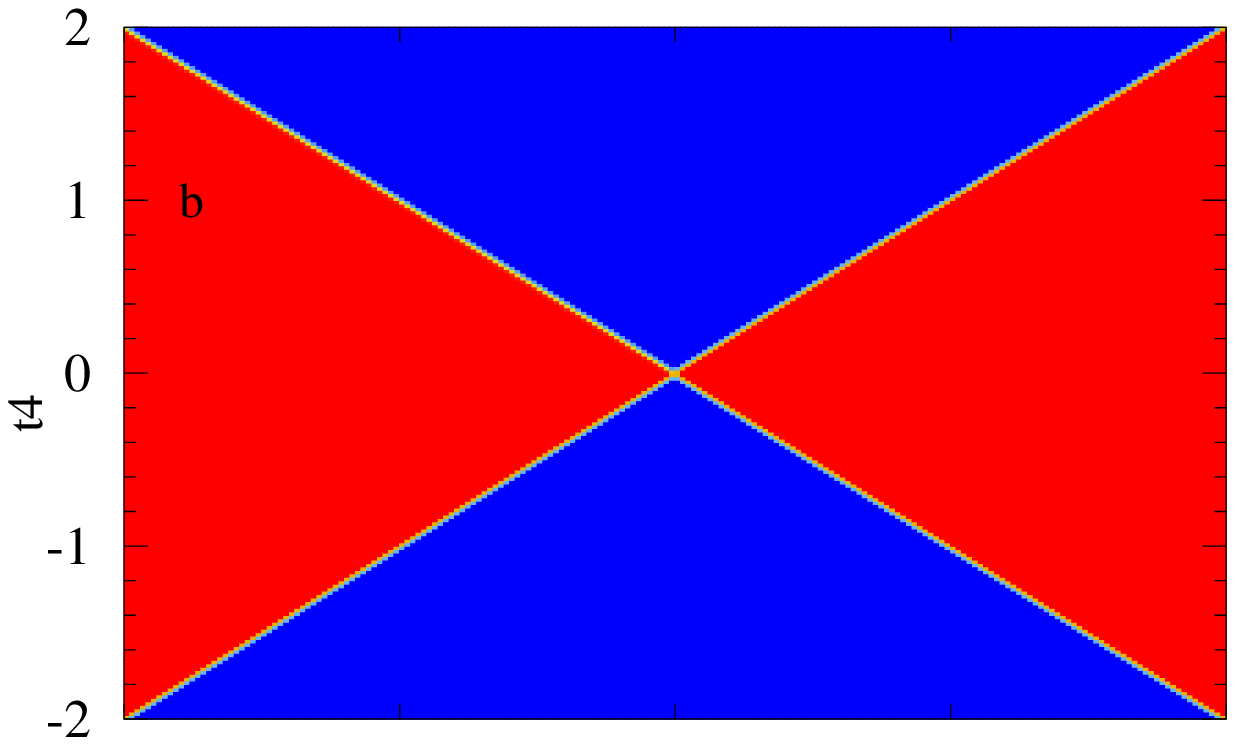}
  \end{minipage}\hskip -0.3cm
  \begin{minipage}{0.25\textwidth}
  \includegraphics[width=110pt,angle=-90]{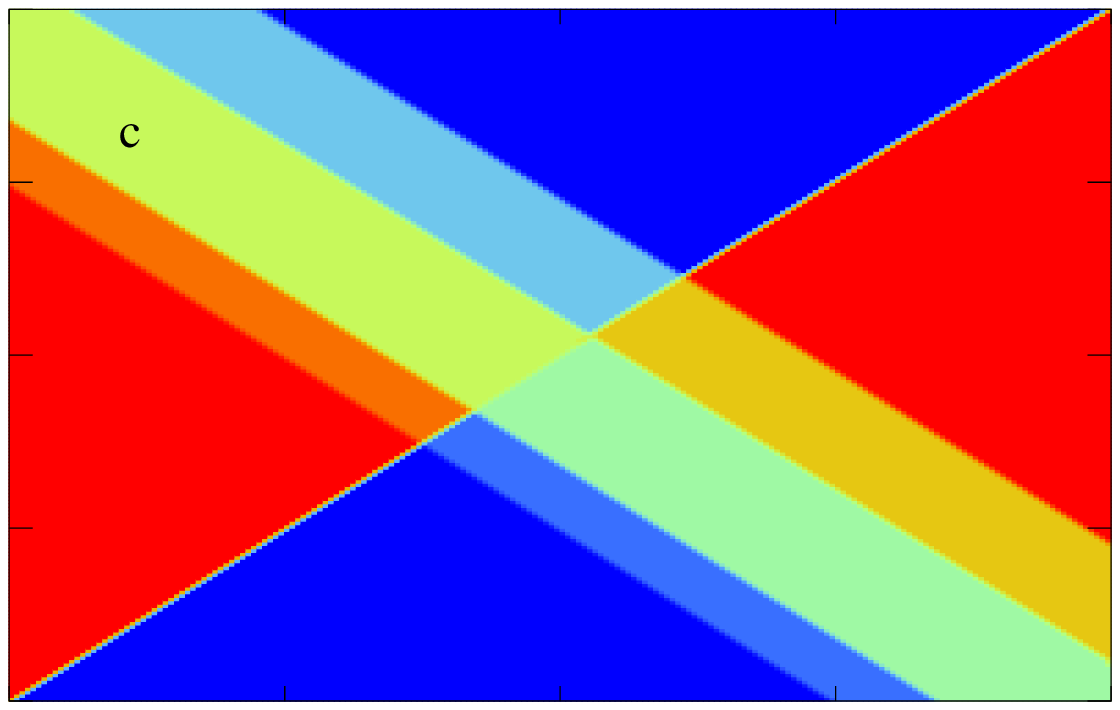}
  \end{minipage}
  \vskip -1.2 cm
  \hskip -0.5 cm
  \begin{minipage}{0.25\textwidth}
  \includegraphics[width=110pt,angle=-90]{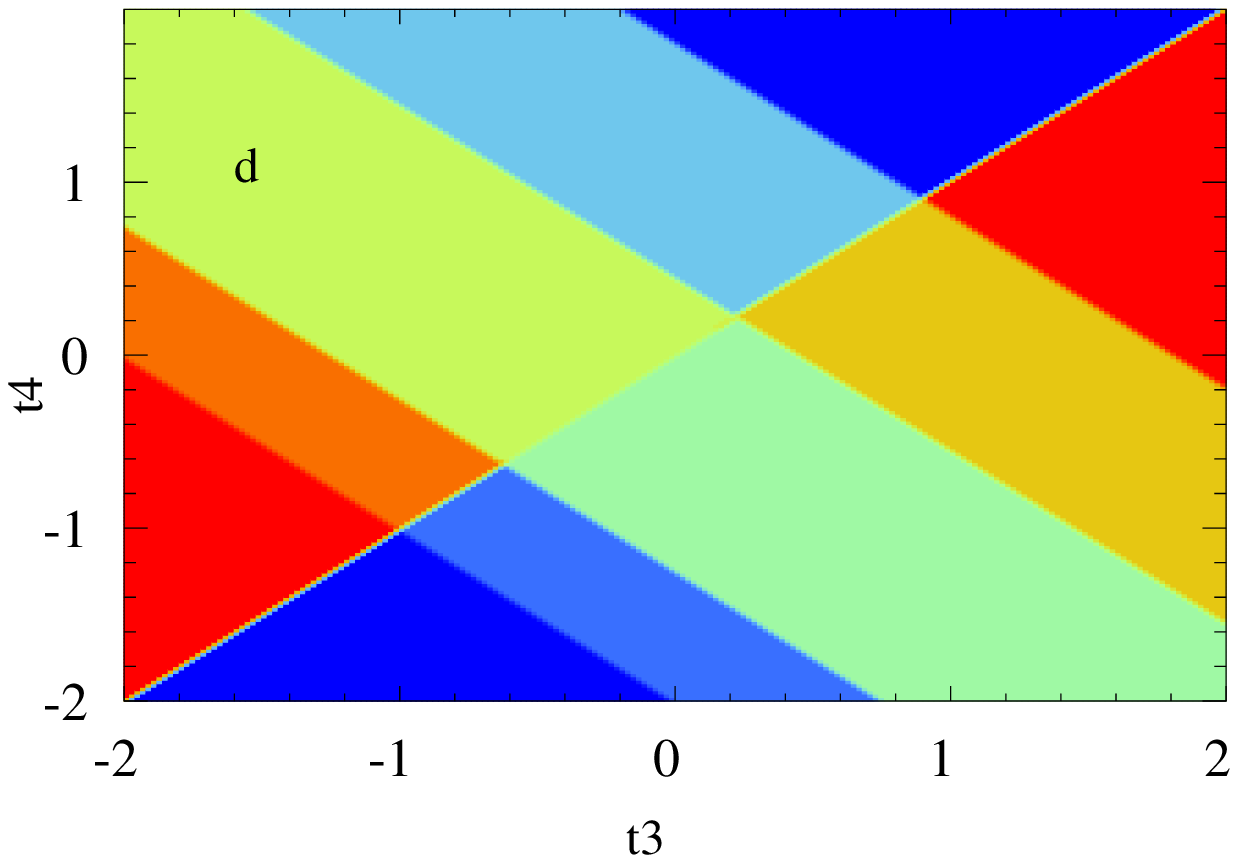}
  \end{minipage}\hskip -0.3cm
  \begin{minipage}{0.25\textwidth}
  \includegraphics[width=110pt,angle=-90]{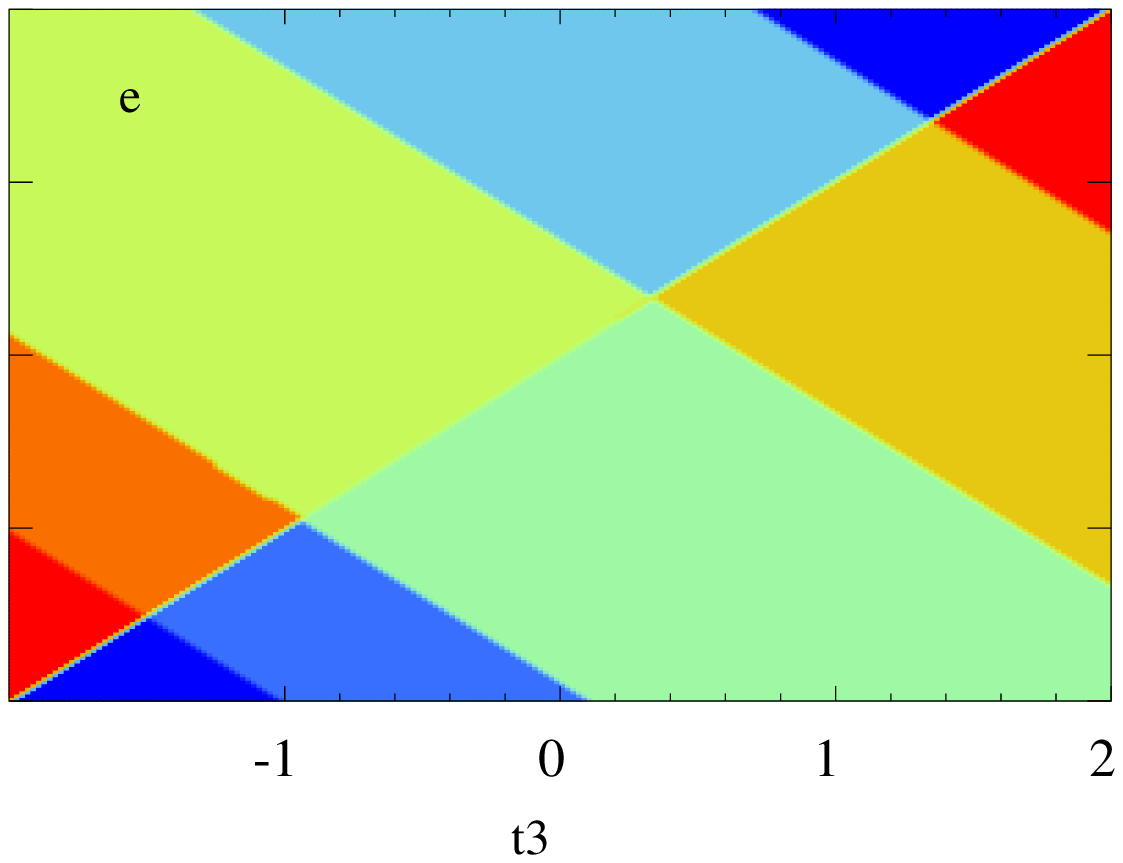}
  \end{minipage}
\caption{(a) Extended SSH model describing the hopping for the
  Hamiltonian, $H_3$. 
  Topological phase diagram of $H_3$ when $t_1=t_2$, for 
  (b) $t_1=0$, (c) $t_1=1/2$, (d) $t_1=1$, and (e) $t_1=3/2$.
Horizontal colorbar indicates the variation of $\nu_3$.}
\label{Extended-SSH-n-3}
\end{figure}
\subsection{Topological phases for $n=3$:}
Similarly, topological phase diagrams for $H_3$ have been shown in
Fig \ref{Extended-SSH-n-3} (b), (c), (d) and (e),
respectively for $t_1=0$, $t_1=1/2$, $t_1=1$, and $t_1=3/2$
when $t_1/t_2=1$, along with the model
in Fig \ref{Extended-SSH-n-3} (a). Seven different topological phases,
$\nu=-3,-2,\cdots,3,4$, along with
the trivial phase $\nu=0$ appear in this case. 
The region of those phases have been noted
in Eq. \ref{H-3}. Two additional nontrivial phases with winding numbers
$\nu=4$ and $\nu=-3$, appear here in comparison to the
previous case for $H_2$. Again, phases with the largest ($\nu=4$) and
the lowest ($\nu=-3$) winding numbers
appear both sides of the diagonal line, $t_3=t_4$, which
corresponds to the Eq. \ref{nun}, for $n=3$, when $t_1=t_2=0$.
Comparing the phase diagrams 
in Fig. \ref{Extended-SSH-n-3}, it is clear that 
with the increase of $t_1$, area of the region
for the intermediate phases increases, at the expense of the
area for phases with extreme values $\nu=4$, and $\nu=-3$.
Besides that, phases with odd
winding numbers ($\nu=3,\,1$ and $-1$) appear below the diagonal 
line and phases with even winding number ($\nu =2,\,-2$) and
the trivial phase ($\nu=0$) appear above that line,
which is opposite to the previous case ($n=2$).
However value of the winding numbers increases with $t_3$ above the
diagonal line, while that of winding numbers decreases with
$t_3$ below the diagonal line. 
Topological phase boundaries on the $t_3$-$t_4$ plane
for $n=3$ can be given by the five equations, (i) $t_3=t_4$, 
(ii) $t_3+t_4=-2t_1$, and $t_3+t_4=t_1(z^3+z^4)$, for
(iii) $z=e^{2\pi i /7}$, (iv) $z=e^{4\pi i /7}$,  and
(v) $z=e^{6\pi i /7}$, when $t_1=t_2$  as shown in Fig. 
\ref{Extended-SSH-n-3}. The lines 
defined by equations (ii), (iii), (iv) and (v) are
parallel to each other but all of them are normal to the
diagonal line (i) in the $t_3$-$t_4$ plane.
Those boundary lines can be obtained by the
equation $E_3(k)=0$.
\bea
\nu_3\!\! \!&=&\!\!\!\!
\begin{cases}
4, &\!\!\! \text{if} \; 
\begin{cases}
t_3 + t_4 < -2t_1, \; \text{and} \; r'<1, \\
t_3 + t_4 > -2t_1\cos{\theta_1}, \; \text{and} \; r'>1,
\end{cases} \\
3, & \!\!\!\text{if} \; -2t_1 < t_3 + t_4 < -2t_1\cos{\theta_2}, \; \text{and} \; r'<1, \\
2, & \!\!\!\text{if} \; -2t_1\cos{\theta_3} < t_3\!+\!t_4 < -2t_1\cos{\theta_1}, \; \text{and} \; r'>1, \\
1, & \!\!\!\text{if} \; -2t_1\cos{\theta_2} < t_3\!+\!t_4 < -2t_1\cos{\theta_3}, \; \text{and} \; r'<1, \\
0, & \!\!\!\text{if} \; -2t_1\cos{\theta_2} < t_3\!+\!t_4 < -2t_1\cos{\theta_3}, \; \text{and} \; r'>1, \\
-1, & \!\!\!\text{if} \; -2t_1\cos{\theta_3} < t_3\!+\!t_4 < -2t_1\cos{\theta_1}, \; \text{and} \; r'<1, \\
-2, & \!\!\!\text{if} \; -2t_1 < t_3 + t_4 < -2t_1\cos{\theta_2}, \; \text{and} \; r'>1, \\
-3, & \!\!\!\text{if} \; 
\begin{cases}
t_3 + t_4 < -2t_1, \; \text{and} \; r'>1, \\
t_3 + t_4 > -2t_1\cos{\theta_1}, \; \text{and} \; r'<1,
\end{cases}
\end{cases}\nonumber \\
&&{\rm where}\, \theta_1=6\pi/7,\, \theta_2=12\pi/7,\,
           {\rm and}\, \theta_3=18\pi/7.
\label{H-3}
\eea
\subsection{Topological phases for $n=4$:}
\begin{figure}[h]
 \psfrag{a}{(a)}
\psfrag{b}{(b)}
\psfrag{c}{(c)}
\psfrag{d}{(d)}
\psfrag{e}{(e)}
\psfrag{A}{\large A}
\psfrag{B}{\large B}
\psfrag{t1}{ $t_1$}
\psfrag{t2}{ $t_2$}
\psfrag{t3}{ $t_3$}
\psfrag{t4}{ $t_4$}
\psfrag{nu}{$\nu_4$}  
\psfrag{0}{$0$}
\psfrag{1}{$1$}
\psfrag{2}{$2$}
\psfrag{-2}{\hskip -0.2cm $-2$}
\psfrag{-1}{\hskip -0.2cm $-1$}
\psfrag{-3}{\hskip -0.0cm $-3$}
\psfrag{0}{$0$}
\psfrag{1}{$1$}
\psfrag{2}{$2$}
\psfrag{-4}{\hskip -0.0cm $-4$}
\psfrag{3}{$3$}
\psfrag{4}{$4$}
\psfrag{5}{$5$}
\psfrag{1.0}{$1.0$}
\psfrag{-1.0}{$-1.0$}
\psfrag{-0.5}{$-0.5$}
\psfrag{-1.5}{$-1.5$}
\psfrag{1.5}{$1.5$}
\psfrag{0.0}{$0.0$}
\psfrag{0.5}{$0.5$}
\psfrag{2.0}{$2.0$}
\psfrag{-2.0}{$-2.0$}
\includegraphics[width=240pt]{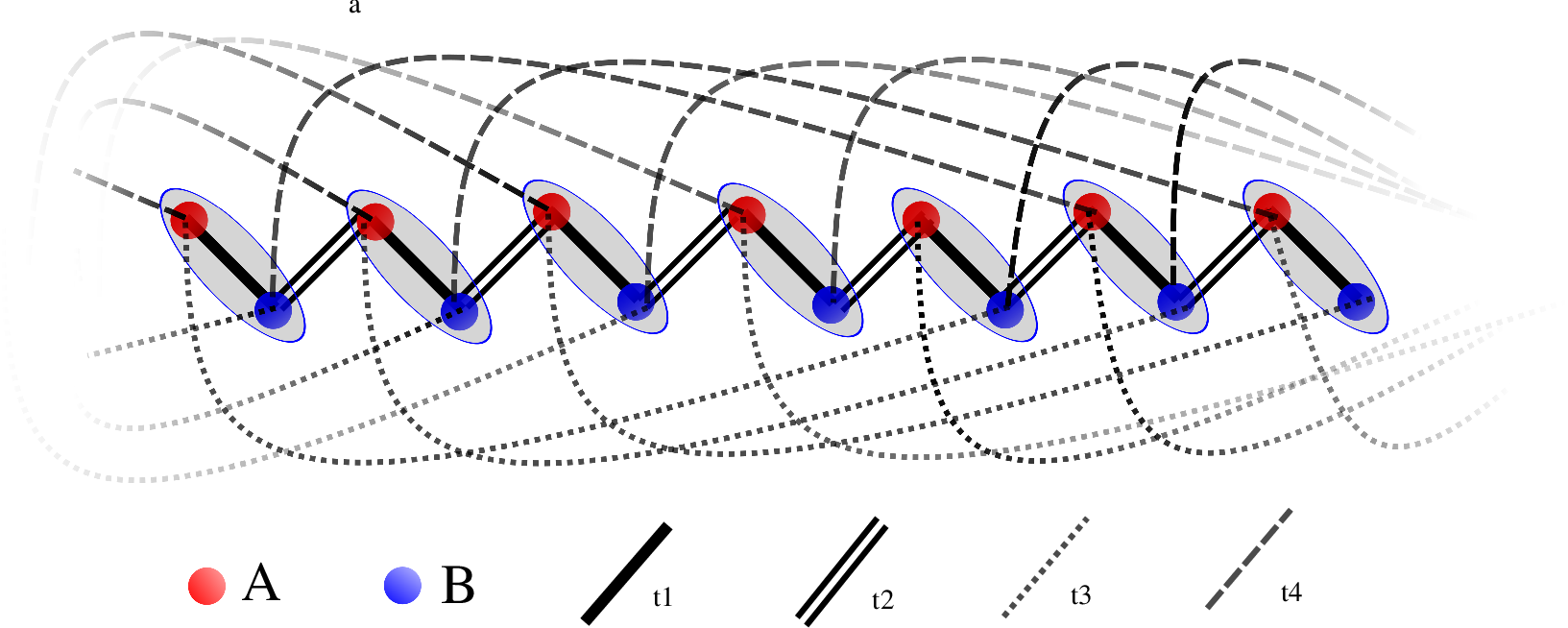}
\vskip 0.2 cm
\includegraphics[width=245 pt,angle=0]{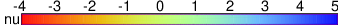}
\vskip -0.45cm
\hskip -.5 cm
\begin{minipage}{0.25\textwidth}
  \includegraphics[width=110pt,angle=-90]{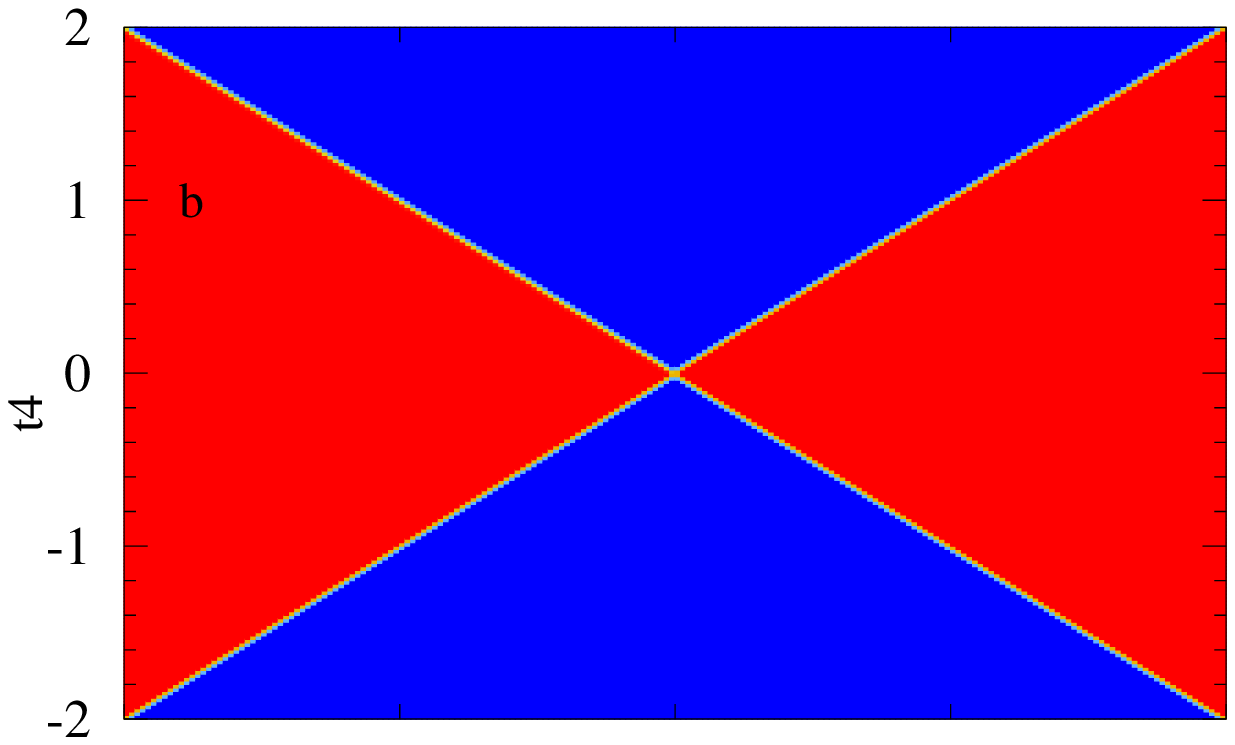}
  \end{minipage}\hskip -0.3cm
  \begin{minipage}{0.25\textwidth}
  \includegraphics[width=110pt,angle=-90]{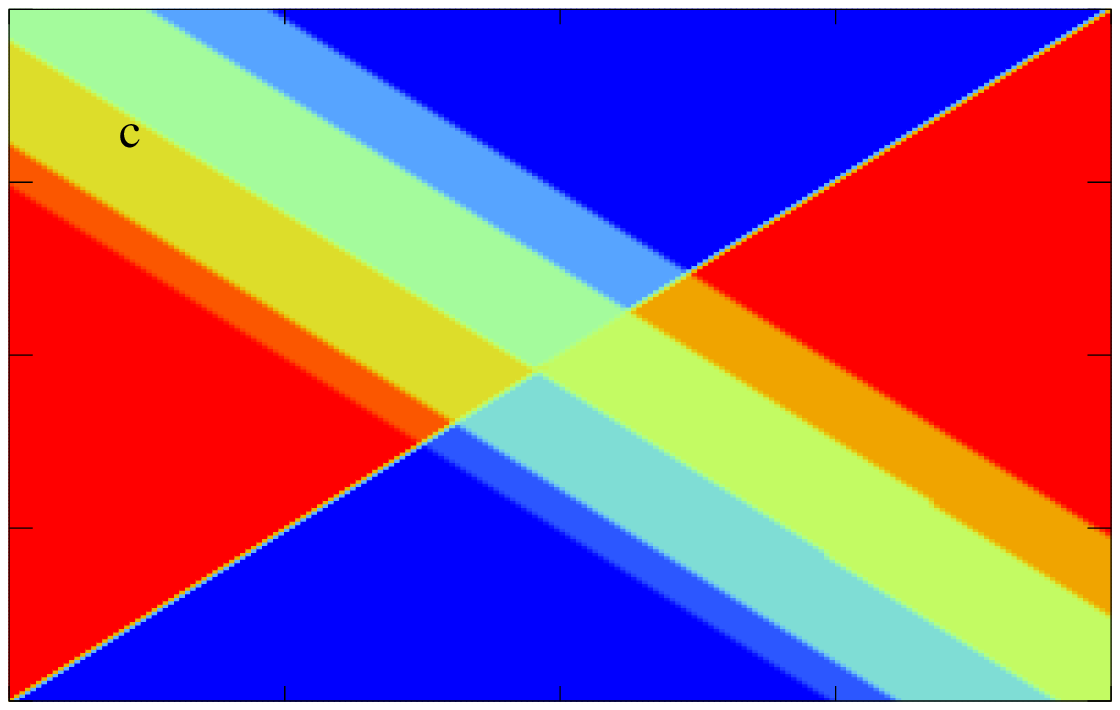}
  \end{minipage}
  \vskip -1.2 cm
  \hskip -0.5 cm
  \begin{minipage}{0.25\textwidth}
  \includegraphics[width=110pt,angle=-90]{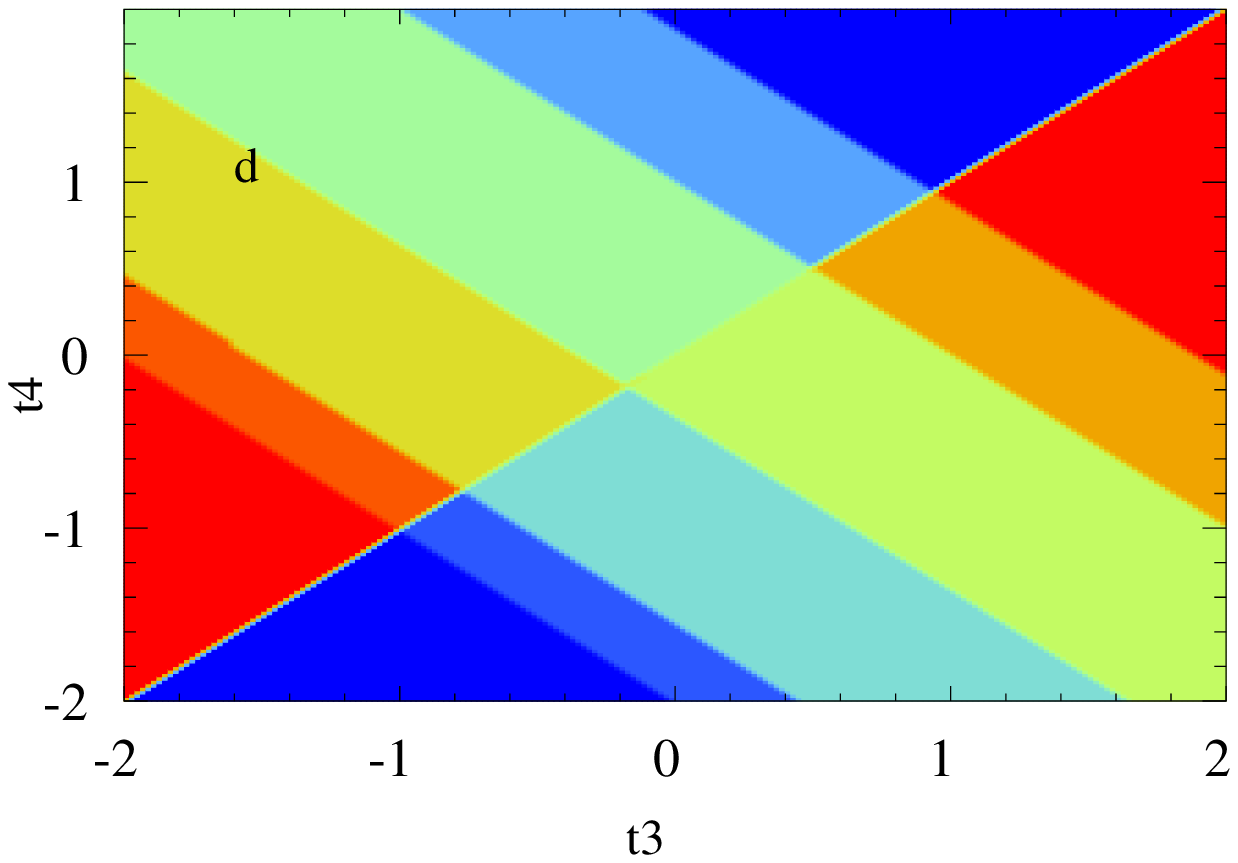}
  \end{minipage}\hskip -0.3cm
  \begin{minipage}{0.25\textwidth}
  \includegraphics[width=110pt,angle=-90]{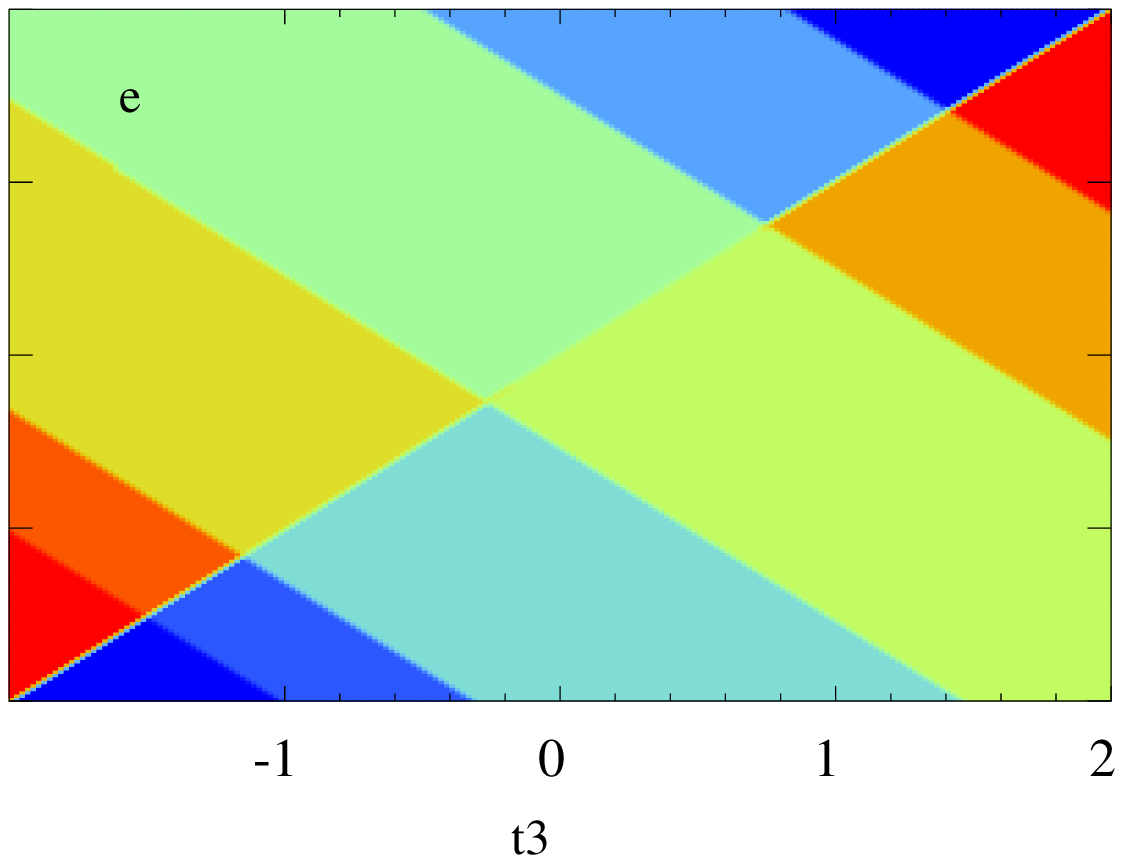}
  \end{minipage}
\caption{(a) Extended SSH model describing the hopping for the
  Hamiltonian, $H_4$. 
  Topological phase diagram of $H_4$ when $t_1=t_2$, for 
  (b) $t_1=0$, (c) $t_1=1/2$, (d) $t_1=1$, and (e) $t_1=3/2$.
Horizontal colorbar indicates the variation of $\nu_4$.}
\label{Extended-SSH-n-4}
\end{figure}

The eSSH model for $n=4$, {\em i.e.}, $H_4$ and its
topological phases for $t_1=0$, $t_1=1/2$, $t_1=1$, and $t_1=3/2$
when $t_1/t_2=1$, have been shown in the 
Figs \ref{Extended-SSH-n-4} (a), (b), (c), (d) and (e), 
respectively. The system hosts nine different
nontrivial phases with $\nu=-4,-3,\cdots,4,5$.
Phases with the largest ($\nu=5$) and
the lowest ($\nu=-4$) winding numbers
appear twice symmetrically around the diagonal line, $t_3=t_4$,
which corresponds to the Eq. \ref{nun}, for $n=4$, when $t_1=t_2=0$.
However, the extent for the phases with $\nu=5$ and
$\nu=-4$ are found to reduce 
as demonstrated in Fig. \ref{Extended-SSH-n-4} (b), (c), (d) and (e),
while the areas for the intermediate phases expand 
with the increase of $t_1$. 
Apart from this fact, phases with odd winding numbers
$\nu=\pm 1,\pm 3$ appear above the diagonal line, while
that with even winding numbers
$\nu=\pm 2,4$ appear below the diagonal line,
which is similar to the case $n=2$.
For the even (odd) value of $n$ trivial phase $\nu=0$
always appears below (above) the diagonal line.
Apart from the diagonal line ($z=-1$),
other phase boundaries can be
given by parallel lines $t_3+t_4=t_1(z^4+z^5)$, where (i) $z=1$
(ii) $z=e^{2\pi i /9}$, (iii) $z=e^{4\pi i /9}$, (iv) $z=e^{6\pi i /9}$,  and
(v) $z=e^{8\pi i /9}$, which can be obtained
from the equation $E_4(k)=0$. 

\bea
\nu_4 \!\!&=&\!\!\!
\begin{cases}
5, & \!\!\! \!\!\! \text{if} \; 
\begin{cases}
t_3 + t_4 < -2t_1, \; \text{and} \; r'<1, \\
t_3 + t_4 > -2t_1\cos{\theta_1}, \; \text{and} \; r'>1,
\end{cases} \\
4, &\!\!\! \!\!\!  \text{if} \; -2t_1 < t_3 + t_4 < -2t_1\cos{\theta_2}, \; \text{and} \; r'<1, \\
3, & \!\!\! \!\!\! \text{if} \; -2t_1\cos{\theta_3} \!<\! t_3 + t_4 \!< \!-2t_1\cos{\theta_1}, \; \text{and} \; r' \!>\! 1, \\
2, & \!\!\! \!\!\! \text{if} \; -2t_1\cos{\theta_2} \!< t_3 + t_4 <\! -2t_1\cos{\theta_4}, \; \text{and} \; r'\! <\! 1, \\
1, & \!\!\! \!\!\! \text{if} \; -2t_1\cos{\theta_4}\! < t_3 + t_4 <\! -2t_1\cos{\theta_3}, \; \text{and} \; r'\! > \!1, \\
0, & \!\!\! \!\!\! \text{if} \; -2t_1\cos{\theta_4}\! < t_3 + t_4 <\!-2t_1\cos{\theta_3},  \; \text{and} \; r'\! <\! 1, \\
-1, &\!\!\! \!\!  \text{if} \; -2t_1\cos{\theta_2}\! < t_3 + t_4 <\!-2t_1\cos{\theta_4}, \; \text{and} \; r' \!>\!1, \\
-2, &\!\!\!  \!\! \text{if} \; -2t_1\cos{\theta_3} \!< t_3 + t_4 < \!-2t_1\cos{\theta_1}, \; \text{and} \; r'\!<\!1, \\
-3, &\!\!\!  \!\! \text{if} \; -2t_1 < t_3 + t_4 < -2t_1\cos{\theta_2}, \; \text{and} \; r'>1, \\
-4, & \!\!\!  \text{if} \; 
\begin{cases}
t_3 + t_4 < -2t_1, \; \text{and} \; r'>1, \\
t_3 + t_4 > -2t_1\cos{\theta_1}, \; \text{and} \; r'<1,
\end{cases}
\end{cases}\nonumber \\
&&{\rm where}\,\, \theta_1=8\pi/9,\, \theta_2=16\pi/9,\,
\theta_3=24\pi/9,\, {\rm and}\nonumber\\
&&\theta_4=32\pi/9.
\label{H-4}
\eea

Regions of different phases have been
identified as given in the Eq \ref{H-4}
for the Hamiltonian $H_4$. The results can be generalized for
arbitrary values of $n$.

\begin{figure}[h]
\psfrag{a}{(a)}
\psfrag{b}{(b)}
\psfrag{A}{\large A}
\psfrag{B}{\large B}
\psfrag{t1}{\large $t_1$}
\psfrag{t2}{\large $t_2$}
\psfrag{t3}{\large $t_3$}
\psfrag{t4}{\large $t_4$}
\psfrag{w}{ $t_3/t_1$}
\psfrag{z}{$t_4/t_1$}
\psfrag{0}{$0$}
\psfrag{1}{$1$}
\psfrag{2}{$2$}
\psfrag{-3}{$-3$}
\psfrag{-2}{$-2$}
\psfrag{-1}{$-1$}
\psfrag{-4}{$-4$}
\psfrag{-5}{$-5$}
\psfrag{3}{$3$}
\psfrag{4}{$4$}
\psfrag{5}{$5$}
\psfrag{1.0}{$1.0$}
\psfrag{-1.0}{$-1.0$}
\psfrag{-0.5}{$-0.5$}
\psfrag{-1.5}{$-1.5$}
\psfrag{1.5}{$1.5$}
\psfrag{0.0}{$0.0$}
\psfrag{0.5}{$0.5$}
\psfrag{2.0}{$2.0$}
\psfrag{-2.0}{$-2.0$}
\includegraphics[width=240pt]{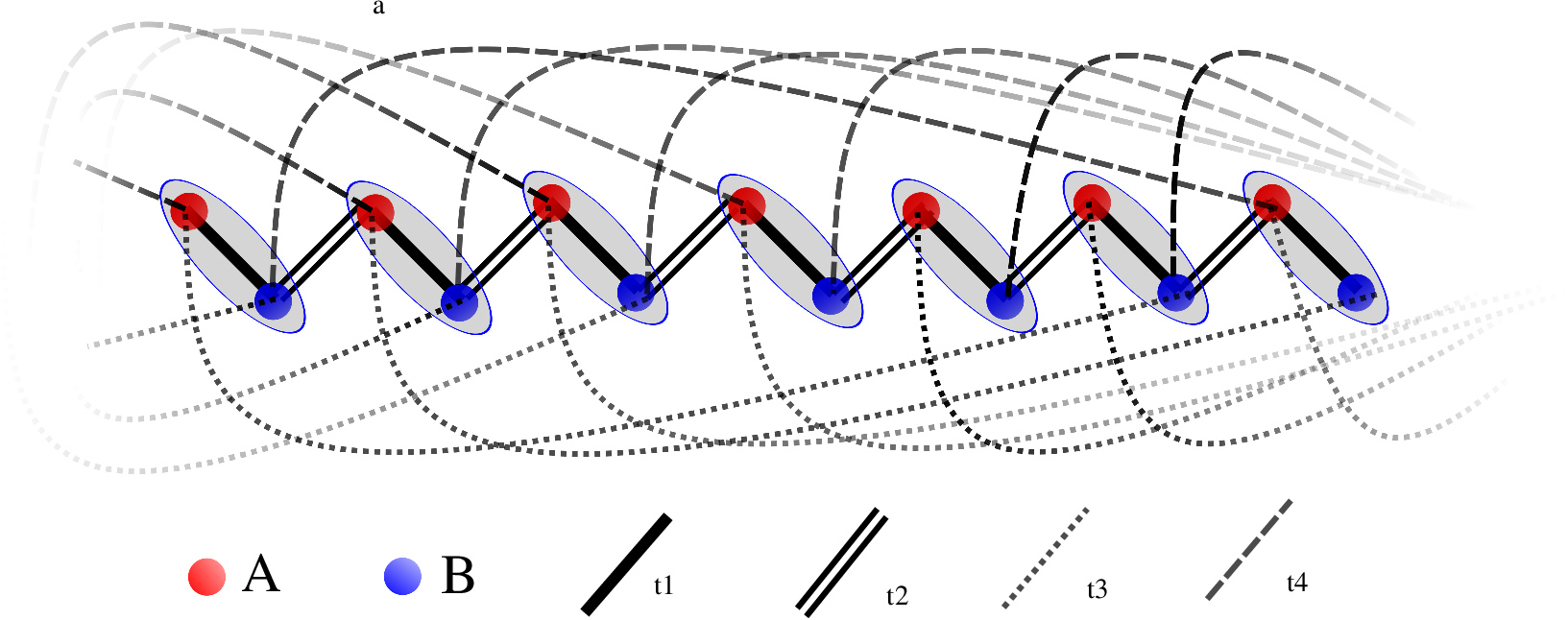}
\vskip -0.3 cm
\includegraphics[width=140pt,angle=-90]{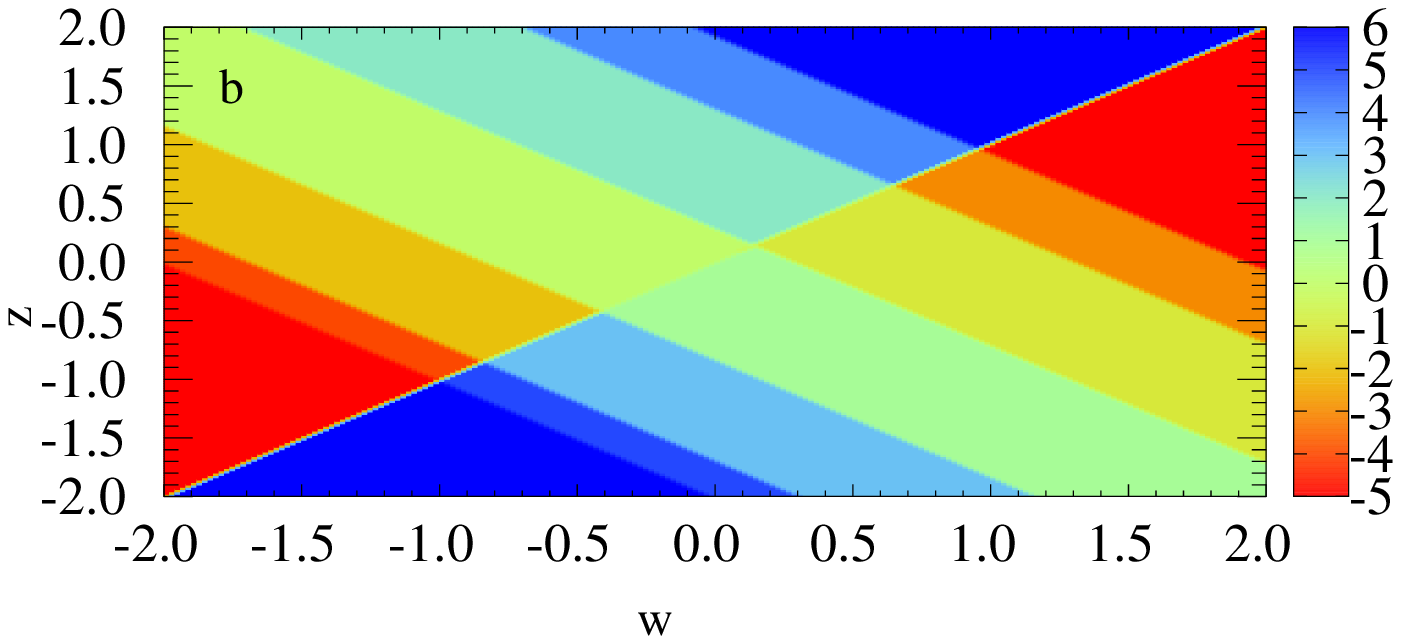}
\caption{(a) Extended SSH model describing the hopping for the
Hamiltonian, $H_5$. (b) Topological phase diagram of $H_5$ when $t_1=t_2$.}
\label{Extended-SSH-n-5}
\end{figure}

Topological phase diagram for $H_5$ have been
shown in Fig. \ref{Extended-SSH-n-5} (b), which hosts eleven different
topological phases, $\nu=-5,-4,\cdots,5,6$.
The relevant model with extended hoppings are shown
in Fig. \ref{Extended-SSH-n-5} (a).
With the increment of $n$ by unity, every time two additional 
topological phases with higher winding numbers
$\nu=n+1$, and $\nu=-n$, appear. 
Locations of the phases for the eSSH model
$H_n$ are given in Eq. \ref{H-n},
however it is valid for odd $n$.
Positions of different phases for even $n$
can be obtained easily from the Eq \ref{H-n}. 

\bea
\nu_n \!\!\!\!&=&\!\!\!\!
\begin{cases}
n+1, & \!\!\! \!\!\! \text{if} \; 
\begin{cases}
t_3 + t_4 < -2t_1, \; \text{and} \; r'<1, \\
t_3 + t_4 > -2t_1\cos{\theta_1}, \; \text{and} \; r'>1,
\end{cases} \\
n, &\!\!\! \!\!\!  \text{if} \; -2t_1 < t_3 + t_4 < -2t_1\cos{\theta_2}, \; \text{and} \; r'<1, \\
n\!-\!1, & \!\!\! \!\!\! \text{if} \; -2t_1\cos{\theta_3} \!<\! t_3 \!+\! t_4 \!< \!-2t_1\cos{\theta_1}, \; \text{and} \; r' \!>\! 1, \\
\quad \vdots &\quad\quad\quad\quad\quad \vdots \\
1, &\! \!\!\! \!\!\! \!\!\! \!\!\!\! \text{if} \; -2t_1\cos{\theta_n}\! < t_3 \!+ \!t_4 <\!-2t_1\cos{\theta_{n\!-\!1}},\,\text{and} \; r'\!<\!1,  \\
0, &\! \!\!\! \!\!\!\!\!\! \!\!\!\!  \text{if} \; -2t_1\cos{\theta_n}\! < t_3 \!+ \!t_4 <\!-2t_1\cos{\theta_{n\!-\!1}},\,\text{and} \; r' \!>\!1,\\
-1, &\! \!\!\! \!\!\!\!\!\! \!\!\! \! \text{if} \; -2t_1\cos{\theta_{n\!-\!2}}\! < t_3 \!+\! t_4 <\!-2t_1\cos{\theta_{n}},\, \text{and} \; r' \!<\!1,\\
\quad\vdots &\quad\quad\quad\quad\quad\vdots\\
-n\!+\!2, &\!\!\!  \!\! \text{if} \; -2t_1\cos{\theta_3} \!< t_3 \!+\! t_4 < \!-2t_1\cos{\theta_1}, \, \text{and} \; r' \!< \!1, \\
-n+1, &\!\!\!  \!\! \text{if} \; -2t_1 < t_3 + t_4 < -2t_1\cos{\theta_2}, \; \text{and} \; r'>1, \\
-n, & \!\!\!  \text{if} \; 
\begin{cases}
t_3 + t_4 < -2t_1, \; \text{and} \; r'>1, \\
t_3 + t_4 > -2t_1\cos{\theta_1}, \; \text{and} \; r'<1,
\end{cases}
\end{cases}\nonumber \\
&&\!\!\!\!\!{\rm where}\,\, \theta_m=2mn\pi/(2n+1),\, {\rm and}\,\,
m=1,2,\cdots,n.
\label{H-n}
\eea

Extending these results it can be concluded
that total number of topological phases for
$H_n$ is $2n+1$, having the winding numbers, 
$\nu_n=n+1,n,n-1,\cdots, -n+1,-n$.
Topological phases with winding number other than these
do not appear for fixed values of $n$.
It shows that value of winding number increases
linearly with $n$. 
Phase boundaries can be identified by the
values $z=\pm 1$ and $z_m=e^{2\pi i m/(2n+1)}$, where
$m=1,2,\cdots,n-1,n$. As a result, total $n+2$
phase boundaries will be there in the phase diagram.
In the $t_3$-$t_4$ parameter space, all the
boundaries are straight lines. Among them
$z=-1$ and $z_m=e^{2\pi i m/(2n+1)}$, are parallel to each other
and at the same time normal to the line $z=1$. 
Positions of the phases with extreme values $\nu=n+1$ and $\nu=-n$ will
remain nearly the same for odd and even values of $n$, while for
$\nu=n,n-1,\cdots, 1,0,1,\cdots, -n+2,-n+1$, 
positions of those phases around the diagonal line will be
different. It means for those particular phases 
$t_3>t_4$ ($t_3<t_4$) for odd $n$ will be
replaced by $t_3<t_4$ ($t_3>t_4$) for even $n$. 
As a result, the trivial phase ($\nu=0$)
along with the nontrivial phases with even values of
winding numbers will appear above (below) the diagonal line
for odd (even) $n$. Moreover,
value of the winding numbers increases with $t_3$ above the
diagonal line, while that of winding numbers decreases with
$t_3$ below the diagonal line. 
In addition, topological phase diagrams for eSSH models with
a single FN hopping term can be reproduced by
equating either $t_3$ or $t_4$ to zero as shown before \cite{Rakesh1}. 
So, this study
extensively demonstrates that topological phase with
any values of winding number can be
generated by introducing suitable FN hopping terms
in the eSSH models and the phase boundaries can be
obtained analytically. In the same way, it can be shown that
Floquet topological phase with any values of winding
numbers can be obtained in the dynamic eSSH models, 
which is described in the following section. 
\section{Floquet engineering topological phases}
\label{Floquet}
Time evolution of a periodically driven system
$H(t)=H(t+T)$
can be studied quantum mechanically in terms of
time-dependent Schr\"odinger equation,
\be
i\partial_t |\Psi(t)\rangle=H(t)|\Psi(t)\rangle,
\ee
with $\hbar=1$. According to the Floquet's theorem, the general solution
can be expressed as \cite{Floquet,Bloch} 
\be
|\Psi(t)\rangle=\sum_{p}c_p|\psi_p(t)\rangle,
\ee
with the coefficients $c_p$, which are independent of time and where
\be
|\psi_p(t)\rangle=e^{-iE_p t}|u_p(t)\rangle.
\ee
Here, $E_p$ is the quasienergy of $p$-th Floquet band and
the Floquet mode $|u_p(t)\rangle$ is again time periodic,
$|u_p(t)\rangle=|u_p(t+T)\rangle $. Quasienergy can be restricted
within the first Floquet zone, $-\omega/2 \le E_p \le \omega/2$,
where $\omega=2\pi/T$ is the frequency.
The Floquet states satisfy the Floquet equation,
\be
(H(t)-i\partial_t )|u_p(t)\rangle=E_p|u_p(t)\rangle.
\ee
Now the matrix elements of the Floquet Hamiltonian, $H_F(t)=H(t)-i\partial_t$
can be given in the Fourier space as
\be
H_F^{pq}=\frac{1}{T}\int_0^TH(t) e^{i(p-q)\omega t}dt -p\omega \delta_{pq},
\label{H-F}
\ee
where $p,q$ are integers.

\begin{figure}[h]
\psfrag{a}{$t_1$}
\psfrag{b}{$t_2$}
\psfrag{c}{$t_3$}
\psfrag{d}{$t_4$}
\psfrag{e}{$t_2'$}
\psfrag{f}{$t_1'$}
\psfrag{g}{$t_4'$}
\psfrag{h}{$t_3'$}
\psfrag{H1}{$H_{11}$}
\psfrag{H2}{$H_{12}$}
\psfrag{t1}{$t=mT$}
\psfrag{t2}{$t=t_p\!+\!mT$}
\includegraphics[width=240pt]{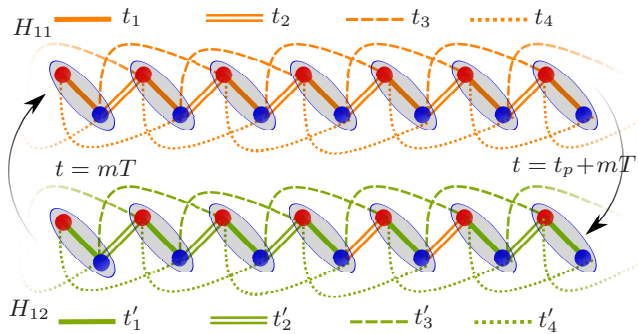}
\caption{Thematic sketch of the periodically driven
  eSSH model over a complete driving protocol.
  In the beginning of the cycle, driving system is described by the Hamiltonian
  $H_{11}$, while after time $t_p$ system is described by
  the Hamiltonian $H_{12}$. The same quenching sequence repeats
periodically with the period $T$.}
\label{Extended-SSH-n-floquet}
\end{figure}

In order to study the Floquet evolution of this class of
eSSH models under piecewise periodic quenching, time-dependent
Hamiltonian $H_n(t)$ has been expressed as
\be
H_n(t)=\left\{\begin{array}{ll}
H_{n1}, & {\rm if}\;mT\le t <t_p+mT,\\[0.3em]
H_{n2},  & {\rm if}\;t_p+mT\le t <(1+m)T,
\end{array}\right.
\label{Ht}
\ee
where $T$ is the driving time period, and
it is split into two arbitrary time intervals $t_p$ and
$T-t_p$, in which $m$ is integers.
The Hamiltonians in two different time intervals
are given by
\bea
H_{nj}\!\!\!\!&=&\!\!\!(\alpha t_1)^{\delta_{j,1}}(\beta t_2')^{\delta_{j,2}}\!\sum_{i=1}^N a_i^\dag b_i
\!+\!(\alpha t_1')^{\delta_{j,2}}(\beta t_2)^{\delta_{j,1}}\!\sum_{i=1}^N a_{i+1}^\dag b_i\nonumber\\
\!\!\!&\!\!\!\!\!\!\!\!\!\!\!\!\!\!\!\!\!\!+&\!\!\!\!\!\!\!\!\!\!\!\!\!
(\gamma t_3)^{\delta_{j,1}}(\delta t_4')^{\delta_{j,2}}\!\sum_{i=1}^N a_i^\dag b_{i+n}
\!+\!(\gamma t_3')^{\delta_{j,2}}(\delta t_4)^{\delta_{j,1}}\!\sum_{i=1}^N a_{i+n+1}^\dag b_i\nonumber \\
&&\qquad\qquad +\,h.c.
\label{H-nj}
\eea
A sketch of this driving protocol is depicted in
Fig. \ref{Extended-SSH-n-floquet}, where the
Hamiltonian with $n=1$ is shown. In the beginning of
the cycle system is described by the Hamiltonian
  $H_{11}$, while after time $t_p$ system is described by
  the Hamiltonian $H_{12}$. The same quenching sequence repeats
  periodically with the period $T$. Similar quenching dynamics
  does hold for higher values of $n$.
Here $t_j,j=1,2,3,4$ are the hopping parameters for the first segment 
of the driving cycle, while $t_j'$ are the same for
the second segment of the cycle. 
Four different sets of parameters have been chosen thereafter
depending on the staggeredness of hopping parameters.
Four additional parameters, say, $\alpha$, $\beta$, $\gamma$ and
$\delta$ are further introduced where $\alpha$ and $\beta$ modify the
NN hopping while $\gamma$ and $\delta$ alter the FN hopping parameters. 

Now substituting Eq \ref{Ht} into Eq \ref{H-F},
elements of Floquet Hamiltonian can be obtained as
\be
H_{nF}^{pq}=\frac{1}{T}\!\left[\int_0^{t_p}\!\!H_{n1} e^{i(p-q)\omega t}dt
+\!\int_{t_p}^T\!\!H_{n2} e^{i(p-q)\omega t}dt\right] -p\omega \delta_{pq},
\nonumber
\ee
where diagonal and off-diagonal matrix elements can be written as 
\be\left\{\begin{array}{l}
 H_{nF}^{pp}=\frac{1}{T}\left[H_{n1}t_p
\!+\!H_{n2} (T-t_p)\right] -p\omega,\\[0.3em]
H_{nF}^{pq}= \frac{-i}{2\pi(p-q)}\left[e^{i(p-q)\omega t}-1 \right](H_{n1}-H_{n2}).
\end{array}\right.
\label{H-F-pq}\ee
The matrix representation of the Floquet Hamiltonian can be
expressed in terms of an infinite dimensional matrix as
\be
H_{nF}=\left[\begin{array}{ccccc}
    \ddots &\vdots&\vdots&\vdots&\iddots\\
  \cdots&  H_n^{-1,-1}&H_n^{0,-1}&H_n^{1,-1}&\cdots\\[0.3em]
   \cdots& H_n^{-1,0}&H_n^{0,0}&H_n^{1,0}&\cdots\\[0.3em]
   \cdots&  H_n^{-1,1}&H_n^{0,1}&H_n^{1,1}&\cdots\\[0.3em]
    \iddots &\vdots&\vdots&\vdots&\ddots
\end{array}\right].
\label{H-nF}\ee
However in order to obtain the quasienergy spectrum
numerically dimension of the Floquet matrix, (Eq \ref{H-nF}) will be truncated. 
Symmetry of the
Floquet matrix as shown in Eq \ref{H-nF} helps to
keep the first Floquet zone in the middle of the
quasienergy spectrum. 
The actual dimension of the
matrix would be determined by the maximum values of $p$, $q$ and
the number of sites, $N$.
In Eq \ref{H-nF}, the matrix is shown for $p(q)=0,\pm 1$. 
Quasienergy spectrum for the Floquet Hamiltonian can be obtained
by diagonalizing the $H_{nF}$ ,
for fixed value of $p$ and $N$ with and without imposing the PBC
on $H_{nF}^{pq}$. However, quasienergy edge states
would be observed in the spectrum only without PBC.
According to the Floquet theorem, quasienergy band diagram
exhibits periodicity along the
energy axis. Hence it is sufficient to draw the
energy band within the first Floquet zone, $-\omega/2 \le E \le \omega/2$.

Due to the presence of translational symmetry both in the
Hamiltonians $H_{n1}$ and $H_{n2}$, they can be transformed to
Bloch Hamiltonians $H_{nj}(k)$ by imposing PBC, while
expressing in terms of the
Bloch wave vector $k$ in the
momentum space as 
\be
H_n(k,t)=\left\{\begin{array}{ll}
H_{n1}(k), & {\rm if}\;mT\le t <t_p+mT,\\[0.3em]
H_{n2}(k),  & {\rm if}\;t_p+mT\le t <(1+m)T.
\end{array}\right.
\label{Hkt}
\ee
The staggered Hamiltonian look like
\be
H_{nj}(k)=h_{njx}(k)\sigma_x+h_{njy}(k)\sigma_y,
\ee
where
\be \left\{\begin{array}{l}
h_{n1x}= \alpha t_1 + \beta t_2 \cos{(k)}+\gamma t_3\cos{(nk)}
+\delta t_4\cos{((n+1)k)},\\[0.3em]
h_{n1y}= \beta t_2 \sin{(k)}-\gamma t_3\sin{(nk)}
+\delta t_4\sin{((n+1)k)},\\[0.3em]
h_{n2x}= \beta t_1' + \alpha t_2' \cos{(k)}+\delta t_3\cos{(nk)}
+\gamma t_4\cos{((n+1)k)},\\[0.3em]
h_{n2y}= \beta t_2' \sin{(k)}-\delta t_3\sin{(nk)}
+\gamma t_4\sin{((n+1)k)},
\end{array}\right.
\label{hn12xy}\ee
and $\sigma_x$ and $\sigma_y$ are the Pauli matrices.
Floquet Hamiltonian in the momentum space can be
constructed like the Eq \ref{H-nF}, by substituting
$H_{nj}(k)$.

In order to study the 
evolution of the system over a complete
cycle the Floquet operator has been introduced as 
\bea
U_n(T)&=&\mathcal T e^{-i\int_{0}^{T}dt H_n(t)},\nonumber\\[0.3em]
&=&e^{iH_{n2}(T-t_p)}\,e^{-iH_{n1}t_p},
\eea
where $\mathcal T$ is the time  ordering operator. 
For the translationally invariant bulk,
Floquet operator becomes
\be
U_n(T,k)=u_{n2}(k)u_{n1}(k),
\ee
where $u_{n1}(k,t_p)=e^{-iH_{n1}(k)t_p}$, and
$u_{n2}(k,T-t_p)=e^{-iH_{n2}(k)(T-t_p)}$.  
It is noteworthy that Floquet operator $U_n(T,k)$, defined in this
way does not preserve the chiral symmetry. 
However, two different Floquet operators, say
$U_{n1}(T,k)=\sqrt{u_{n1}}u_{n2}\sqrt{u_{n1}}$, and
$U_{n2}(T,k)=\sqrt{u_{n2}}u_{n1}\sqrt{u_{n2}}$ can be constructed
out of $U_n(T,k)$ by two different similarity transformations \cite{Zhou2}. 
As they are related by the similarity transformations,
all the Floquet operators, $U_n(T,k)$, $U_{n1}(T,k)$, $U_{n2}(T,k)$,  
must have the common quasienergy spectra. 
It can be shown that both $U_{n1}(T,k)$ and $U_{n2}(T,k)$
preserve the chiral symmetry.
For this purpose, two different time segments
have been expressed as $\tau_1=t_p/2$, and $\tau_2=(T-t_p)/2$.
As a result, $U_{n1}(T,k)=u_{n1}(\tau_1)u_{n2}(2\tau_2)u_{n1}(\tau_1)$, 
and $U_{n2}(T,k)=u_{n2}(\tau_2)u_{n1}(2\tau_1)u_{n2}(\tau_2)$. 

\begin{figure}[h]
  \psfrag{wz}{\hskip -0.15 cm $w_{10}$}
  \psfrag{wp}{\hskip -0.15 cm $w_{1\pi}$}
  \psfrag{00}{\hskip 0.14 cm $0.0$}
  \psfrag{2.0}{\hskip -0.2 cm $2.0$}
  \psfrag{1.5}{\hskip -0.2 cm $1.5$}
  \psfrag{3}{\hskip -0.15 cm $3$}
  \psfrag{2}{\hskip -0.15 cm $2$}
  \psfrag{1}{\hskip -0.15 cm $1$}
  \psfrag{-1}{\hskip -0.25 cm $-1$}
  \psfrag{0}{\hskip -0.15 cm $0$}
\psfrag{1.0}{\hskip -0.25 cm $1.0$}
\psfrag{0.0}{\hskip -0.18 cm $0.0$}
\psfrag{0.5}{\hskip -0.25 cm $0.5$}
\psfrag{0.50}{\hskip 0.1 cm $0.5$}
\psfrag{-2}{\hskip -0.25 cm $-2$}
\psfrag{0.25}{\hskip 0.1 cm $0.25$}
\psfrag{-0.5}{\hskip -0.43 cm $-0.5$}
\psfrag{-1.0}{\hskip -0.43 cm $-1.0$}
\psfrag{y}{\large k}
\psfrag{E}{\hskip -0.5 cm Energy}
\psfrag{a}{\hskip -0.06 cm (a)}
\psfrag{b}{\hskip -0.06 cm (b)}
\psfrag{c}{\hskip -0.06 cm (c)}
\psfrag{d}{\hskip -0.06 cm (d)}
\psfrag{e}{\hskip -0.06 cm (e)}
\psfrag{f}{\hskip -0.06 cm (f)}
\psfrag{g}{\hskip -0.06 cm (g)}
\psfrag{h}{\hskip -0.06 cm (h)}
\psfrag{z}{$ET/\pi$}
\psfrag{I}{\hskip 0.35 cm $I_{pr}$}
\psfrag{t}{ $\theta/\pi$}
\includegraphics[width=245 pt,angle=0]{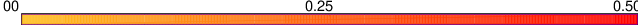}
\vskip -0.05cm
\hskip -1.0 cm
\begin{minipage}{0.20\textwidth}
  \includegraphics[width=134pt,angle=-90]{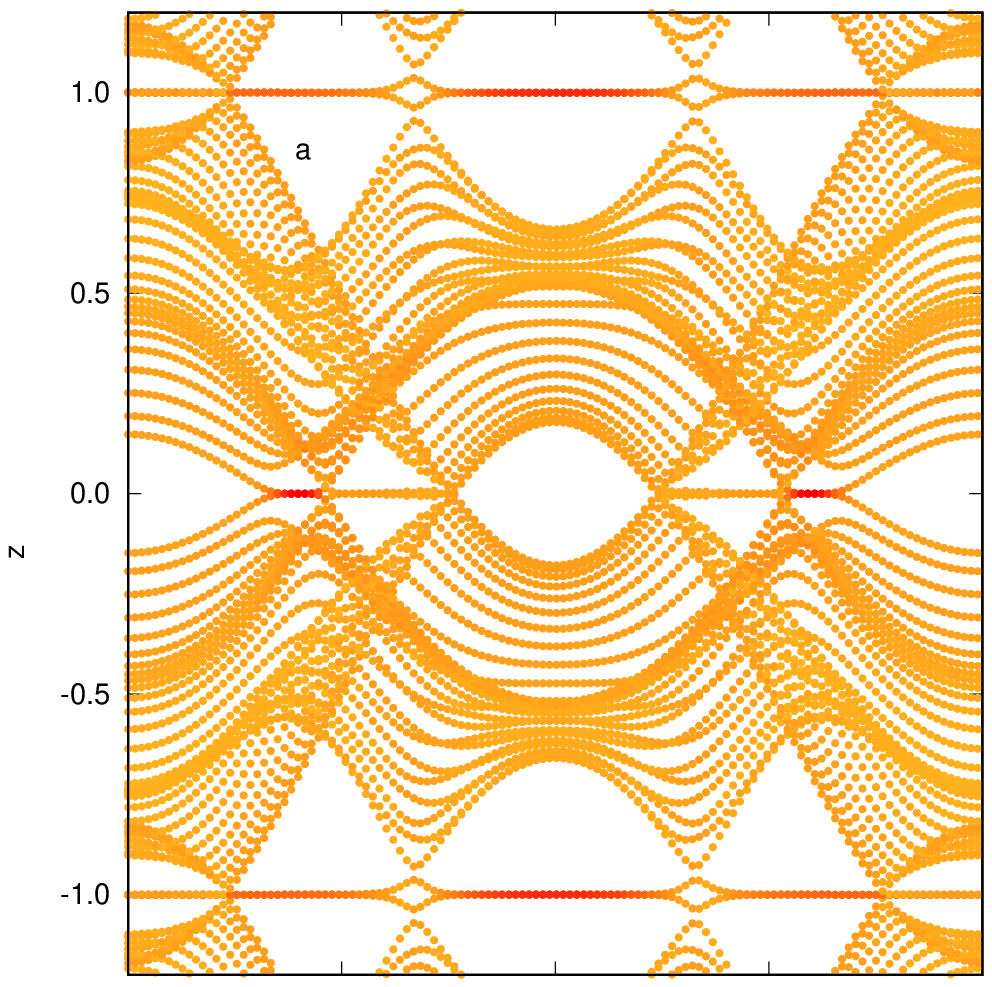}
  \end{minipage}\hskip 0.8cm
  \begin{minipage}{0.2\textwidth}
  \includegraphics[width=134pt,angle=-90]{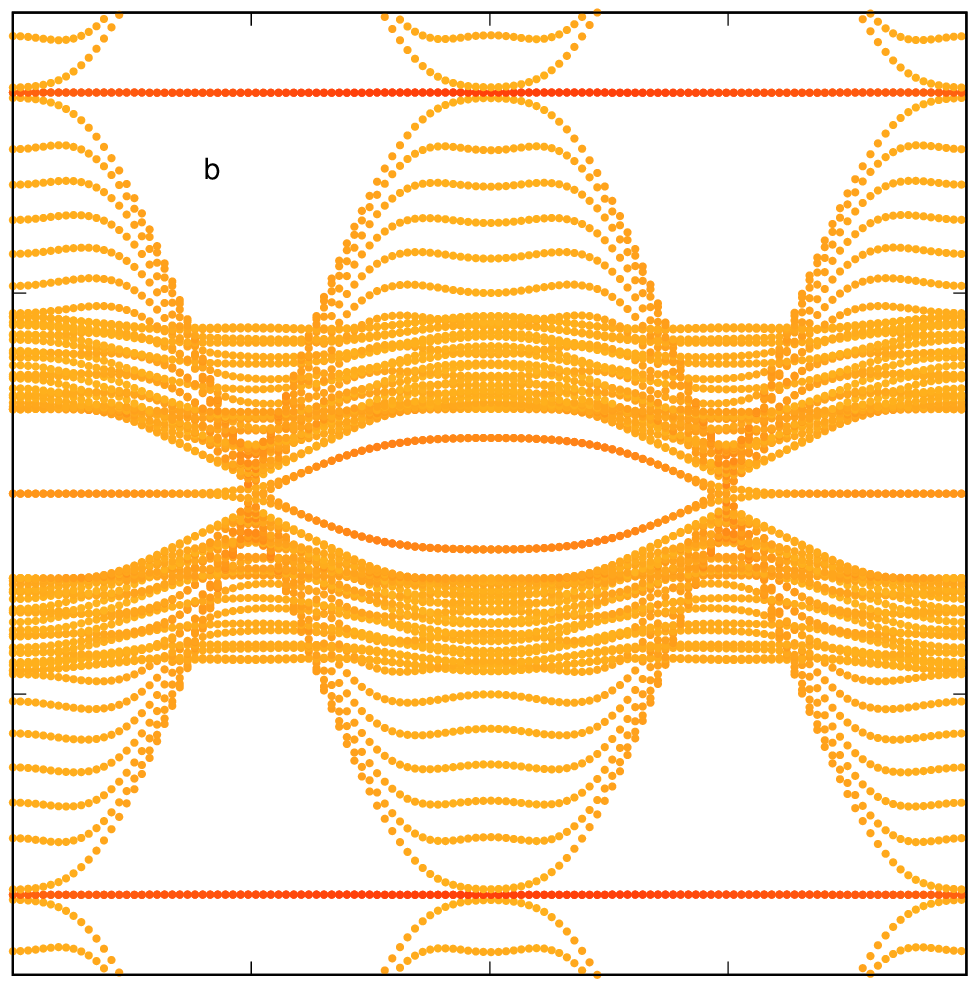}
  \end{minipage}
  \vskip -0.3 cm
  \hskip -1.0 cm
  \begin{minipage}{0.20\textwidth}
  \includegraphics[width=134pt,angle=-90]{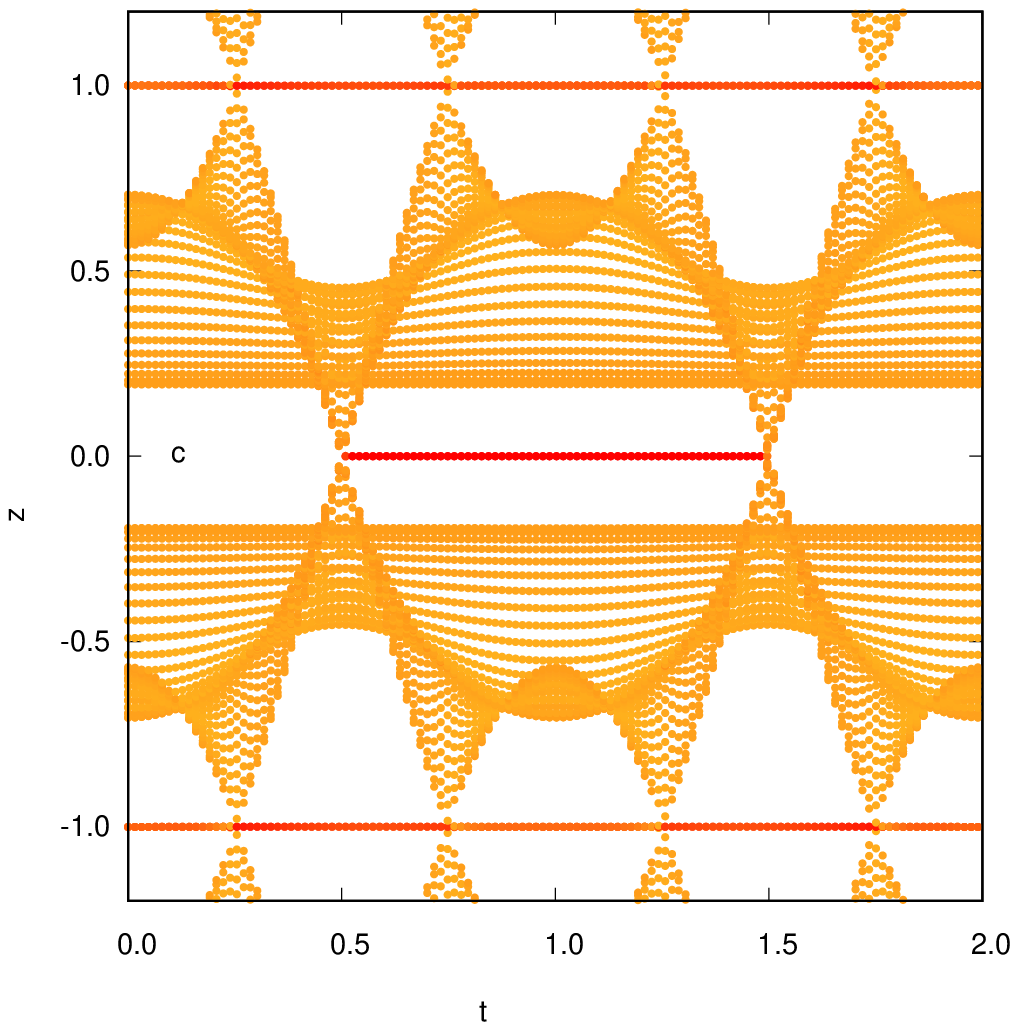}
  \end{minipage}\hskip 0.8cm
  \begin{minipage}{0.2\textwidth}
  \includegraphics[width=134pt,angle=-90]{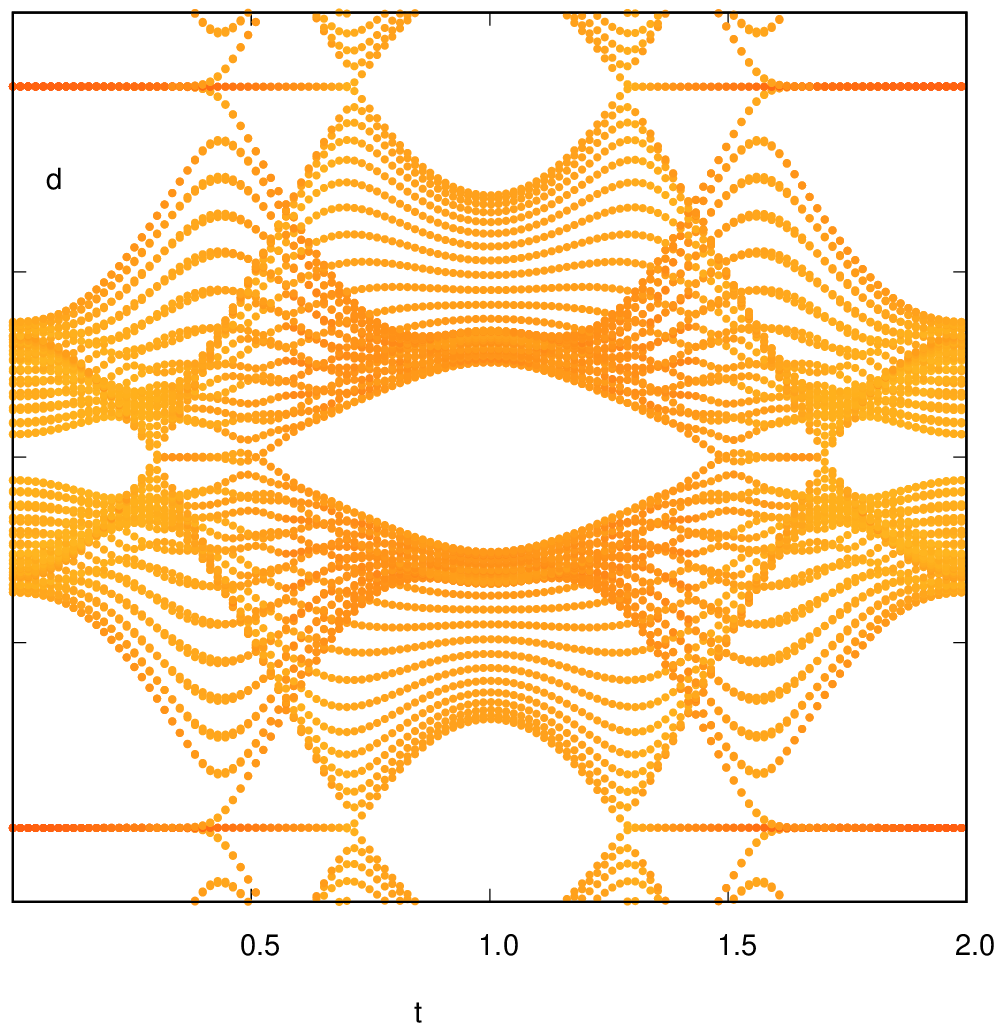}
  \end{minipage}
  \vskip 0.1 cm
   \hskip 0.04 cm
  \begin{minipage}{0.23\textwidth}\hskip 0.15cm
  \includegraphics[width=118pt,angle=0]{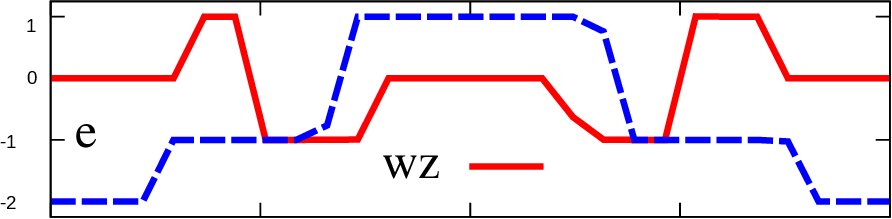}
  \end{minipage}\hskip 0.31cm
  \begin{minipage}{0.23\textwidth}
  \vskip 0.0cm
  \includegraphics[width=118pt,angle=0]{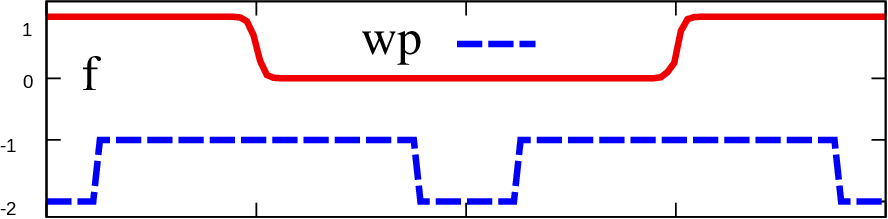}
  \end{minipage}
   \vskip 0.02 cm
   \hskip 0.04 cm
  \begin{minipage}{0.23\textwidth}\hskip 0.15cm
  \includegraphics[width=118pt,angle=0]{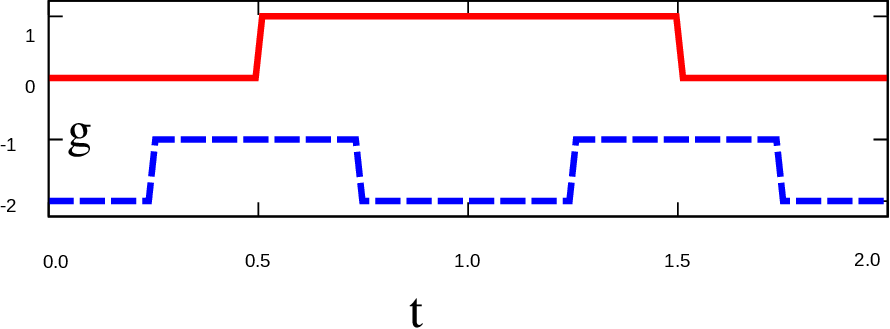}
  \end{minipage}\hskip 0.33cm
  \begin{minipage}{0.23\textwidth}
  \vskip 0.02cm
  \includegraphics[width=118pt,angle=0]{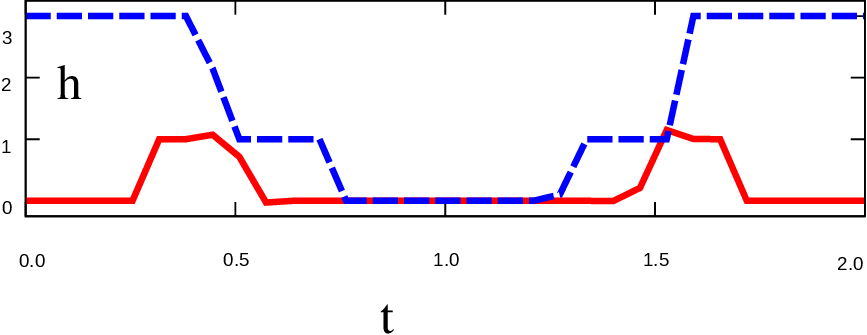}
  \end{minipage}
  \caption{Quasienergy spectra for open systems when $n=1$ are shown in (a) and (b)
    with respect to $\theta$. Variation of winding numbers with respect to
    $\theta$  is shown in (c) and (d). Edge states are indicated by the higher value of
  $I_{pr}$. }
\label{edge-states-floquet-m-1}
\end{figure}

\begin{figure}[h]
  \psfrag{00}{\hskip 0.14 cm $0.0$}
  \psfrag{2.0}{\hskip -0.2 cm $2.0$}
  \psfrag{1.5}{\hskip -0.2 cm $1.5$}
  \psfrag{3}{\hskip -0.15 cm $3$}
  \psfrag{2}{\hskip -0.15 cm $2$}
  \psfrag{1}{\hskip -0.15 cm $1$}
  \psfrag{-1}{\hskip -0.25 cm $-1$}
  \psfrag{0}{{\scriptsize $0$}}
  \psfrag{10}{{\scriptsize  $10$}}
  \psfrag{20}{{\scriptsize $20$}}
  \psfrag{30}{{\scriptsize $30$}}
  \psfrag{40}{{\scriptsize $40$}}
  \psfrag{50}{{\scriptsize $50$}}
  \psfrag{60}{{\scriptsize $60$}}
  \psfrag{70}{{\scriptsize $70$}}
  \psfrag{80}{\hskip -0.05 cm{\scriptsize $80$}}
\psfrag{1.0}{\hskip -0.15 cm {\scriptsize $1.0$}}
\psfrag{0.0}{\hskip -0.15 cm {\scriptsize $0.0$}}
\psfrag{0.2}{\hskip -0.15 cm {\scriptsize $0.2$}}
\psfrag{0.5}{\hskip -0.15 cm {\scriptsize $0.5$}}
\psfrag{0.50}{\hskip 0.1 cm $0.5$}
\psfrag{-2}{\hskip -0.25 cm $-2$}
\psfrag{0.25}{\hskip 0.1 cm $0.25$}
\psfrag{-0.5}{\hskip -0.33 cm {\scriptsize $-0.5$}}
\psfrag{-1.0}{\hskip -0.33 cm {\scriptsize $-1.0$}}
\psfrag{y}{\large k}
\psfrag{E}{\hskip -0.5 cm Energy}
\psfrag{a}{\hskip -0.06 cm (a)}
\psfrag{b}{\hskip -0.06 cm (b)}
\psfrag{c}{\hskip -0.06 cm (c)}
\psfrag{d}{\hskip -0.06 cm (d)}
\psfrag{e}{\hskip -0.06 cm (e)}
\psfrag{f}{\hskip -0.06 cm (f)}
\psfrag{g}{\hskip -0.06 cm (g)}
\psfrag{h}{\hskip -0.06 cm (h)}
\psfrag{z}{{\scriptsize $ET/\pi$}}
\psfrag{p}{{\scriptsize $|\psi|^2$}}
\psfrag{ze}{{\scriptsize $0$-energy}}
\psfrag{pi}{{\scriptsize $\pi$-energy}}
\psfrag{I}{\hskip 0.35 cm $I_{pr}$}
\psfrag{t}{ $\theta/\pi$}
\psfrag{s}{sites}
\includegraphics[width=245 pt,angle=0]{top-color-box.eps}
\vskip -0.05cm
\hskip -1.0 cm
\begin{minipage}{0.20\textwidth}
  \includegraphics[width=46pt,angle=-90]{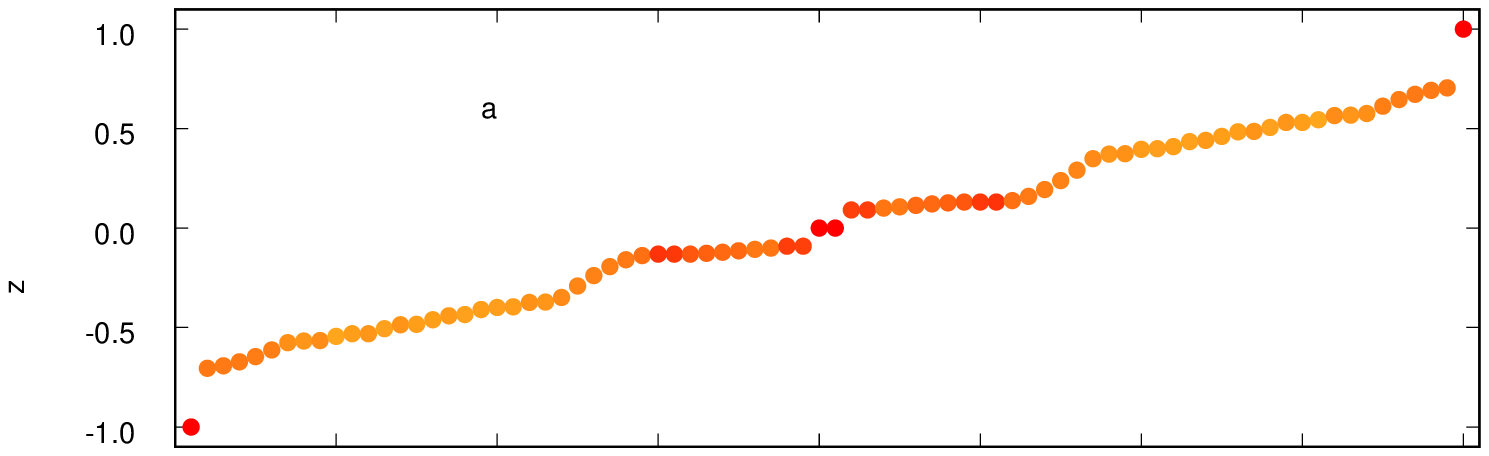}
  \end{minipage}\hskip 0.8cm
  \begin{minipage}{0.20\textwidth}
  \includegraphics[width=46pt,angle=-90]{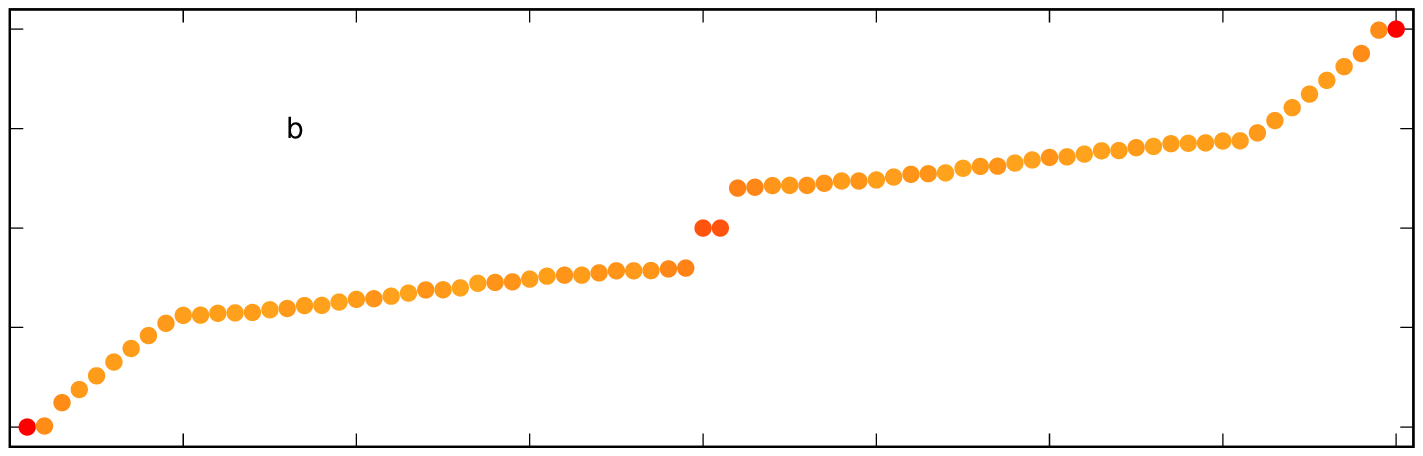}
  \end{minipage}
  \vskip -0.2 cm
  \hskip 0.1 cm
  \begin{minipage}{0.20\textwidth}\hskip -1.2 cm
  \includegraphics[width=46pt,angle=-90]{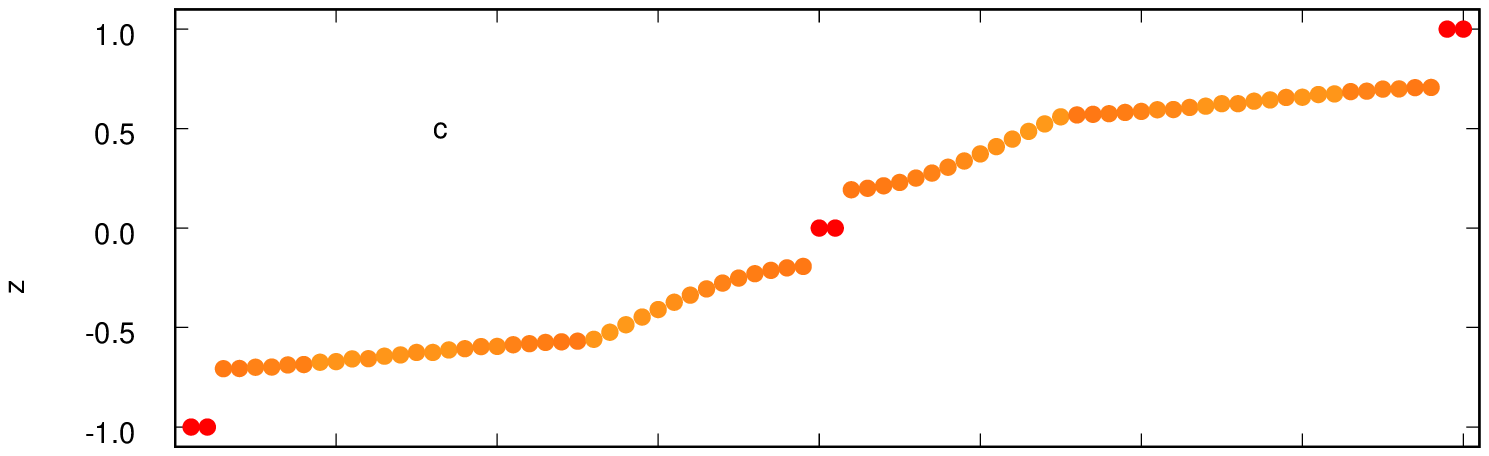}
  \end{minipage}\hskip -0.3cm
  \begin{minipage}{0.20\textwidth}
     \vskip -0.4cm
     \hskip -0.0 cm
  \includegraphics[width=46pt,angle=-90]{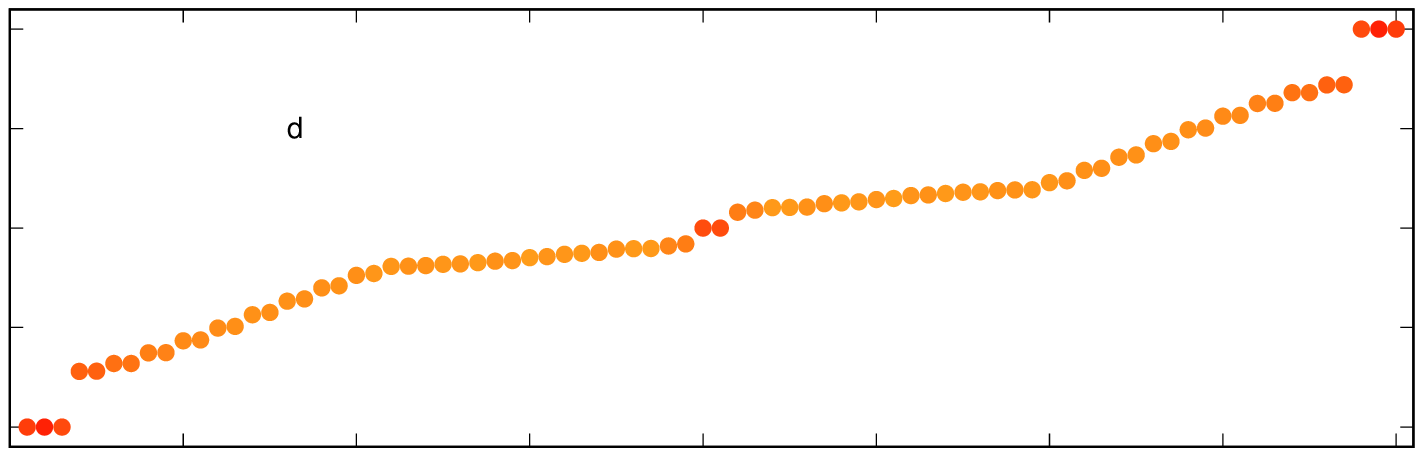}
  \end{minipage}
  \vskip -0.2 cm
   \hskip -0.3 cm
  \begin{minipage}{0.24\textwidth}\hskip -0.1 cm
  \includegraphics[width=128 pt,angle=0]{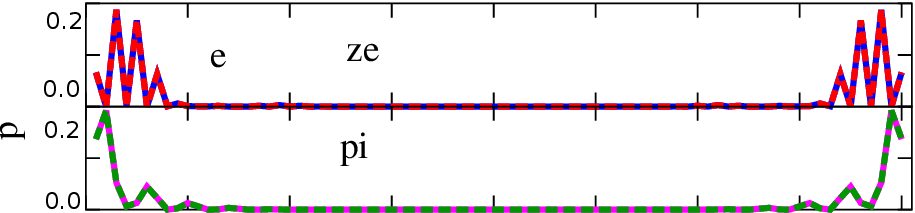}
  \end{minipage}\hskip 0.22cm
  \begin{minipage}{0.23\textwidth}
  \vskip 0.1cm
  \includegraphics[width=122pt,angle=0]{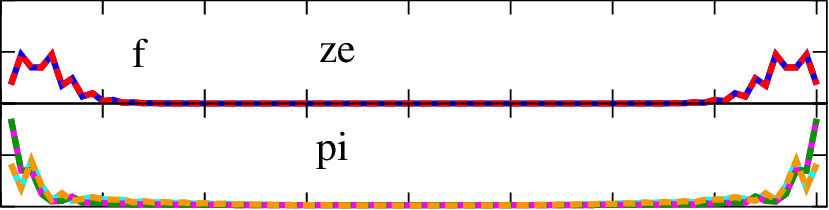}
  \end{minipage}
   \vskip -0.0 cm
   \hskip -0.31 cm
  \begin{minipage}{0.23\textwidth}\hskip 0.0cm
  \includegraphics[width=130pt,angle=0]{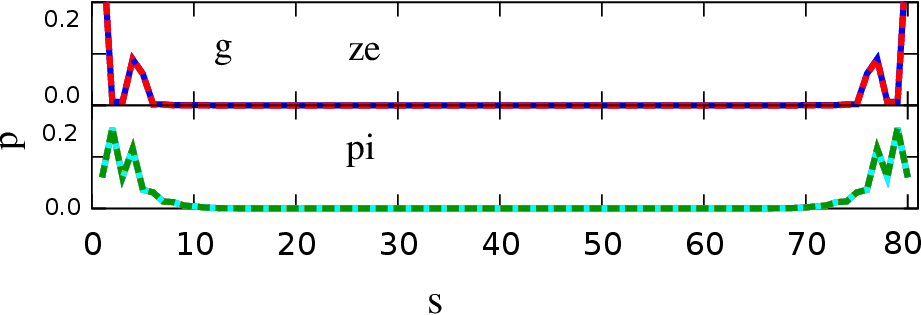}
  \end{minipage}\hskip 0.5cm
  \begin{minipage}{0.23\textwidth}
  \vskip 0.0cm
  \includegraphics[width=120pt,angle=0]{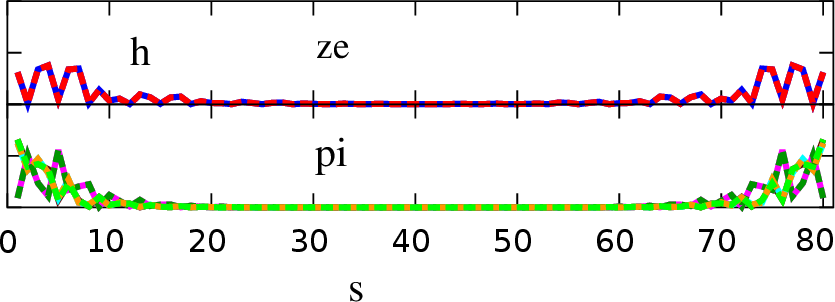}
  \end{minipage}
  \caption{Quasienergy spectra for open systems when $n=1$ are shown in (a) and (b)
    with respect to $\theta$. Variation of winding numbers with respect to
    $\theta$  is shown in (c) and (d). Edge states are indicated by the higher value of
  $I_{pr}$. }
\label{edge-states-floquet-fixed-q-m-1}
\end{figure}

In order to preserve the chiral symmetry, the unitary Floquet
operators, 
$U_{nj}(T,k),j=1,2$, must satisfy the relation
$\Gamma_j U_{nj}(T,k) \Gamma_j=U_{nj}(T,k)^\dag$, with 
the chiral operator $\Gamma_j$ such that $\Gamma_j^2=\mathcal I$, 
and $\mathcal I$ is a $2\times 2$ unit matrix.
 For the simplification of $U_{nj}(T,k)$, the $2\times 2$ unitary
operator, $u_{nj}(k,t)=e^{-iH_{nj}(k)t}$, has been
expanded as
\[e^{-iH_{nj}(k)t}=\cos{(E_{nj} t)}\mathcal I-i\sin{(E_{nj} t)} \boldsymbol {\hat h}_{nj}
\cdot \boldsymbol \sigma,\]
where $E_{nj}=h_{nj}$, $\hat h_{nj}={\boldsymbol h}_{nj}/ h_{nj}$, 
and ${\boldsymbol h}_{nj}(k)=h_{njx}(k)\hat x+h_{njy}(k)\hat y$.
Ultimately multiplying the three $2\times 2$ unitary matrices 
successively, it can be shown that
\be
U_{nj}(T,k)=g_{nj0}\mathcal I +i g_{njx}\sigma_x +i g_{njy}\sigma_y,
\ee
where
\be \left\{\begin{array}{l}
g_{nj0}=X'_{nj}\left(X_{nj}^2-Y_{nj}^2\right)
-2X_{nj}Y_{nj}Y'_{nj}\left(\hat n_1\cdot\hat n_2\right),\\[0.3em]
g_{nj\alpha}=Y_{nj}^2Y'_{nj}\left[\left(\hat n_1\cdot\hat n_2\right)n_{j\alpha}
\!+\!(-1)^{\delta_{j2}}\epsilon_{\alpha\beta z}n_{j\beta}
\left(\hat n_1\!\times\!\hat n_2\right)_z\right]\\[0.3em]
\quad\quad\quad\;\;
-X_{nj}^2Y'_{nj}\left(n_{2\alpha}\delta_{j1}+n_{1\alpha}\delta_{j2} \right)
-2X_{nj}X'_{nj}Y_{nj}n_{j\alpha},\\[0.3em]
     Y_{nj}/X_{nj}=\tan{(h_{nj}\tau_j)},\, 
     Y'_{n1}/X'_{n1}=\tan{(2h_{n2}\tau_2)},\\[0.3em]
       Y'_{n2}/X'_{n2}=\tan{(2h_{n1}\tau_1)},\,j=1,2.
\end{array}\right.
\label{U-parameters}\ee
As a result, $\Gamma_j=\sigma_z$ serves as the chiral operator.
In this case, a pair of winding numbers, $\nu_{nj}$
act as the topological invariants for the chiral-symmetric
Floquet operators, $U_{nj}(T,k)$
which are defined as \cite{Asboth1,Asboth2} 
\be
\nu_{nj}=\frac{1}{2\pi}\int_{-\pi}^{\pi} \frac{\partial \phi_{nj}(k)}{\partial k}dk,
\label{Nu11}
\ee
where 
$\phi_{nj}(k)=\arctan{(g_{njy}/g_{njx})}$. 
Another pair of winding numbers $(w_{n0},w_{n\pi})$,
can be defined in terms of $\nu_{nj}$, which are given by 
\be
w_{n0}=\frac{\nu_{n1}+\nu_{n2}}{2}, \quad
w_{n\pi}=\frac{\nu_{n1}-\nu_{n2}}{2}.
\ee
$(w_{n0},w_{n\pi})$ always pick integral values and they are
fully capable to determine the topological properties
of bulk states of nonchiral $U_{n}(T,k)$ \cite{Asboth1,Asboth2}. 

\begin{figure}[h]
   \psfrag{wz}{\hskip -0.15 cm $w_{20}$}
  \psfrag{wp}{\hskip -0.15 cm $w_{2\pi}$}
  \psfrag{00}{\hskip 0.14 cm $0.0$}
  \psfrag{2.0}{\hskip -0.2 cm $2.0$}
  \psfrag{1.5}{\hskip -0.2 cm $1.5$}
  \psfrag{4}{\hskip -0.15 cm $4$}
  \psfrag{3}{\hskip -0.15 cm $3$}
  \psfrag{2}{\hskip -0.15 cm $2$}
  \psfrag{1}{\hskip -0.15 cm $1$}
  \psfrag{-1}{\hskip -0.25 cm $-1$}
  \psfrag{0}{\hskip -0.15 cm $0$}
\psfrag{1.0}{\hskip -0.25 cm $1.0$}
\psfrag{0.0}{\hskip -0.18 cm $0.0$}
\psfrag{0.5}{\hskip -0.25 cm $0.5$}
\psfrag{0.50}{\hskip 0.1 cm $0.5$}
\psfrag{-2}{\hskip -0.25 cm $-2$}
\psfrag{-3}{\hskip -0.25 cm $-3$}
\psfrag{-4}{\hskip -0.25 cm $-4$}
\psfrag{0.25}{\hskip 0.1 cm $0.25$}
\psfrag{-0.5}{\hskip -0.43 cm $-0.5$}
\psfrag{-1.0}{\hskip -0.43 cm $-1.0$}
\psfrag{y}{\large k}
\psfrag{E}{\hskip -0.5 cm Energy}
\psfrag{a}{\hskip -0.06 cm (a)}
\psfrag{b}{\hskip -0.06 cm (b)}
\psfrag{c}{\hskip -0.06 cm (c)}
\psfrag{d}{\hskip -0.06 cm (d)}
\psfrag{e}{\hskip -0.06 cm (e)}
\psfrag{f}{\hskip -0.06 cm (f)}
\psfrag{g}{\hskip -0.06 cm (g)}
\psfrag{h}{\hskip -0.06 cm (h)}
\psfrag{z}{$ET/\pi$}
\psfrag{I}{\hskip 0.35 cm $I_{pr}$}
\psfrag{t}{ $\theta/\pi$}
\includegraphics[width=245 pt,angle=0]{top-color-box.eps}
\vskip -0.05cm
\hskip -1.0 cm
\begin{minipage}{0.20\textwidth}
  \includegraphics[width=134pt,angle=-90]{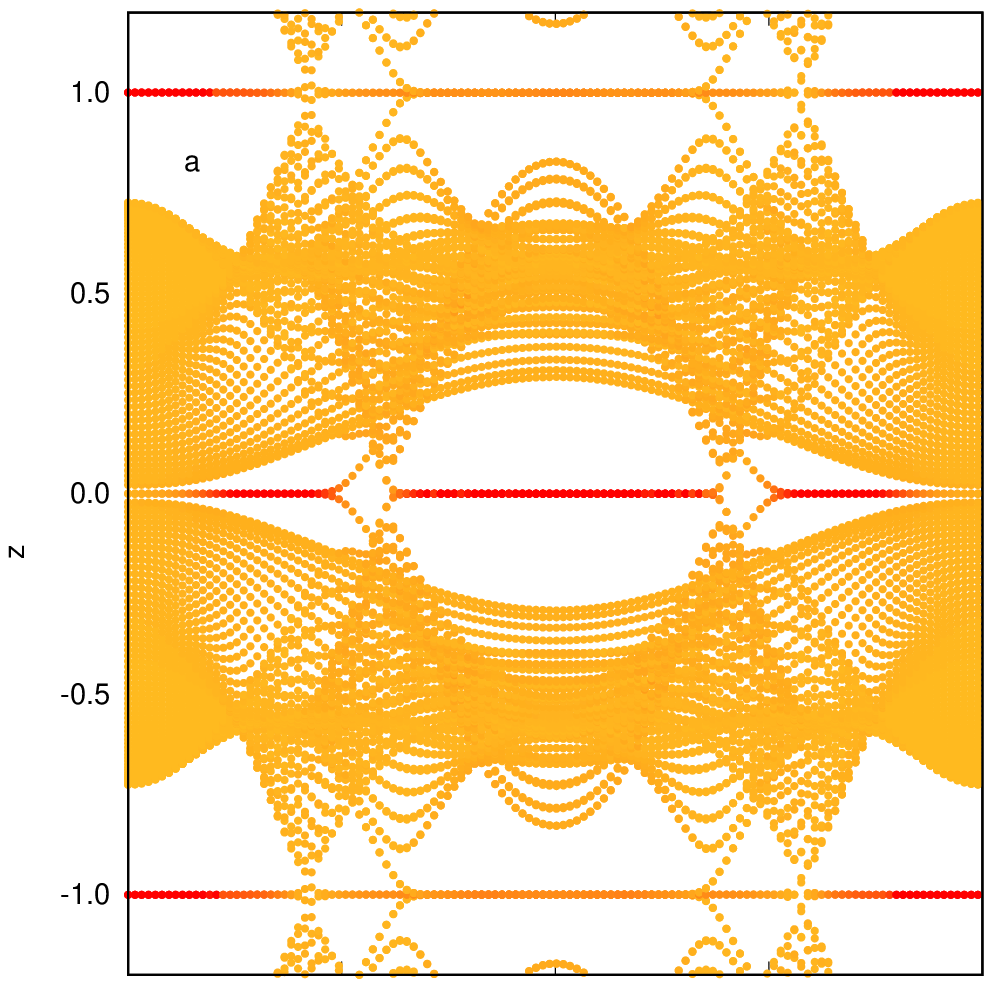}
  \end{minipage}\hskip 0.8cm
  \begin{minipage}{0.2\textwidth}
  \includegraphics[width=134pt,angle=-90]{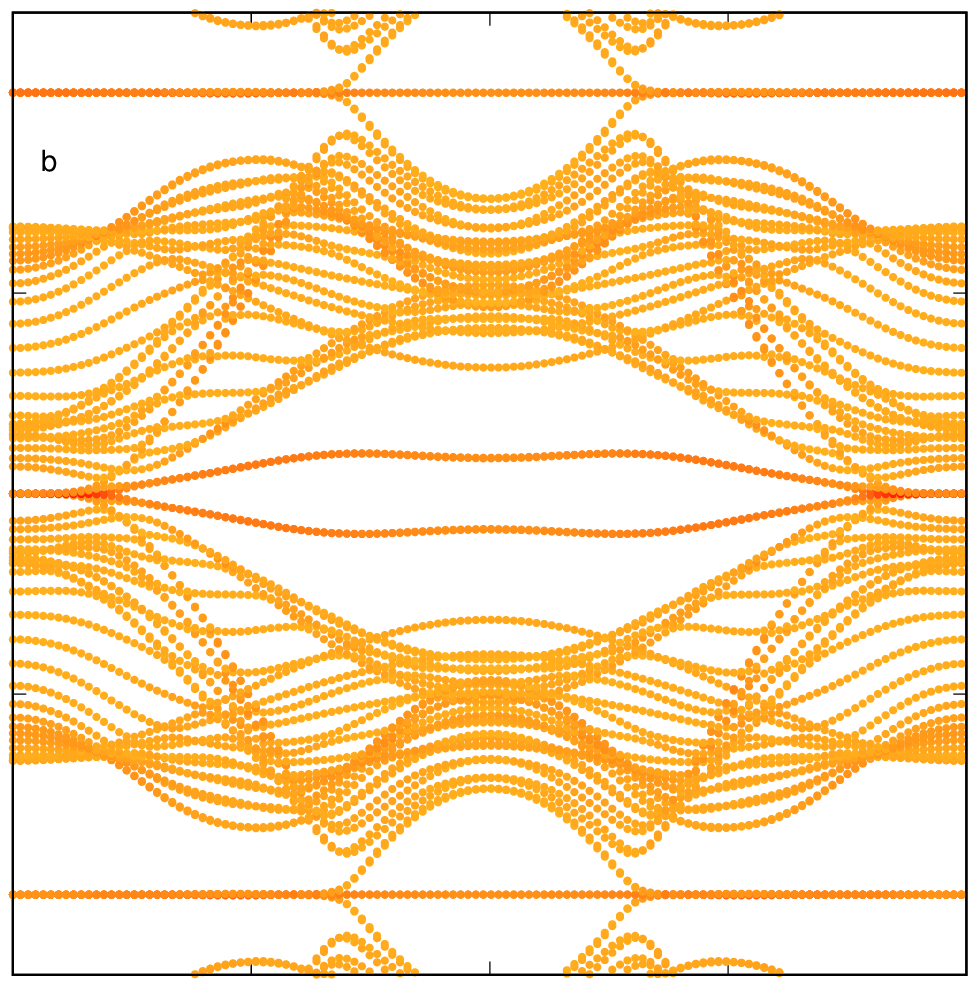}
  \end{minipage}
  \vskip -0.3 cm
  \hskip -1.0 cm
  \begin{minipage}{0.20\textwidth}
  \includegraphics[width=134pt,angle=-90]{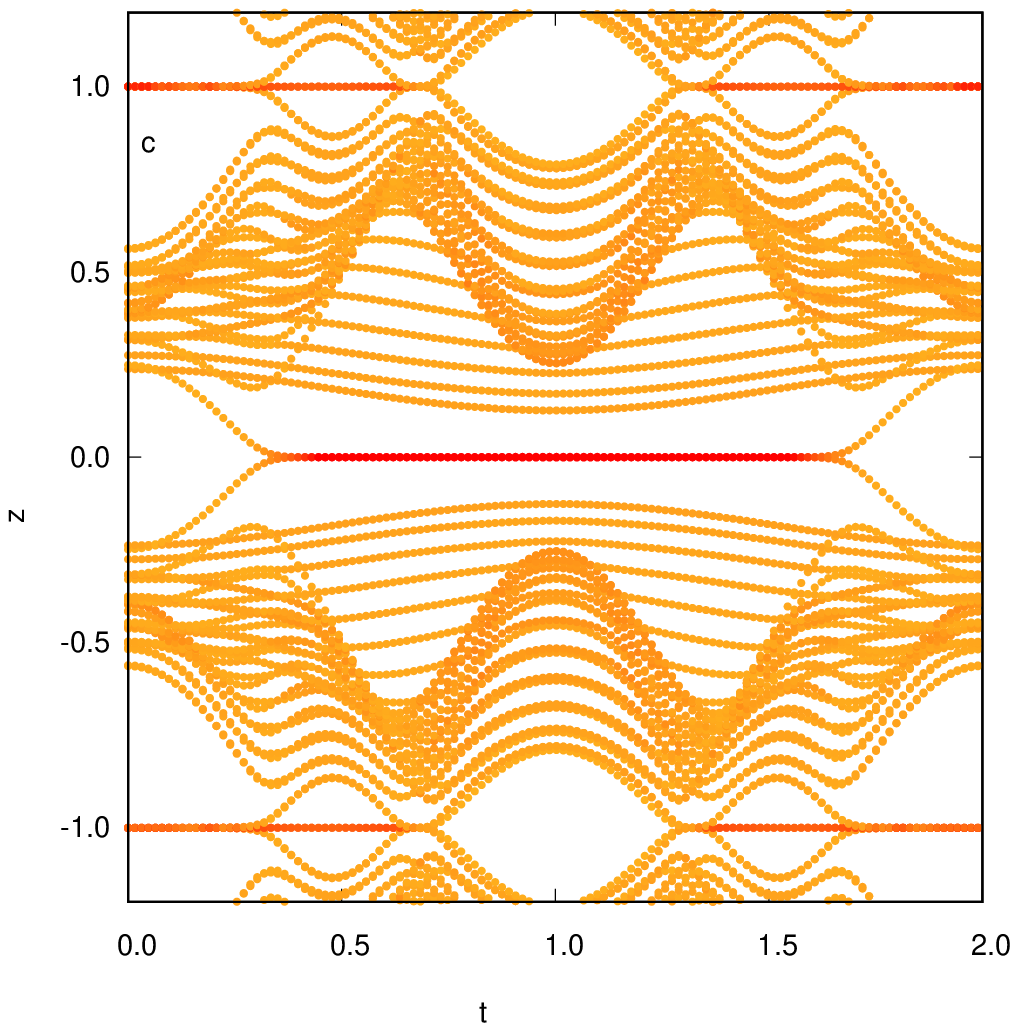}
  \end{minipage}\hskip 0.8cm
  \begin{minipage}{0.2\textwidth}
  \includegraphics[width=134pt,angle=-90]{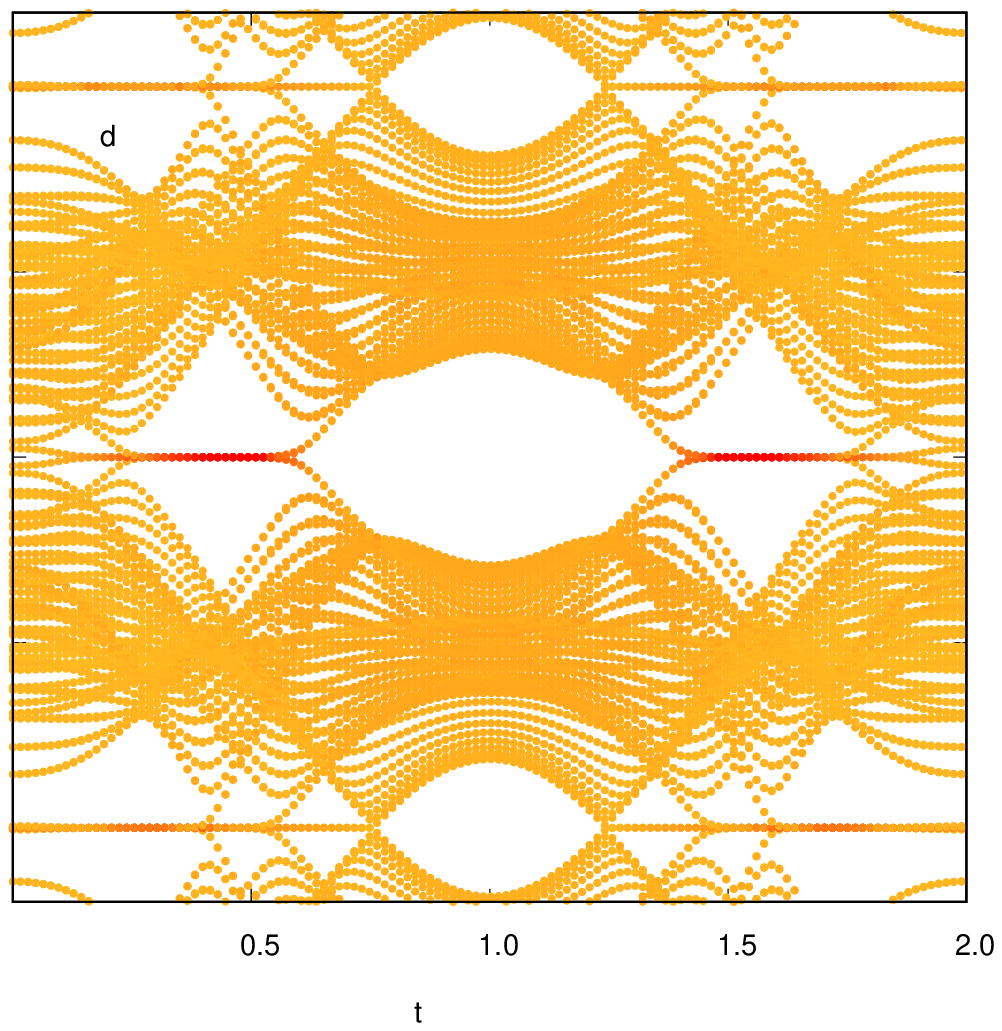}
  \end{minipage}
  \vskip 0.1 cm
   \hskip 0.04 cm
  \begin{minipage}{0.23\textwidth}\hskip 0.15cm
  \includegraphics[width=118pt,angle=0]{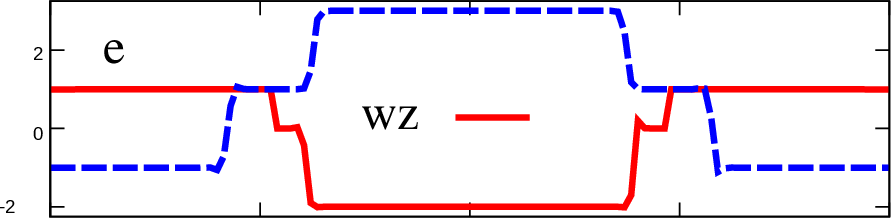}
  \end{minipage}\hskip 0.31cm
  \begin{minipage}{0.23\textwidth}
  \vskip 0.0cm
  \includegraphics[width=118pt,angle=0]{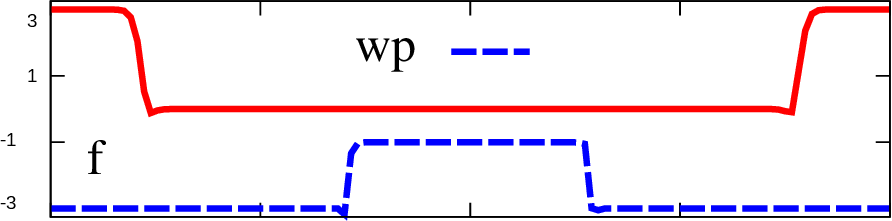}
  \end{minipage}
   \vskip 0.02 cm
   \hskip 0.04 cm
  \begin{minipage}{0.23\textwidth}\hskip 0.15cm
  \includegraphics[width=118pt,angle=0]{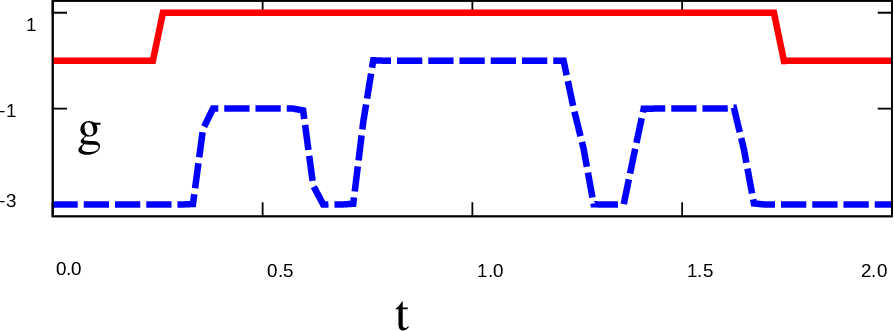}
  \end{minipage}\hskip 0.33cm
  \begin{minipage}{0.23\textwidth}
  \vskip 0.02cm
  \includegraphics[width=118pt,angle=0]{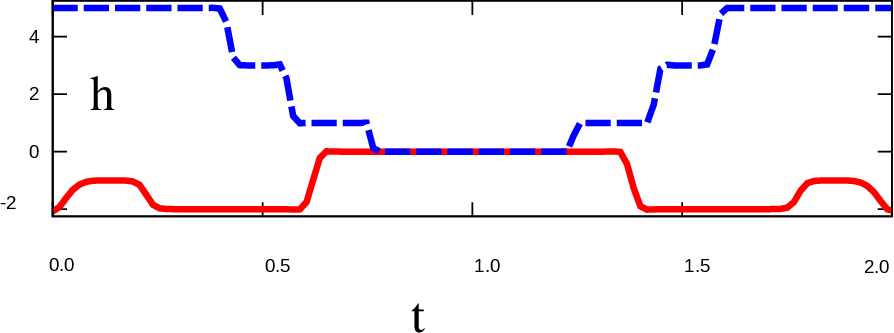}
  \end{minipage}
  \caption{Quasienergy spectra for open systems when $n=2$ are shown in (a) and (b)
    with respect to $\theta$. Variation of winding numbers with respect to
    $\theta$  is shown in (c) and (d). Edge states are indicated by the higher value of
  $I_{pr}$. }
\label{edge-states-floquet-m-2}
\end{figure}

\begin{figure}[h]
  \psfrag{00}{\hskip 0.14 cm $0.0$}
  \psfrag{2.0}{\hskip -0.2 cm $2.0$}
  \psfrag{1.5}{\hskip -0.2 cm $1.5$}
  \psfrag{3}{\hskip -0.15 cm $3$}
  \psfrag{2}{\hskip -0.15 cm $2$}
  \psfrag{1}{\hskip -0.15 cm $1$}
  \psfrag{-1}{\hskip -0.25 cm $-1$}
  \psfrag{0}{{\scriptsize $0$}}
  \psfrag{10}{{\scriptsize  $10$}}
  \psfrag{20}{{\scriptsize $20$}}
  \psfrag{30}{{\scriptsize $30$}}
  \psfrag{40}{{\scriptsize $40$}}
  \psfrag{50}{{\scriptsize $50$}}
  \psfrag{60}{{\scriptsize $60$}}
  \psfrag{70}{{\scriptsize $70$}}
  \psfrag{80}{\hskip -0.05 cm{\scriptsize $80$}}
\psfrag{1.0}{\hskip -0.15 cm {\scriptsize $1.0$}}
\psfrag{0.0}{\hskip -0.15 cm {\scriptsize $0.0$}}
\psfrag{0.2}{\hskip -0.15 cm {\scriptsize $0.2$}}
\psfrag{0.5}{\hskip -0.15 cm {\scriptsize $0.5$}}
\psfrag{0.50}{\hskip 0.1 cm $0.5$}
\psfrag{-2}{\hskip -0.25 cm $-2$}
\psfrag{0.25}{\hskip 0.1 cm $0.25$}
\psfrag{-0.5}{\hskip -0.33 cm {\scriptsize $-0.5$}}
\psfrag{-1.0}{\hskip -0.33 cm {\scriptsize $-1.0$}}
\psfrag{y}{\large k}
\psfrag{E}{\hskip -0.5 cm Energy}
\psfrag{a}{\hskip -0.06 cm (a)}
\psfrag{b}{\hskip -0.06 cm (b)}
\psfrag{c}{\hskip -0.06 cm (c)}
\psfrag{d}{\hskip -0.06 cm (d)}
\psfrag{e}{\hskip -0.06 cm (e)}
\psfrag{f}{\hskip -0.06 cm (f)}
\psfrag{g}{\hskip -0.06 cm (g)}
\psfrag{h}{\hskip -0.06 cm (h)}
\psfrag{z}{{\scriptsize $ET/\pi$}}
\psfrag{p}{{\scriptsize $|\psi|^2$}}
\psfrag{ze}{{\scriptsize $0$-energy}}
\psfrag{pi}{{\scriptsize $\pi$-energy}}
\psfrag{I}{\hskip 0.35 cm $I_{pr}$}
\psfrag{t}{ $\theta/\pi$}
\psfrag{s}{sites}
\includegraphics[width=245 pt,angle=0]{top-color-box.eps}
\vskip -0.05cm
\hskip -1.0 cm
\begin{minipage}{0.20\textwidth}
  \includegraphics[width=46pt,angle=-90]{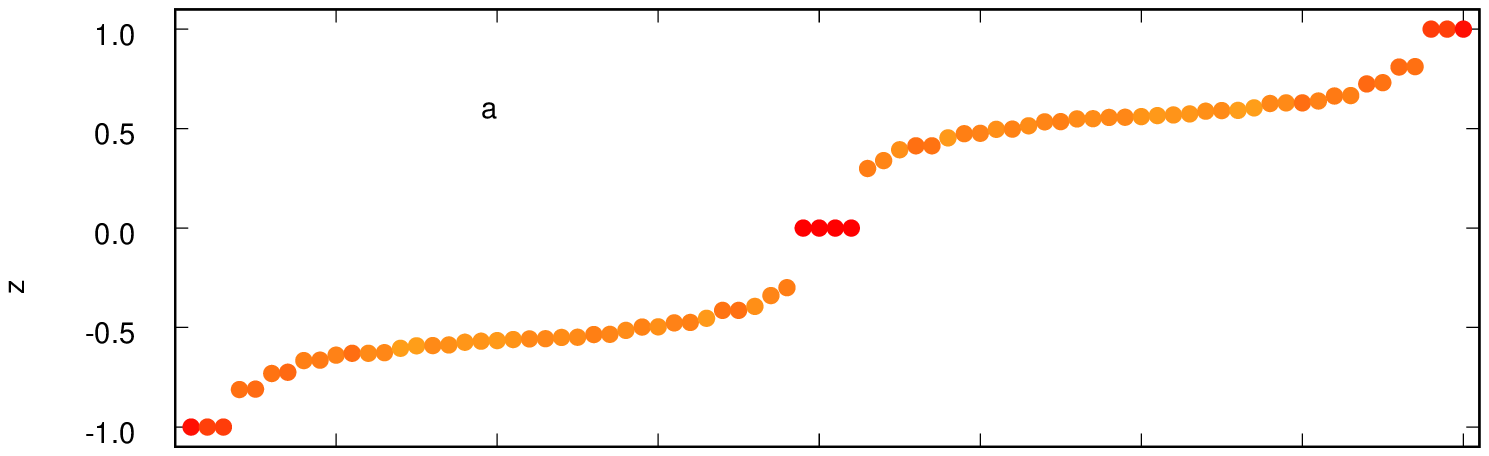}
  \end{minipage}\hskip 0.8cm
  \begin{minipage}{0.20\textwidth}
  \includegraphics[width=46pt,angle=-90]{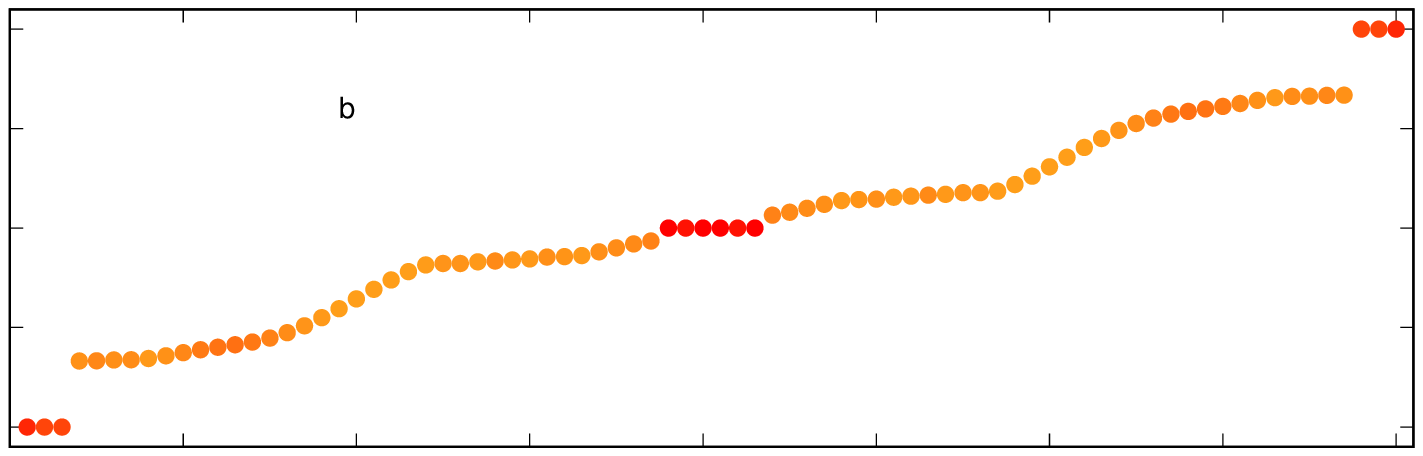}
  \end{minipage}
  \vskip -0.2 cm
  \hskip 0.1 cm
  \begin{minipage}{0.20\textwidth}\hskip -1.2 cm
  \includegraphics[width=46pt,angle=-90]{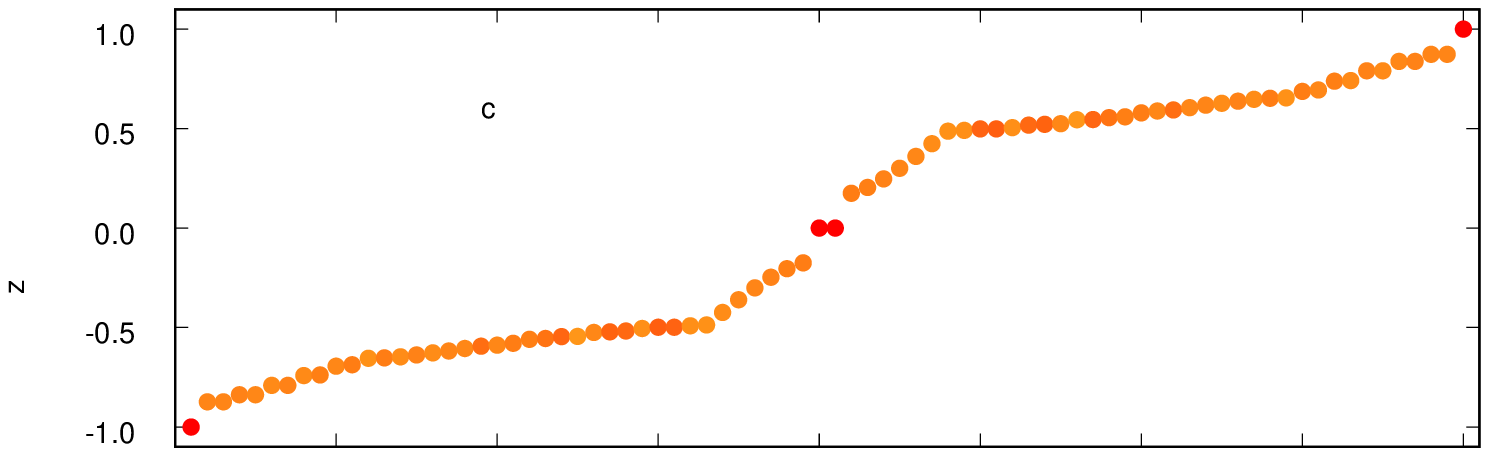}
  \end{minipage}\hskip -0.3cm
  \begin{minipage}{0.20\textwidth}
     \vskip -0.4cm
     \hskip -0.0 cm
  \includegraphics[width=46pt,angle=-90]{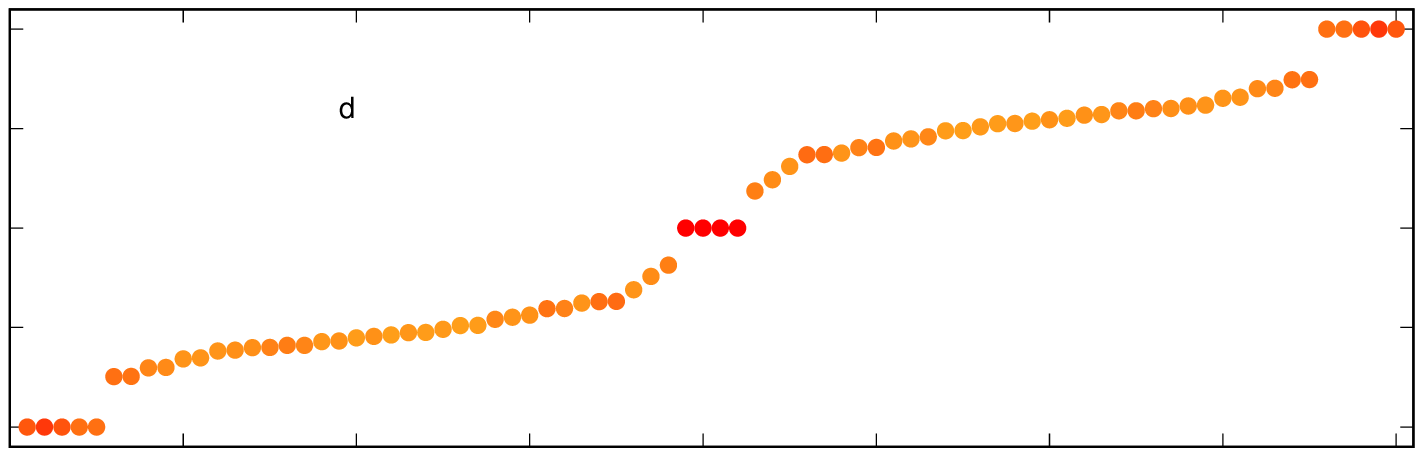}
  \end{minipage}
  \vskip -0.2 cm
   \hskip -0.3 cm
  \begin{minipage}{0.24\textwidth}\hskip -0.1 cm
  \includegraphics[width=128 pt,angle=0]{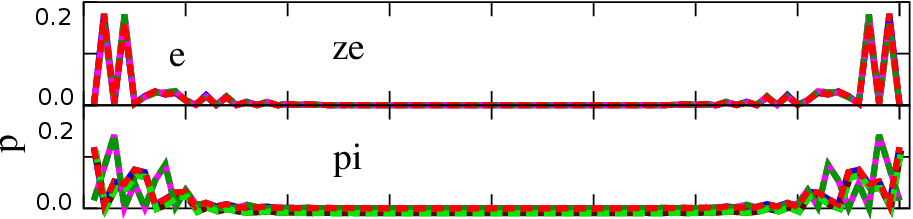}
  \end{minipage}\hskip 0.22cm
  \begin{minipage}{0.23\textwidth}
  \vskip 0.1cm
  \includegraphics[width=122pt,angle=0]{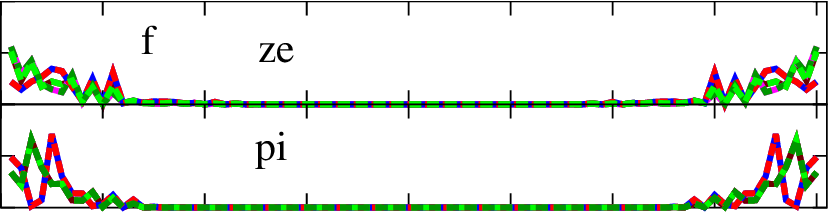}
  \end{minipage}
   \vskip -0.0 cm
   \hskip -0.31 cm
  \begin{minipage}{0.23\textwidth}\hskip 0.0cm
  \includegraphics[width=130pt,angle=0]{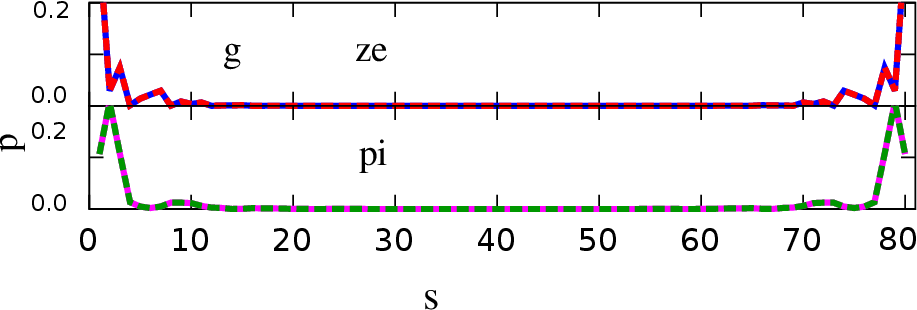}
  \end{minipage}\hskip 0.5cm
  \begin{minipage}{0.23\textwidth}
  \vskip 0.0cm
  \includegraphics[width=120pt,angle=0]{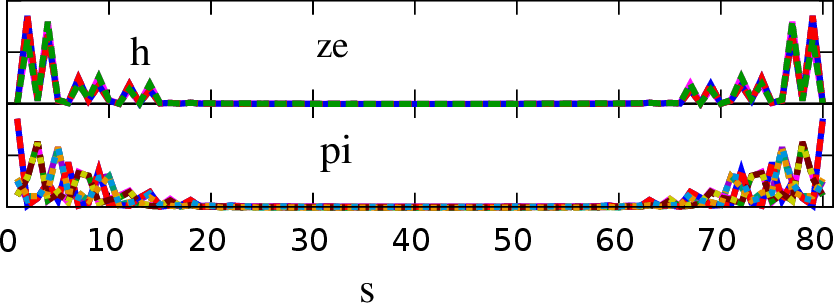}
  \end{minipage}
  \caption{Quasienergy spectra for open systems when $n=2$ are shown in (a) and (b)
    with respect to $\theta$. Variation of winding numbers with respect to
    $\theta$  is shown in (c) and (d). Edge states are indicated by the higher value of
  $I_{pr}$. }
\label{edge-states-floquet-fixed-q-m-2}
\end{figure}
\section{Numerical results}
\label{Floquet-numerical}
In order to investigate the Floquet topological phases
of staggered eSSH models, NN and FN hopping terms
have been parameterized in four different ways depending on the
staggerdness of those hopping terms.
For examples, in the first case NN and FN hopping parameters
are defined by
\be \left\{\begin{array}{l}
t_1=t_3=1+t_0\cos{\theta},\\[0.3em]
t_2=t_4=1-t_0\cos{\theta},\\[0.3em]
t'_1=t'_3=1+t_0\cos{(\theta+\phi)},\\[0.3em]
t'_2=t'_4=1-t_0\cos{(\theta+\phi)}.
\end{array}\right.
\label{Stag-1}\ee
Here $t_0<1$, which means all the hopping terms are
assumed positive. Topological phase diagram
will be prepared with respect to the
angular parameter $\theta$ in every case, where
$0\le\theta\le 2\pi$. Another angular parameter
$\phi \ne 0$ has been introduced in order to
induce further dissimilarities among the hopping terms. 
For the set given in Eq. \ref{Stag-1} (PI), all the NN (or FN) parameters are
different, whereas NN and FN hopping strengths 
are assumed the same for the respective
Hamiltonians. In this sense, it is fully staggered.

In contrast, another parametrization given in Eq. \ref{Stag-2} (PII),
has been introduced where all the NN and FN
hopping terms are separately equal whereas NN terms are
always different from FN. So, for this case staggeredness is
absolutely absent. It is worth mentioning that the effective
hopping parameters in Hamiltonians, $H_{nj}$
preserve staggeredness since they are defined by $\alpha t_1$, $\beta t_2$,
and etc as shown in Eq. \ref{H-nj}, where values of
$\alpha,\, \beta,\, \gamma $, and $\delta$ are always different. 
\be \left\{\begin{array}{l}
t_1=t_2=1+t_0\cos{\theta},\\[0.3em]
t_3=t_4=1+t_0\cos{(\theta+\phi)},\\[0.3em]
t'_1=t'_2=1-t_0\cos{\theta},\\[0.3em]
t'_3=t'_4=1-t_0\cos{(\theta+\phi)},
\end{array}\right.
\label{Stag-2}\ee

However, NN hopping is staggered, while FN
hopping is non-staggered for both $H_{n1}$
and $H_{n2}$ as given in Eq. \ref{Stag-3} (PIII). 
\be \left\{\begin{array}{l}
t_1=t'_1=1+t_0\cos{\theta},\\[0.3em]
t_2=t'_2=1-t_0\cos{\theta},\\[0.3em]
t_3=t_4=1+t_0\cos{(\theta+\phi)},\\[0.3em]
t'_3=t'_4=1-t_0\cos{(\theta+\phi)},
\end{array}\right.
\label{Stag-3}\ee
In contrast, for another set given in Eq. \ref{Stag-4} (PIV),
NN hopping is non-staggered, while FN hopping is staggered.
So, the parametrization given in Eq. \ref{Stag-3}
and Eq. \ref{Stag-4} can be referred as partially staggered.
\be \left\{\begin{array}{l}
t_1=t_2=1+t_0\cos{\theta},\\[0.3em]
t_3=t'_4=1+t_0\cos{(\theta+\phi)},\\[0.3em]
t'_1=t'_2=1-t_0\cos{\theta},\\[0.3em]
t_4=t'_3=1-t_0\cos{(\theta+\phi)},
\end{array}\right.
\label{Stag-4}\ee

\begin{figure}[h]
   \psfrag{wz}{\hskip -0.15 cm $w_{30}$}
  \psfrag{wp}{\hskip -0.15 cm $w_{3\pi}$}
  \psfrag{00}{\hskip 0.14 cm $0.0$}
  \psfrag{2.0}{\hskip -0.2 cm $2.0$}
  \psfrag{1.5}{\hskip -0.2 cm $1.5$}
  \psfrag{4}{\hskip -0.15 cm $4$}
  \psfrag{5}{\hskip -0.15 cm $5$}
  \psfrag{7}{\hskip -0.15 cm $7$}
  \psfrag{3}{\hskip -0.15 cm $3$}
  \psfrag{2}{\hskip -0.15 cm $2$}
  \psfrag{1}{\hskip -0.15 cm $1$}
  \psfrag{-1}{\hskip -0.25 cm $-1$}
  \psfrag{-3}{\hskip -0.25 cm $-3$}
  \psfrag{-4}{\hskip -0.25 cm $-4$}
  \psfrag{0}{\hskip -0.15 cm $0$}
\psfrag{1.0}{\hskip -0.25 cm $1.0$}
\psfrag{0.0}{\hskip -0.18 cm $0.0$}
\psfrag{0.5}{\hskip -0.25 cm $0.5$}
\psfrag{0.50}{\hskip 0.1 cm $0.5$}
\psfrag{-2}{\hskip -0.25 cm $-2$}
\psfrag{0.25}{\hskip 0.1 cm $0.25$}
\psfrag{-0.5}{\hskip -0.43 cm $-0.5$}
\psfrag{-1.0}{\hskip -0.43 cm $-1.0$}
\psfrag{y}{\large k}
\psfrag{E}{\hskip -0.5 cm Energy}
\psfrag{a}{\hskip -0.06 cm (a)}
\psfrag{b}{\hskip -0.06 cm (b)}
\psfrag{c}{\hskip -0.06 cm (c)}
\psfrag{d}{\hskip -0.06 cm (d)}
\psfrag{e}{\hskip -0.06 cm (e)}
\psfrag{f}{\hskip -0.06 cm (f)}
\psfrag{g}{\hskip -0.06 cm (g)}
\psfrag{h}{\hskip -0.06 cm (h)}
\psfrag{z}{$ET/\pi$}
\psfrag{I}{\hskip 0.35 cm $I_{pr}$}
\psfrag{t}{ $\theta/\pi$}
\includegraphics[width=245 pt,angle=0]{top-color-box.eps}
\vskip -0.05cm
\hskip -1.0 cm
\begin{minipage}{0.20\textwidth}
  \includegraphics[width=134pt,angle=-90]{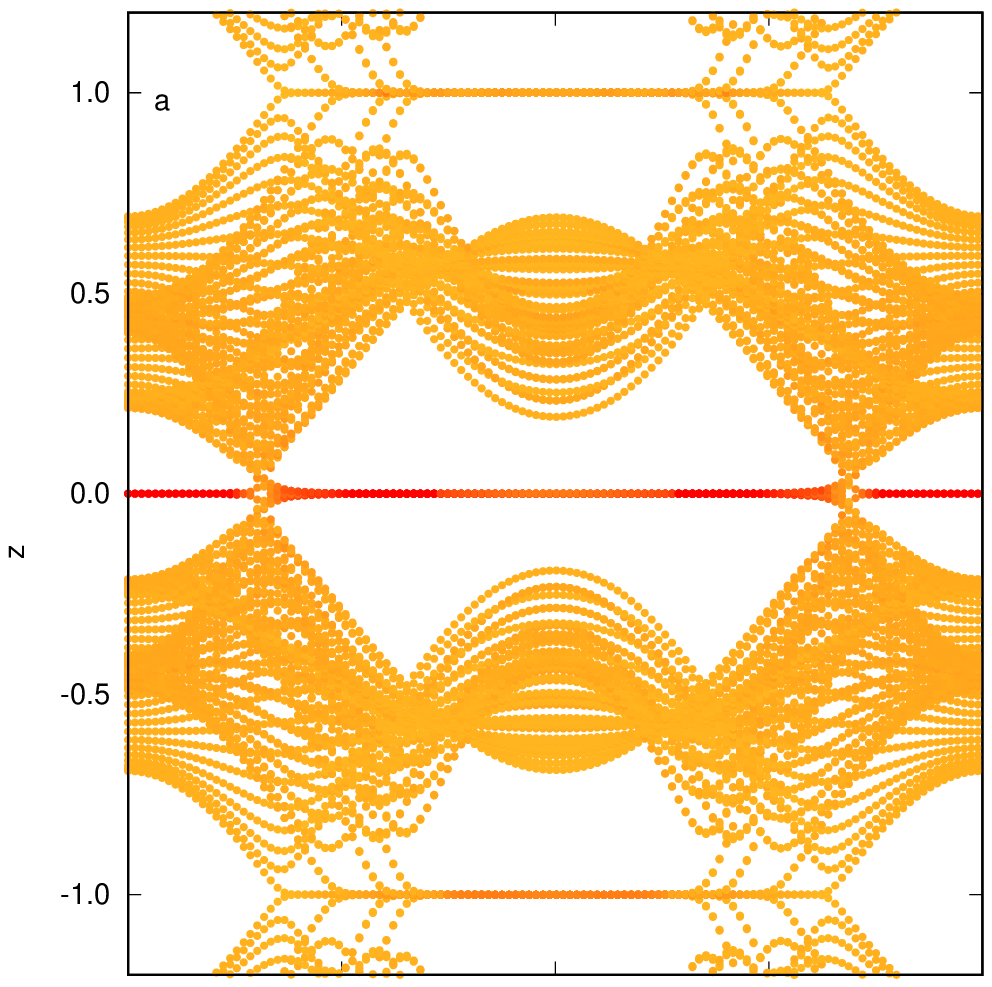}
  \end{minipage}\hskip 0.8cm
  \begin{minipage}{0.2\textwidth}
  \includegraphics[width=134pt,angle=-90]{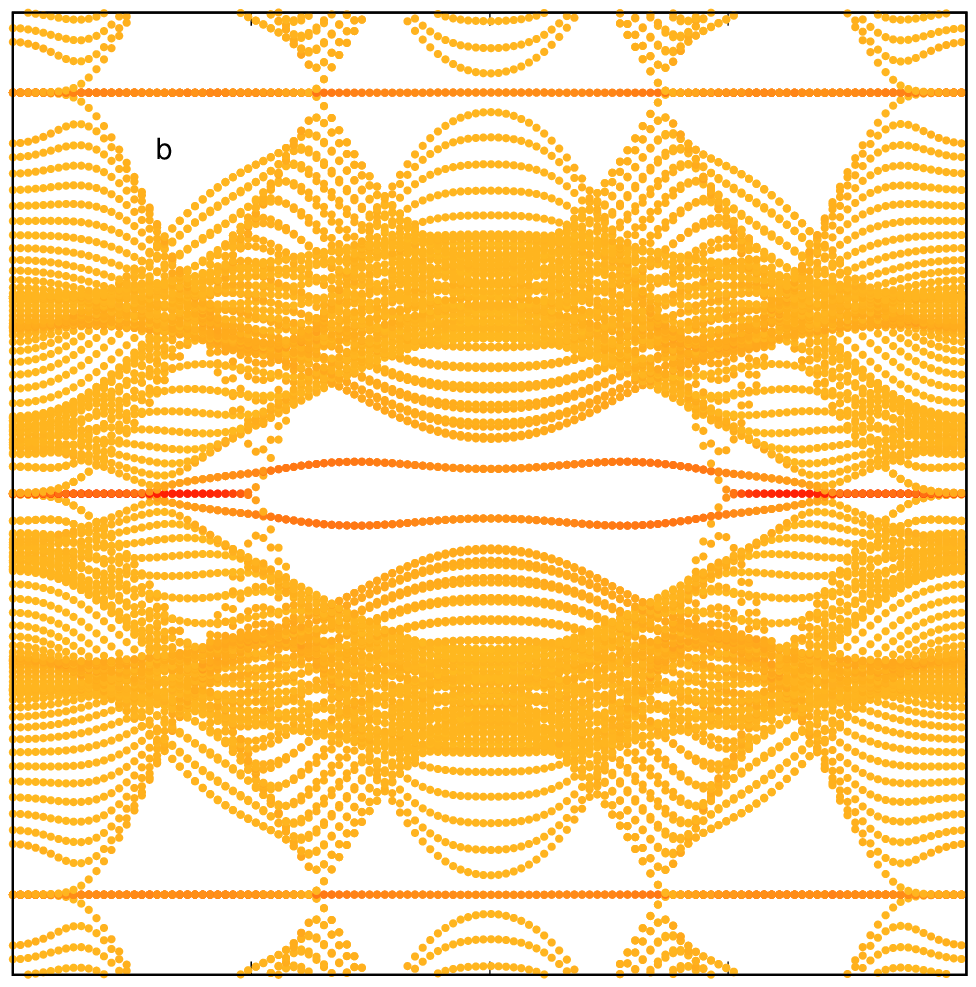}
  \end{minipage}
  \vskip -0.3 cm
  \hskip -1.0 cm
  \begin{minipage}{0.20\textwidth}
  \includegraphics[width=134pt,angle=-90]{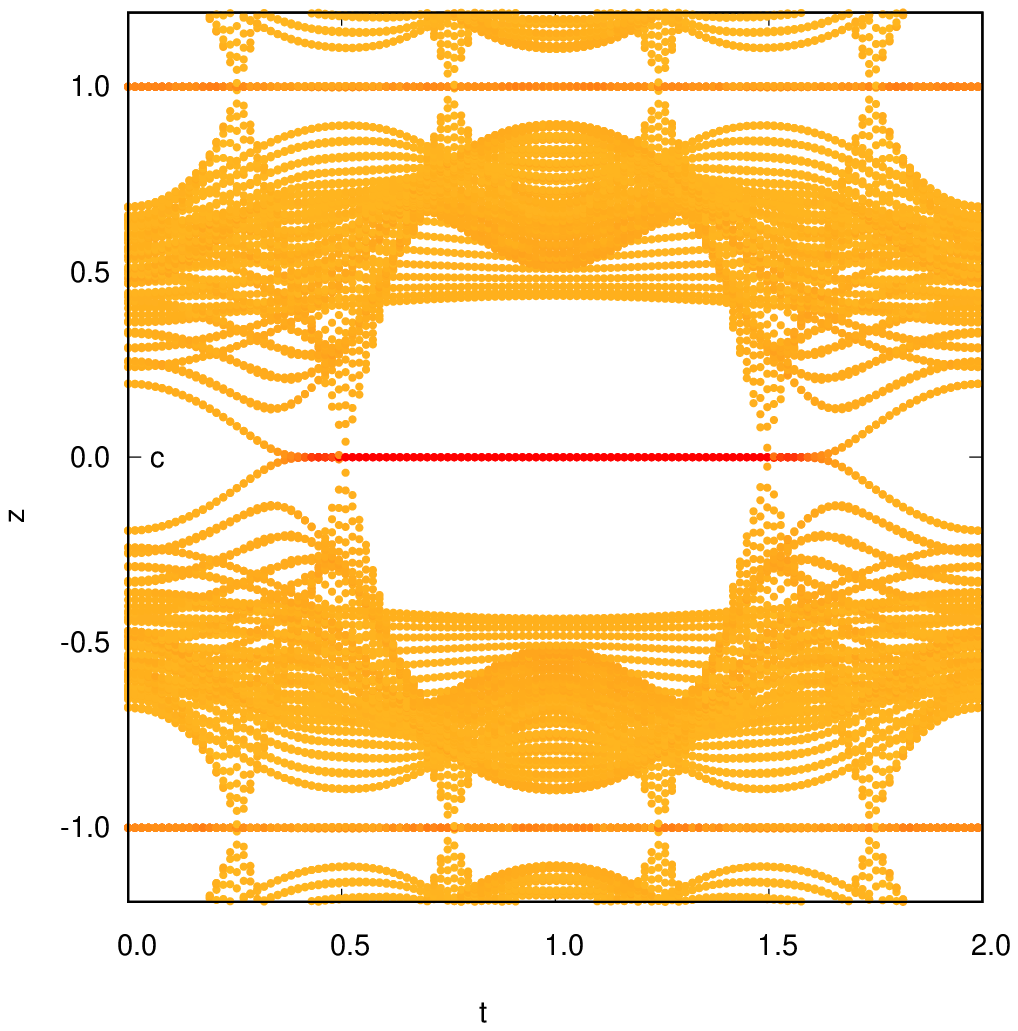}
  \end{minipage}\hskip 0.8cm
  \begin{minipage}{0.2\textwidth}
  \includegraphics[width=134pt,angle=-90]{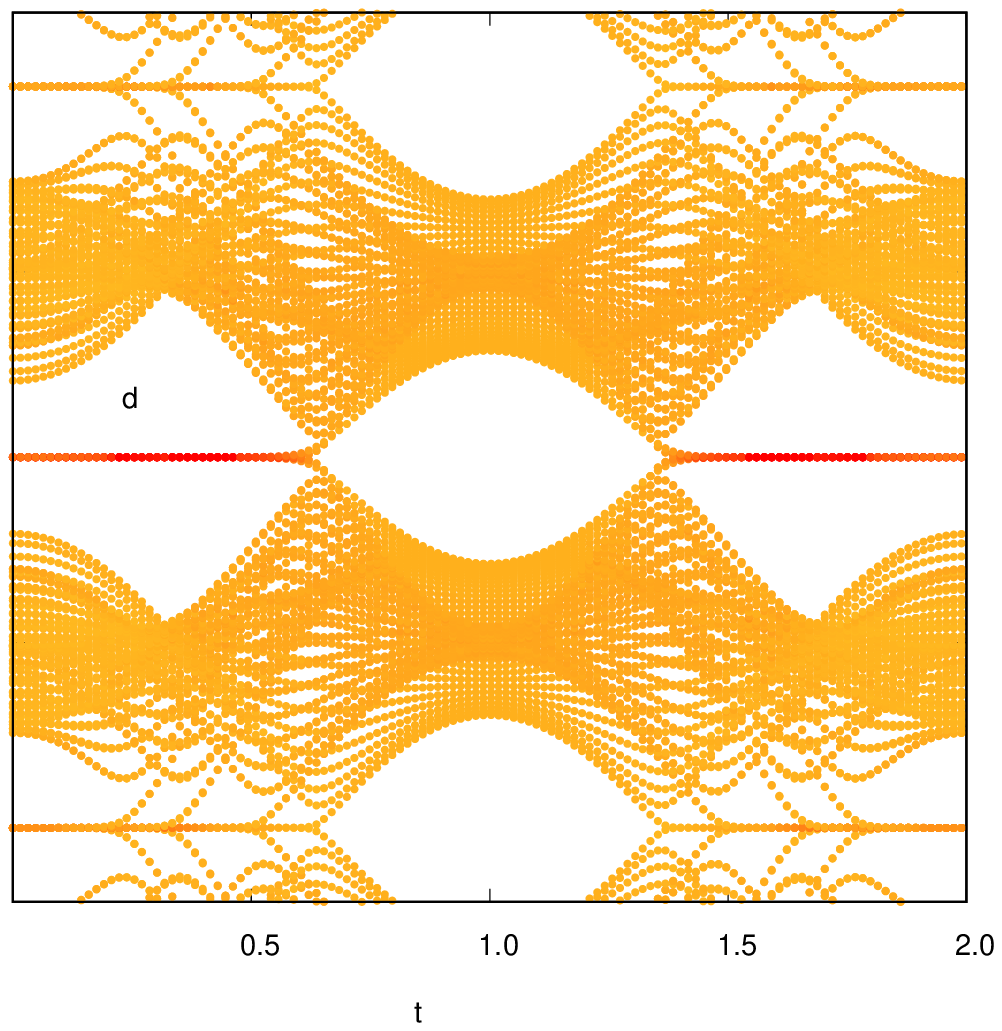}
  \end{minipage}
  \vskip 0.1 cm
   \hskip 0.04 cm
  \begin{minipage}{0.23\textwidth}\hskip 0.15cm
  \includegraphics[width=118pt,angle=0]{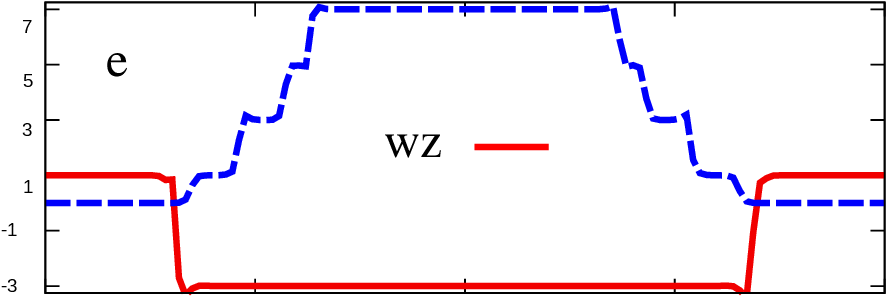}
  \end{minipage}\hskip 0.31cm
  \begin{minipage}{0.23\textwidth}
  \vskip 0.0cm
  \includegraphics[width=118pt,angle=0]{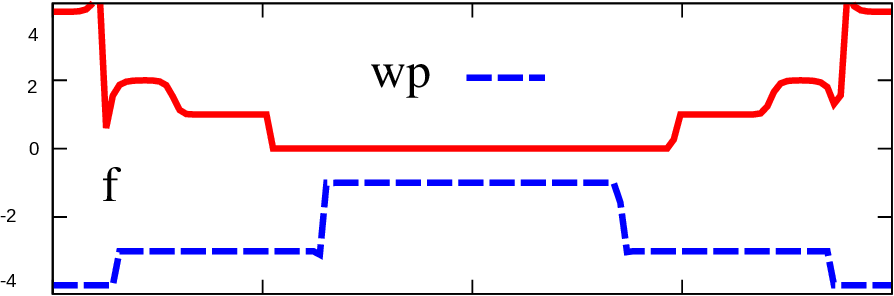}
  \end{minipage}
   \vskip 0.02 cm
   \hskip 0.04 cm
  \begin{minipage}{0.23\textwidth}\hskip 0.15cm
  \includegraphics[width=118pt,angle=0]{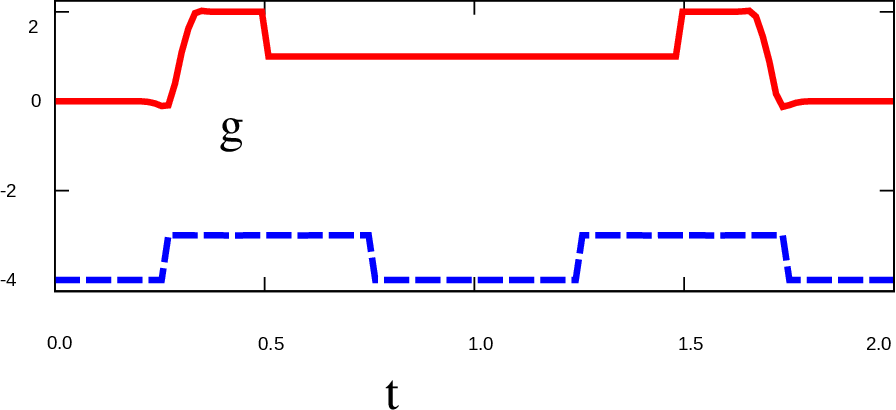}
  \end{minipage}\hskip 0.33cm
  \begin{minipage}{0.23\textwidth}
  \vskip 0.02cm
  \includegraphics[width=118pt,angle=0]{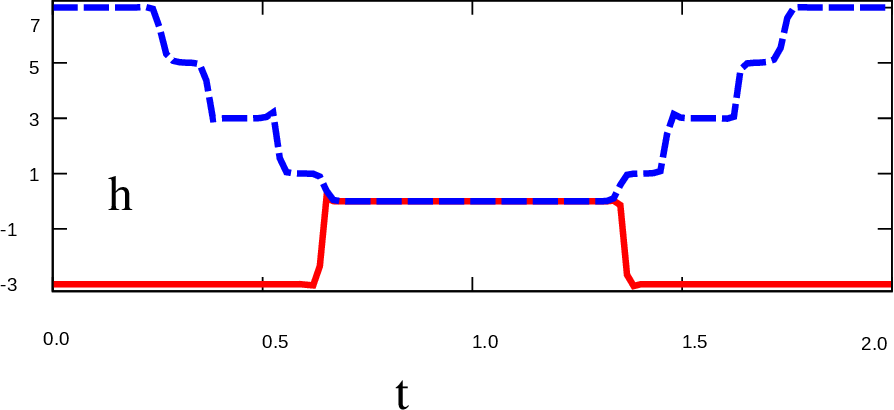}
  \end{minipage}
  \caption{Quasienergy spectra for open systems when $n=3$ are shown in (a) and (b)
    with respect to $\theta$. Variation of winding numbers with respect to
    $\theta$  is shown in (c) and (d). Edge states are indicated by the higher value of
  $I_{pr}$. }
\label{edge-states-floquet-m-3}
\end{figure}

\begin{figure}[h]
  \psfrag{00}{\hskip 0.14 cm $0.0$}
  \psfrag{2.0}{\hskip -0.2 cm $2.0$}
  \psfrag{1.5}{\hskip -0.2 cm $1.5$}
  \psfrag{3}{\hskip -0.15 cm $3$}
  \psfrag{2}{\hskip -0.15 cm $2$}
  \psfrag{1}{\hskip -0.15 cm $1$}
  \psfrag{-1}{\hskip -0.25 cm $-1$}
  \psfrag{0}{{\scriptsize $0$}}
  \psfrag{10}{{\scriptsize  $10$}}
  \psfrag{20}{{\scriptsize $20$}}
  \psfrag{30}{{\scriptsize $30$}}
  \psfrag{40}{{\scriptsize $40$}}
  \psfrag{50}{{\scriptsize $50$}}
  \psfrag{60}{{\scriptsize $60$}}
  \psfrag{70}{{\scriptsize $70$}}
  \psfrag{80}{\hskip -0.05 cm{\scriptsize $80$}}
\psfrag{1.0}{\hskip -0.15 cm {\scriptsize $1.0$}}
\psfrag{0.0}{\hskip -0.15 cm {\scriptsize $0.0$}}
\psfrag{0.2}{\hskip -0.15 cm {\scriptsize $0.2$}}
\psfrag{0.5}{\hskip -0.15 cm {\scriptsize $0.5$}}
\psfrag{0.50}{\hskip 0.1 cm $0.5$}
\psfrag{-2}{\hskip -0.25 cm $-2$}
\psfrag{0.25}{\hskip 0.1 cm $0.25$}
\psfrag{-0.5}{\hskip -0.33 cm {\scriptsize $-0.5$}}
\psfrag{-1.0}{\hskip -0.33 cm {\scriptsize $-1.0$}}
\psfrag{y}{\large k}
\psfrag{E}{\hskip -0.5 cm Energy}
\psfrag{a}{\hskip -0.06 cm (a)}
\psfrag{b}{\hskip -0.06 cm (b)}
\psfrag{c}{\hskip -0.06 cm (c)}
\psfrag{d}{\hskip -0.06 cm (d)}
\psfrag{e}{\hskip -0.06 cm (e)}
\psfrag{f}{\hskip -0.06 cm (f)}
\psfrag{g}{\hskip -0.06 cm (g)}
\psfrag{h}{\hskip -0.06 cm (h)}
\psfrag{z}{{\scriptsize $ET/\pi$}}
\psfrag{p}{{\scriptsize $|\psi|^2$}}
\psfrag{ze}{{\scriptsize $0$-energy}}
\psfrag{pi}{{\scriptsize $\pi$-energy}}
\psfrag{I}{\hskip 0.35 cm $I_{pr}$}
\psfrag{t}{ $\theta/\pi$}
\psfrag{s}{sites}
\includegraphics[width=245 pt,angle=0]{top-color-box.eps}
\vskip -0.05cm
\hskip -1.0 cm
\begin{minipage}{0.20\textwidth}
  \includegraphics[width=46pt,angle=-90]{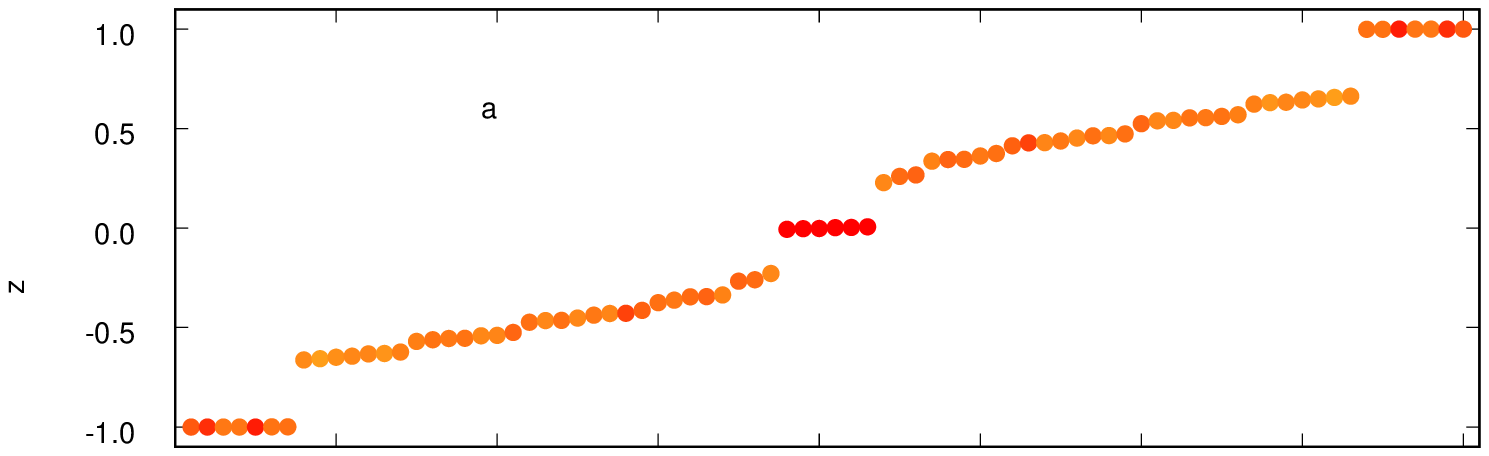}
  \end{minipage}\hskip 0.8cm
  \begin{minipage}{0.20\textwidth}
  \includegraphics[width=46pt,angle=-90]{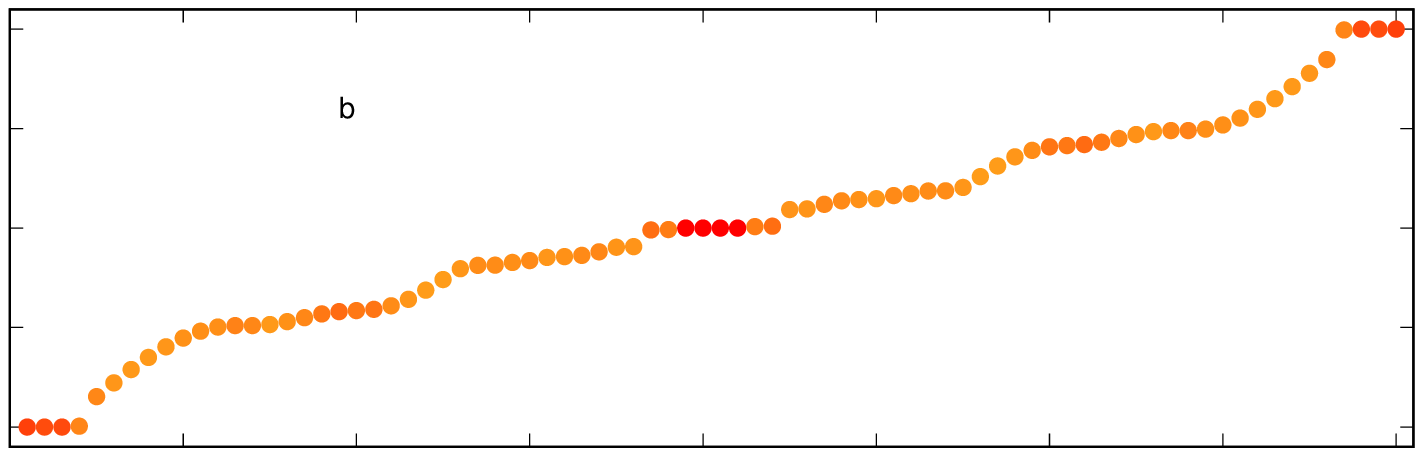}
  \end{minipage}
  \vskip -0.2 cm
  \hskip 0.1 cm
  \begin{minipage}{0.20\textwidth}\hskip -1.2 cm
  \includegraphics[width=46pt,angle=-90]{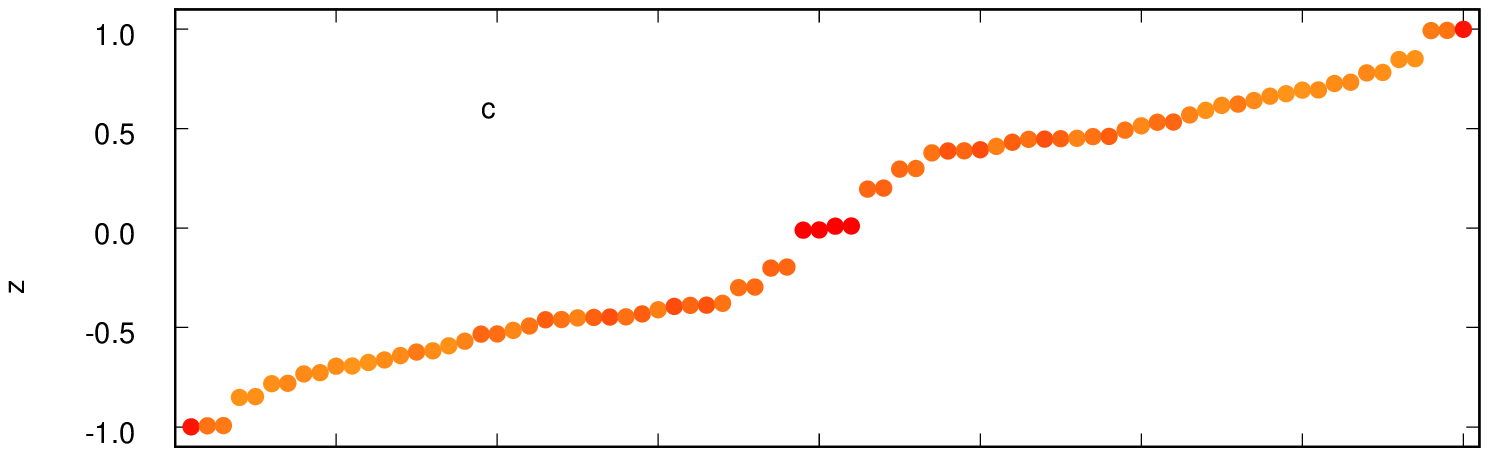}
  \end{minipage}\hskip -0.3cm
  \begin{minipage}{0.20\textwidth}
     \vskip -0.4cm
     \hskip -0.0 cm
  \includegraphics[width=46pt,angle=-90]{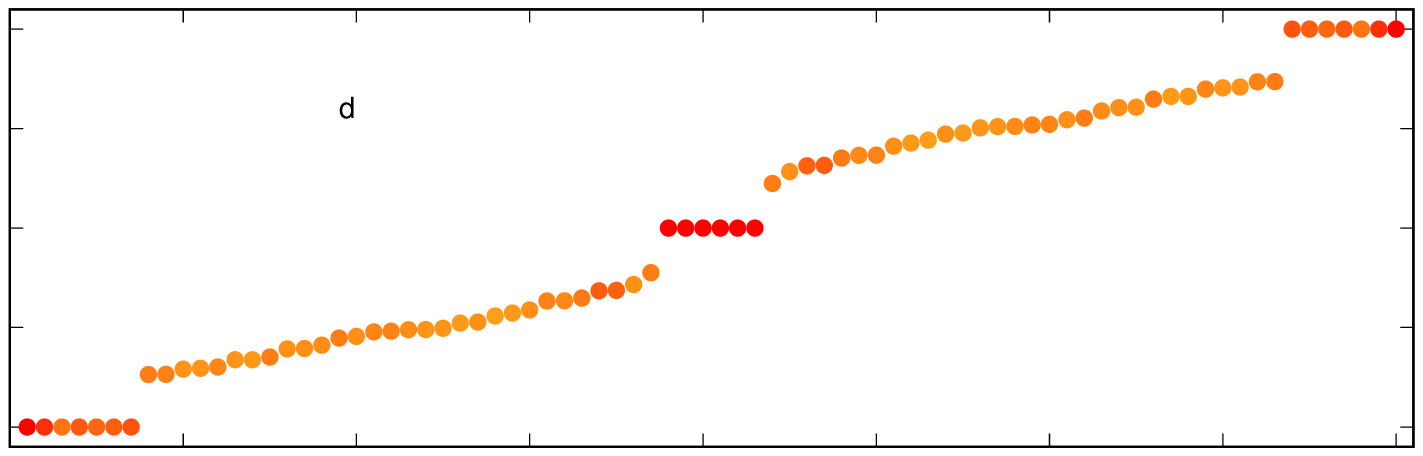}
  \end{minipage}
  \vskip -0.2 cm
   \hskip -0.3 cm
  \begin{minipage}{0.24\textwidth}\hskip -0.1 cm
  \includegraphics[width=128 pt,angle=0]{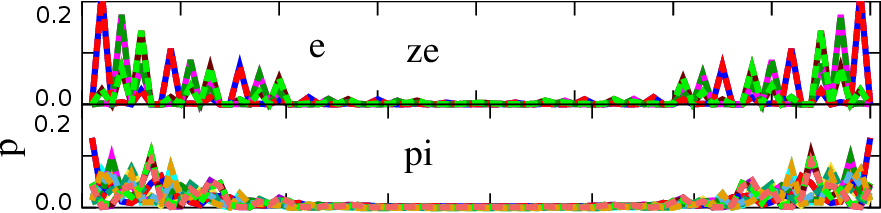}
  \end{minipage}\hskip 0.22cm
  \begin{minipage}{0.23\textwidth}
  \vskip 0.1cm
  \includegraphics[width=122pt,angle=0]{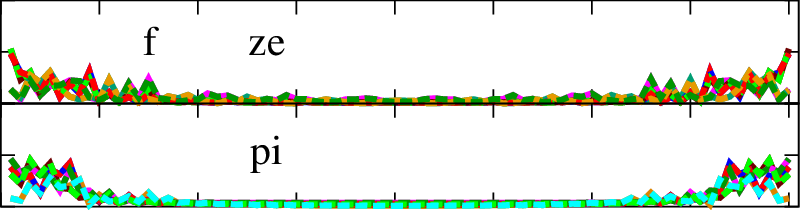}
  \end{minipage}
   \vskip -0.1 cm
   \hskip -0.31 cm
  \begin{minipage}{0.23\textwidth}\hskip 0.0cm
  \includegraphics[width=130pt,angle=0]{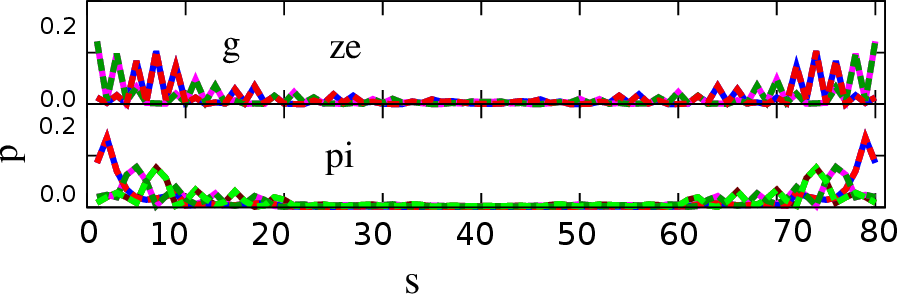}
  \end{minipage}\hskip 0.5cm
  \begin{minipage}{0.23\textwidth}
  \vskip 0.1cm
  \includegraphics[width=120pt,angle=0]{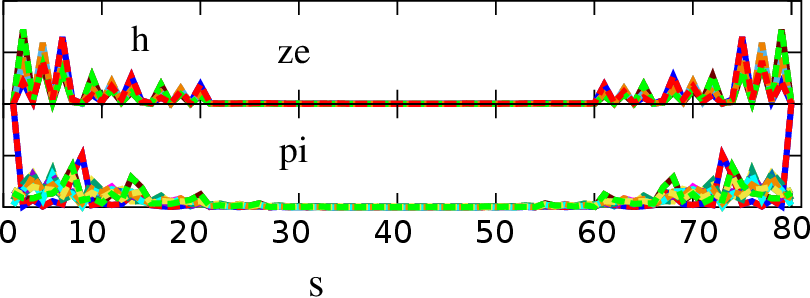}
  \end{minipage}
  \caption{Quasienergy spectra for open systems when $n=3$ are shown in (a) and (b)
    with respect to $\theta$. Variation of winding numbers with respect to
    $\theta$  is shown in (c) and (d). Edge states are indicated by the higher value of
  $I_{pr}$. }
\label{edge-states-floquet-fixed-q-m-3}
\end{figure}

In this study properties of Floquet topological phases with higher winding
will be investigated on four eSSH models defined by $n=1,2,3,4$,
where value of $n$ measure the extent of hopping
path in the FN terms as discussed before.
The systems exhibit phases of higher winding numbers
with the increase of $n$ for four different types of
parametrization of hopping terms.
Here four different parameter sets are denoted by PI, PII, PIII and
PIV, respectively. Floquet topological phases for four
different values of $n=1,2,3,4$, have been discussed
in the following subsections. 
\subsection{Floquet topological phases for $n=1$:}
Quasienergy band diagrams for $n=1$ with respect to $\theta$ 
have been shown in Fig. \ref{edge-states-floquet-m-1}
(a), (b), (c) and (d), respectively for
the sets PI, PII, PIII and PIV. In the same diagram,
variation of winding numbers $w_{10}$, and $w_{1\pi}$
for the sets PI-PIV are shown in 
(e), (f), (g) and (h), respectively, where $0\le \theta \le 2\pi$.
Those figures are plotted for $\phi=\pi$, and $T=\pi$.
However, values of other parameters are different for the
four cases. For examples, Fig. \ref{edge-states-floquet-m-1} (a) and
(e) are drawn for PI where
$t_0=3/4, \alpha=1,\beta=0.8,\gamma=0.7,\delta=0,t_p=0.3T$
It reveals that the system hosts topological phases with
$w_{10}=1,-1$, and $w_{1\pi}=1,-1,-2$, with the variation of $\theta$.
System exhibits band gap in the
topological regions, while band gap vanishes at the transition points. 

\begin{figure}[h]
  \psfrag{wz}{\hskip -0.15 cm $w_{40}$}
  \psfrag{wp}{\hskip -0.15 cm $w_{4\pi}$}  
  \psfrag{00}{\hskip 0.14 cm $0.0$}
  \psfrag{2.0}{\hskip -0.2 cm $2.0$}
  \psfrag{1.5}{\hskip -0.2 cm $1.5$}
  \psfrag{4}{\hskip -0.15 cm $4$}
  \psfrag{5}{\hskip -0.15 cm $5$}
  \psfrag{7}{\hskip -0.15 cm $7$}
  \psfrag{8}{\hskip -0.15 cm $8$}
  \psfrag{3}{\hskip -0.15 cm $3$}
  \psfrag{2}{\hskip -0.15 cm $2$}
  \psfrag{1}{\hskip -0.15 cm $1$}
  \psfrag{-1}{\hskip -0.25 cm $-1$}
  \psfrag{0}{\hskip -0.15 cm $0$}
\psfrag{1.0}{\hskip -0.25 cm $1.0$}
\psfrag{0.0}{\hskip -0.18 cm $0.0$}
\psfrag{0.5}{\hskip -0.25 cm $0.5$}
\psfrag{0.50}{\hskip 0.1 cm $0.5$}
\psfrag{-2}{\hskip -0.25 cm $-2$}
\psfrag{-3}{\hskip -0.25 cm $-3$}
\psfrag{-4}{\hskip -0.25 cm $-4$}
\psfrag{-5}{\hskip -0.25 cm $-5$}
\psfrag{0.25}{\hskip 0.1 cm $0.25$}
\psfrag{-0.5}{\hskip -0.43 cm $-0.5$}
\psfrag{-1.0}{\hskip -0.43 cm $-1.0$}
\psfrag{y}{\large k}
\psfrag{E}{\hskip -0.5 cm Energy}
\psfrag{a}{\hskip -0.06 cm (a)}
\psfrag{b}{\hskip -0.06 cm (b)}
\psfrag{c}{\hskip -0.06 cm (c)}
\psfrag{d}{\hskip -0.06 cm (d)}
\psfrag{e}{\hskip -0.06 cm (e)}
\psfrag{f}{\hskip -0.06 cm (f)}
\psfrag{g}{\hskip -0.06 cm (g)}
\psfrag{h}{\hskip -0.06 cm (h)}
\psfrag{z}{$ET/\pi$}
\psfrag{I}{\hskip 0.35 cm $I_{pr}$}
\psfrag{t}{ $\theta/\pi$}
\includegraphics[width=245 pt,angle=0]{top-color-box.eps}
\vskip -0.05cm
\hskip -1.0 cm
\begin{minipage}{0.20\textwidth}
  \includegraphics[width=134pt,angle=-90]{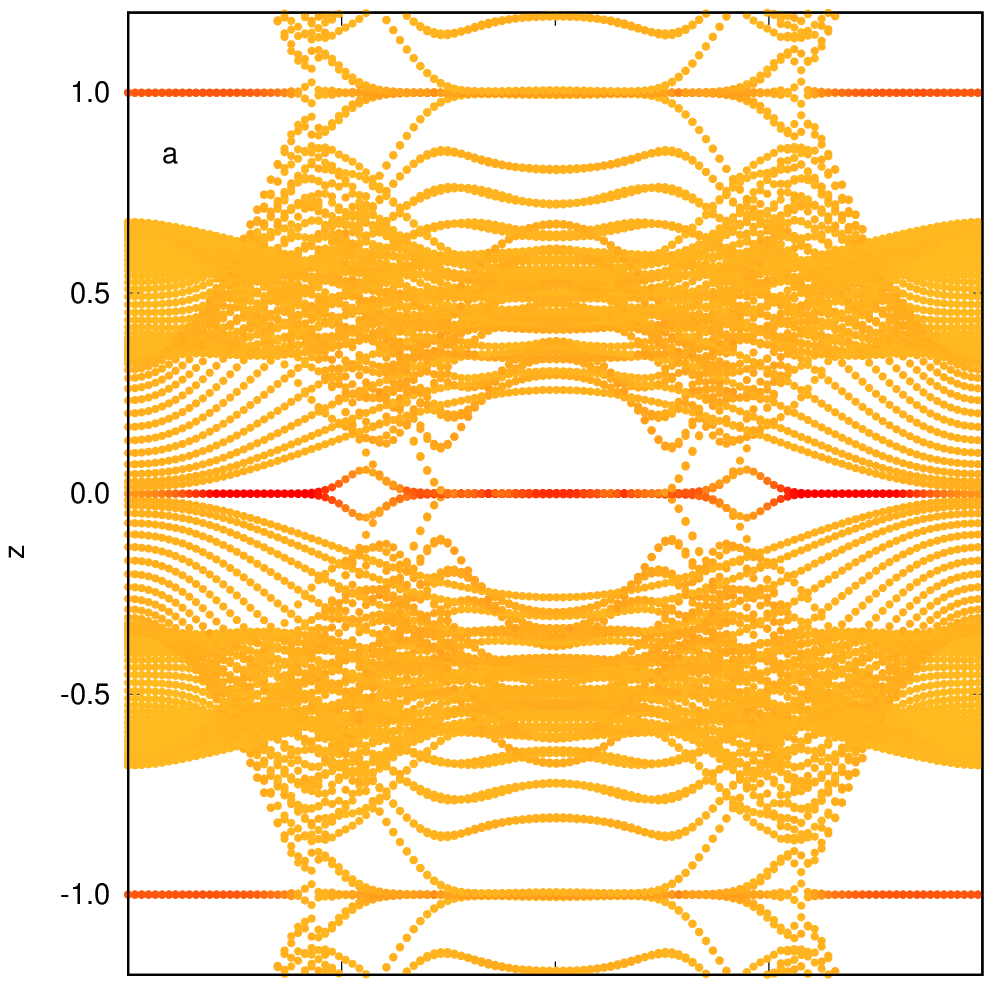}
  \end{minipage}\hskip 0.8cm
  \begin{minipage}{0.2\textwidth}
  \includegraphics[width=134pt,angle=-90]{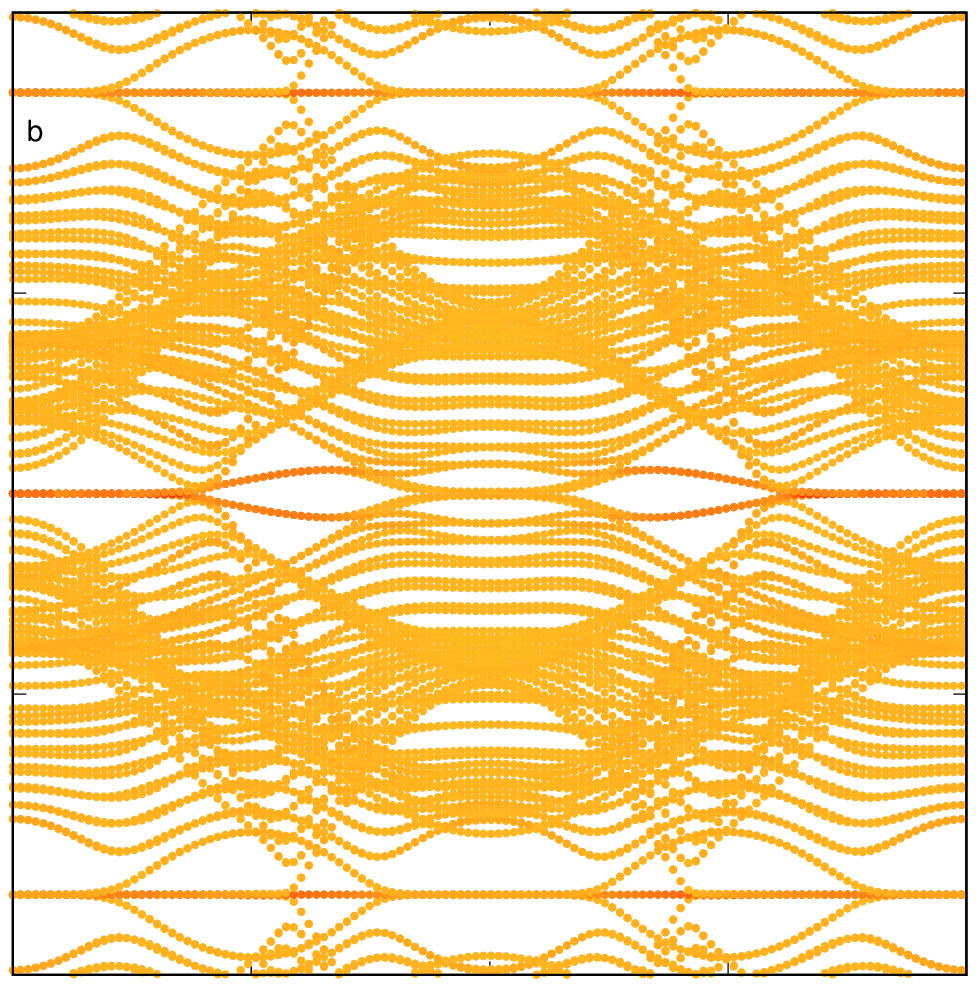}
  \end{minipage}
  \vskip -0.3 cm
  \hskip -1.0 cm
  \begin{minipage}{0.20\textwidth}
  \includegraphics[width=134pt,angle=-90]{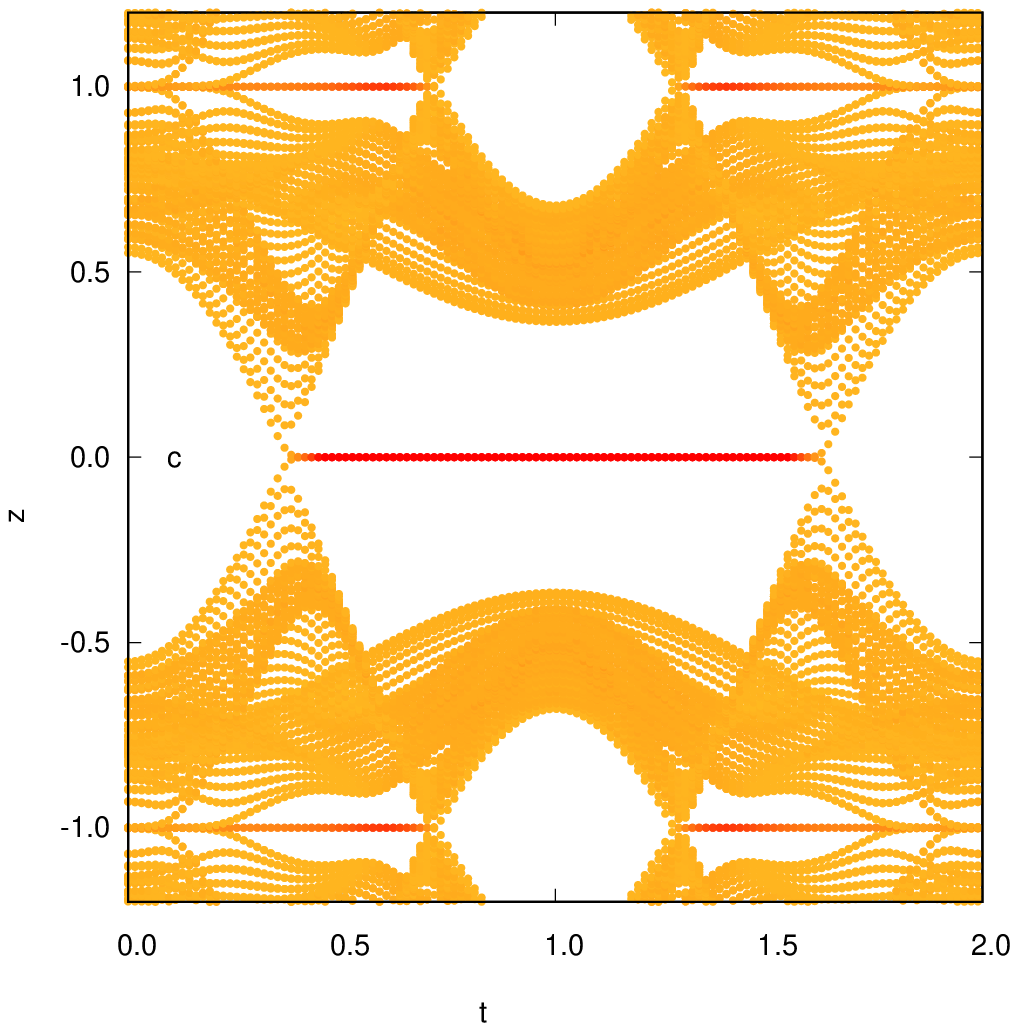}
  \end{minipage}\hskip 0.8cm
  \begin{minipage}{0.2\textwidth}
  \includegraphics[width=134pt,angle=-90]{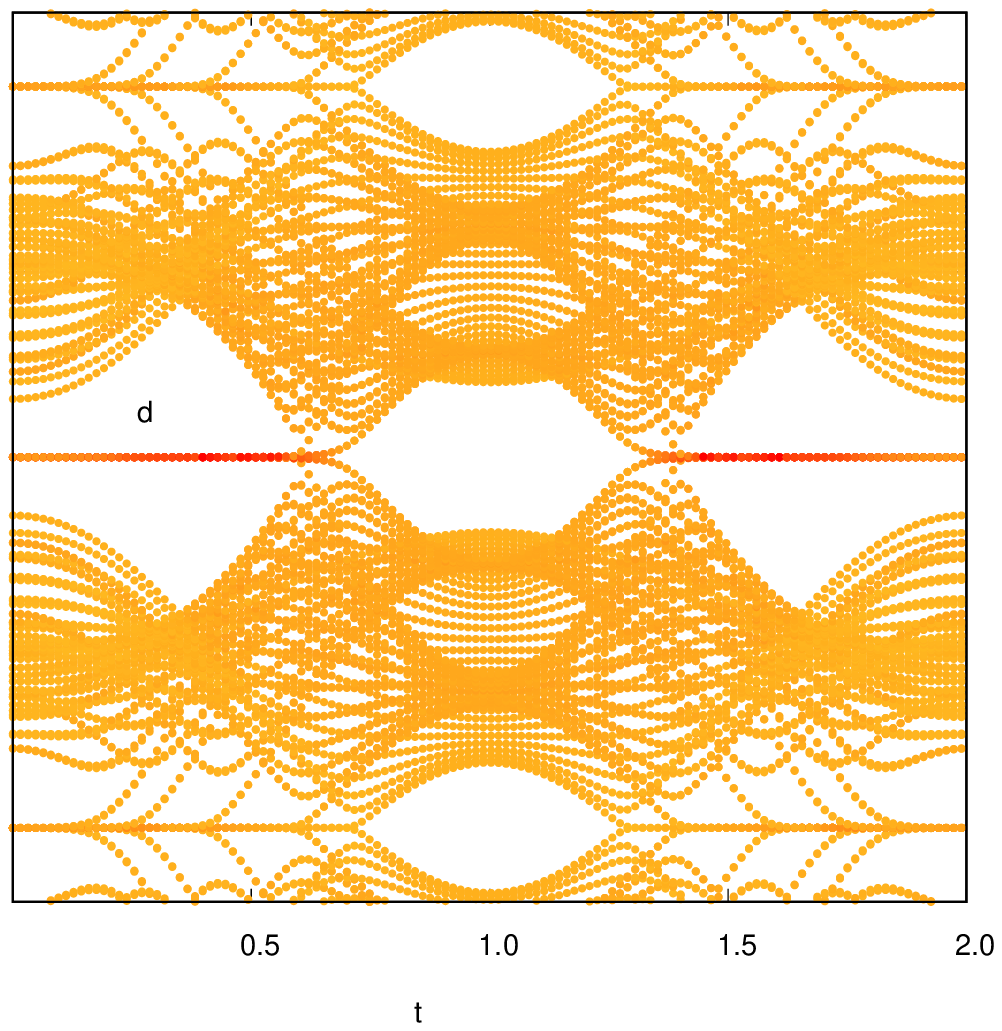}
  \end{minipage}
  \vskip 0.1 cm
   \hskip 0.04 cm
  \begin{minipage}{0.23\textwidth}\hskip 0.15cm
  \includegraphics[width=118pt,angle=0]{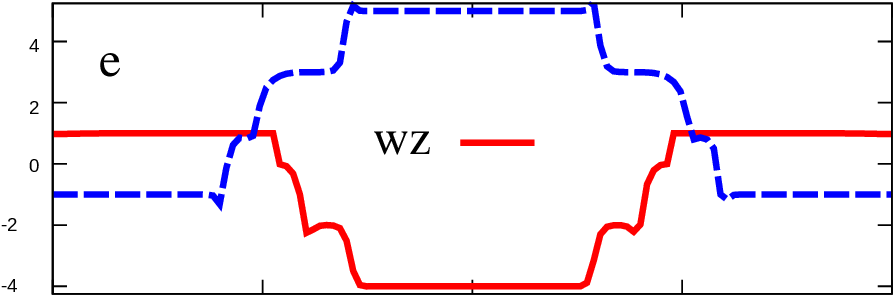}
  \end{minipage}\hskip 0.31cm
  \begin{minipage}{0.23\textwidth}
  \vskip 0.0cm
  \includegraphics[width=118pt,angle=0]{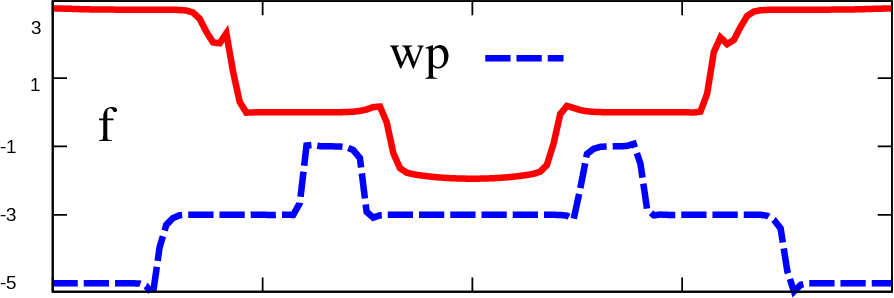}
  \end{minipage}
   \vskip 0.02 cm
   \hskip 0.04 cm
  \begin{minipage}{0.23\textwidth}\hskip 0.15cm
  \includegraphics[width=118pt,angle=0]{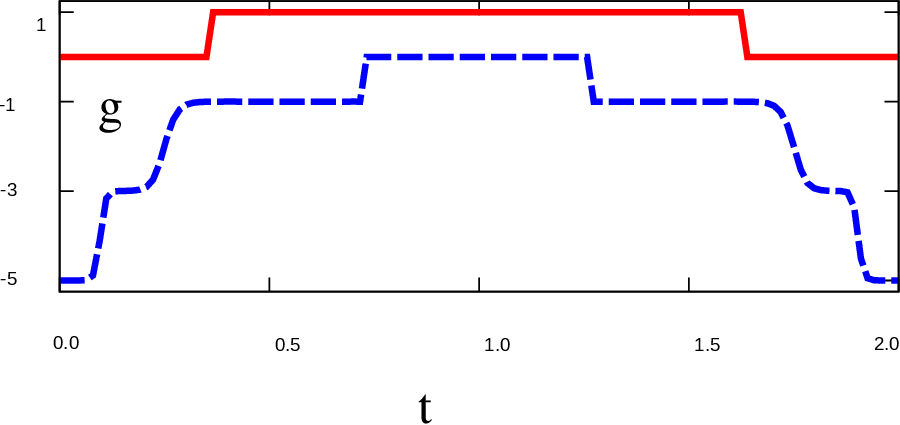}
  \end{minipage}\hskip 0.33cm
  \begin{minipage}{0.23\textwidth}
  \vskip 0.02cm
  \includegraphics[width=118pt,angle=0]{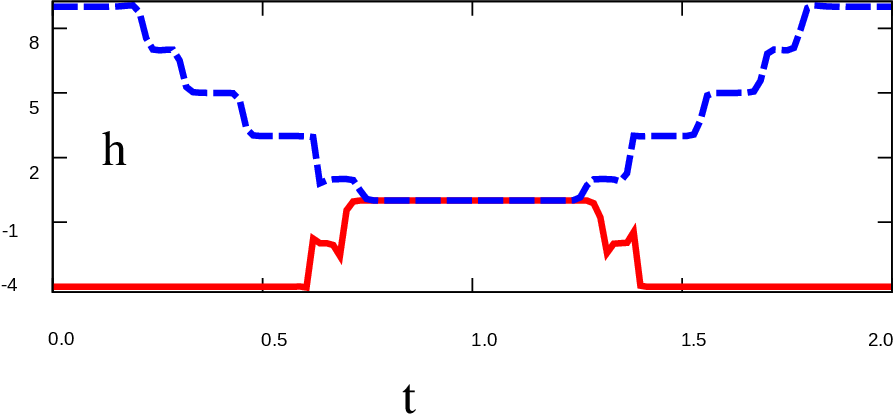}
  \end{minipage}
  \caption{Quasienergy spectra for open systems when $n=4$ are shown in (a) and (b)
    with respect to $\theta$. Variation of winding numbers with respect to
    $\theta$  is shown in (c) and (d). Edge states are indicated by the higher value of
  $I_{pr}$. }
\label{edge-states-floquet-m-4}
\end{figure}

\begin{figure}[h]
  \psfrag{wz}{$\omega_0$}
  \psfrag{wp}{$\omega_\pi$}
  \psfrag{00}{\hskip 0.14 cm $0.0$}
  \psfrag{2.0}{\hskip -0.2 cm $2.0$}
  \psfrag{1.5}{\hskip -0.2 cm $1.5$}
  \psfrag{3}{\hskip -0.15 cm $3$}
  \psfrag{2}{\hskip -0.15 cm $2$}
  \psfrag{1}{\hskip -0.15 cm $1$}
  \psfrag{-1}{\hskip -0.25 cm $-1$}
  \psfrag{0}{{\scriptsize $0$}}
  \psfrag{10}{{\scriptsize  $10$}}
  \psfrag{20}{{\scriptsize $20$}}
  \psfrag{30}{{\scriptsize $30$}}
  \psfrag{40}{{\scriptsize $40$}}
  \psfrag{50}{{\scriptsize $50$}}
  \psfrag{60}{{\scriptsize $60$}}
  \psfrag{70}{{\scriptsize $70$}}
  \psfrag{80}{\hskip -0.05 cm{\scriptsize $80$}}
\psfrag{1.0}{\hskip -0.15 cm {\scriptsize $1.0$}}
\psfrag{0.0}{\hskip -0.15 cm {\scriptsize $0.0$}}
\psfrag{0.2}{\hskip -0.15 cm {\scriptsize $0.2$}}
\psfrag{0.5}{\hskip -0.15 cm {\scriptsize $0.5$}}
\psfrag{0.50}{\hskip 0.1 cm $0.5$}
\psfrag{-2}{\hskip -0.25 cm $-2$}
\psfrag{0.25}{\hskip 0.1 cm $0.25$}
\psfrag{-0.5}{\hskip -0.33 cm {\scriptsize $-0.5$}}
\psfrag{-1.0}{\hskip -0.33 cm {\scriptsize $-1.0$}}
\psfrag{y}{\large k}
\psfrag{E}{\hskip -0.5 cm Energy}
\psfrag{a}{\hskip -0.06 cm (a)}
\psfrag{b}{\hskip -0.06 cm (b)}
\psfrag{c}{\hskip -0.06 cm (c)}
\psfrag{d}{\hskip -0.06 cm (d)}
\psfrag{e}{\hskip -0.06 cm (e)}
\psfrag{f}{\hskip -0.06 cm (f)}
\psfrag{g}{\hskip -0.06 cm (g)}
\psfrag{h}{\hskip -0.06 cm (h)}
\psfrag{z}{{\scriptsize $ET/\pi$}}
\psfrag{p}{{\scriptsize $|\psi|^2$}}
\psfrag{ze}{{\scriptsize $0$-energy}}
\psfrag{pi}{{\scriptsize $\pi$-energy}}
\psfrag{I}{\hskip 0.35 cm $I_{pr}$}
\psfrag{t}{ $\theta/\pi$}
\psfrag{s}{sites}
\includegraphics[width=245 pt,angle=0]{top-color-box.eps}
\vskip -0.05cm
\hskip -1.0 cm
\begin{minipage}{0.20\textwidth}
  \includegraphics[width=46pt,angle=-90]{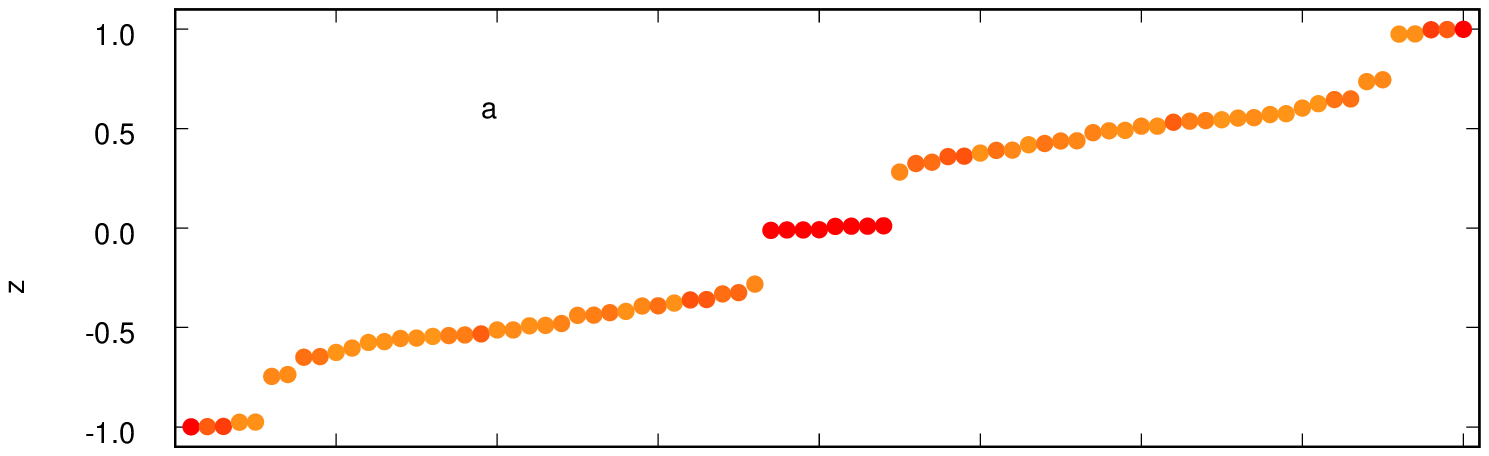}
  \end{minipage}\hskip 0.8cm
  \begin{minipage}{0.20\textwidth}
  \includegraphics[width=46pt,angle=-90]{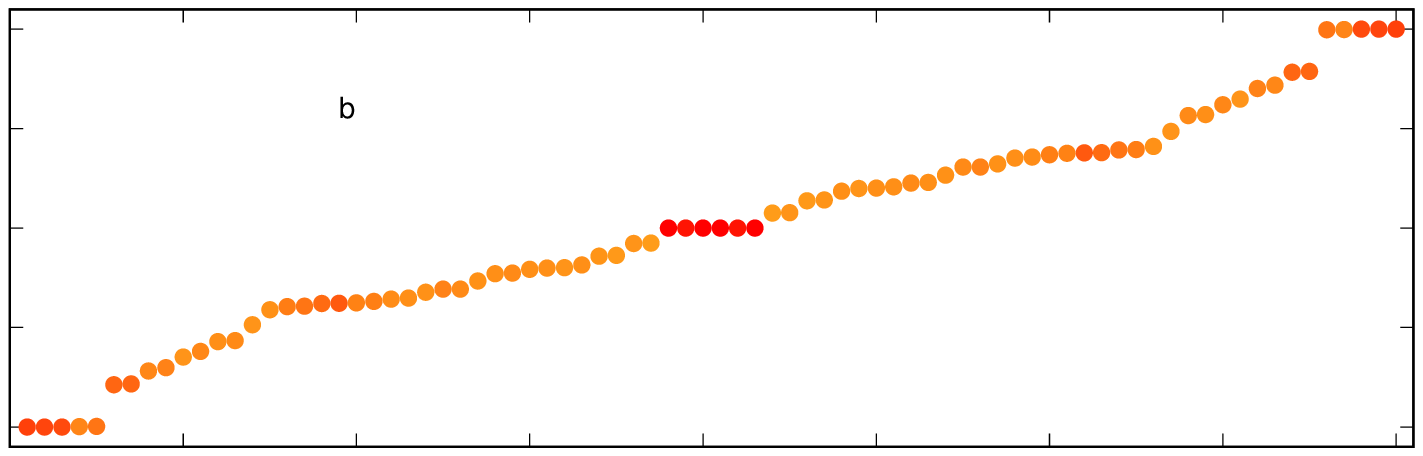}
  \end{minipage}
  \vskip -0.2 cm
  \hskip 0.1 cm
  \begin{minipage}{0.20\textwidth}\hskip -1.2 cm
  \includegraphics[width=46pt,angle=-90]{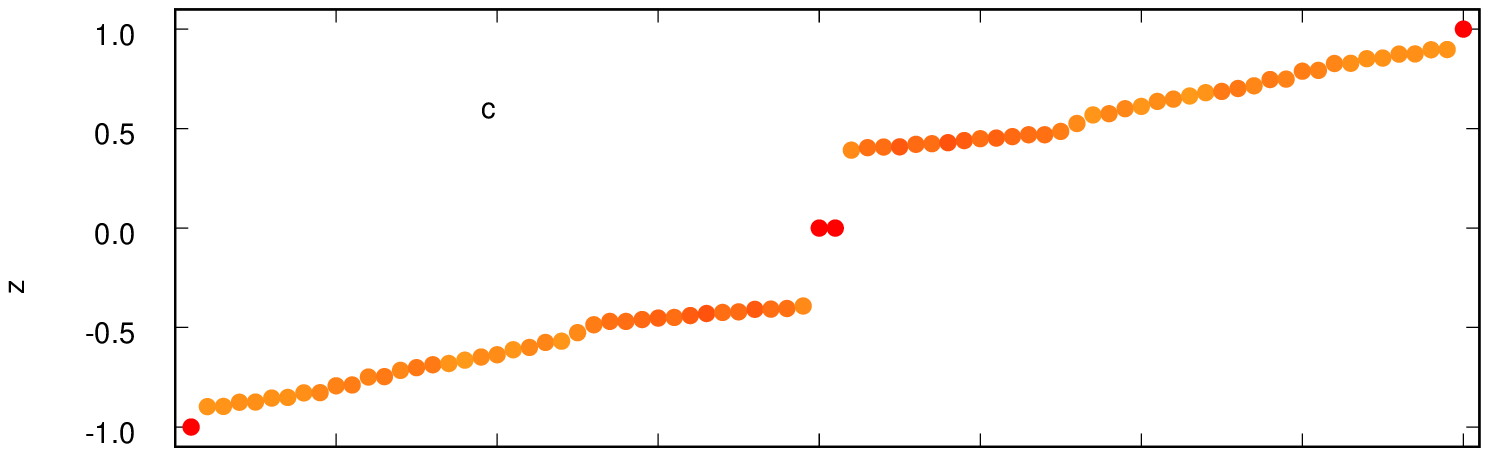}
  \end{minipage}\hskip -0.3cm
  \begin{minipage}{0.20\textwidth}
     \vskip -0.4cm
     \hskip -0.0 cm
  \includegraphics[width=46pt,angle=-90]{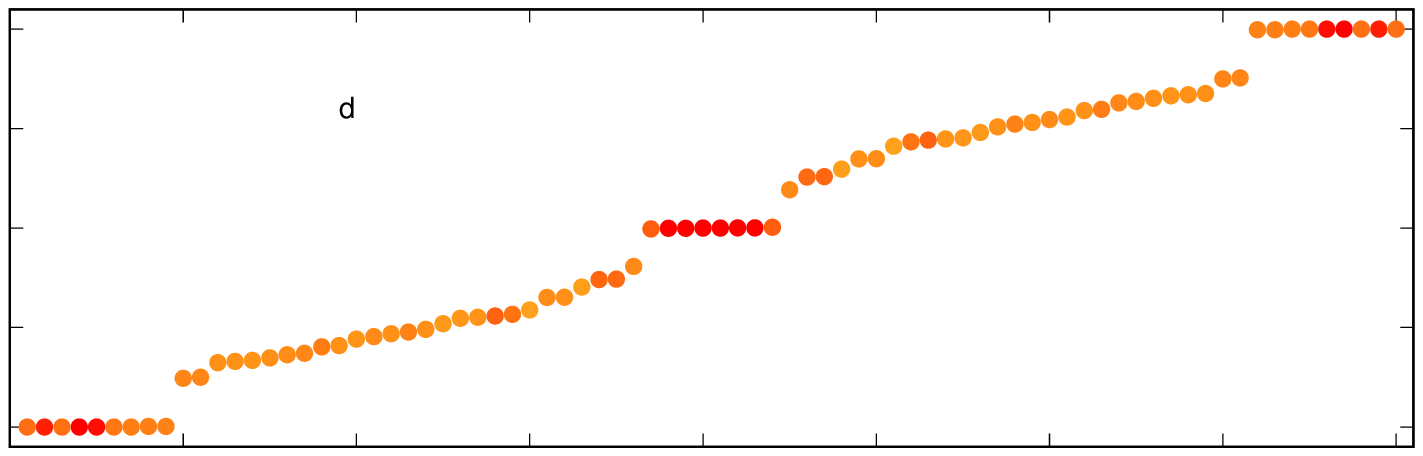}
  \end{minipage}
  \vskip -0.2 cm
   \hskip -0.3 cm
  \begin{minipage}{0.24\textwidth}\hskip -0.1 cm
  \includegraphics[width=128 pt,angle=0]{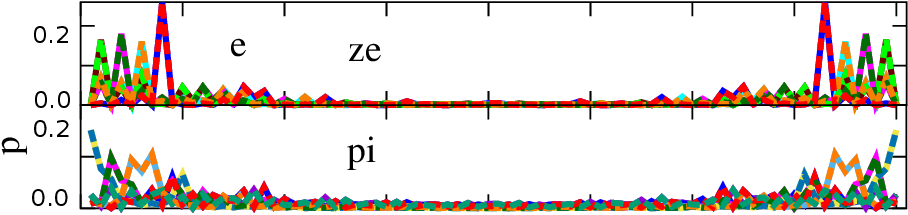}
  \end{minipage}\hskip 0.22cm
  \begin{minipage}{0.23\textwidth}
  \vskip 0.1cm
  \includegraphics[width=122pt,angle=0]{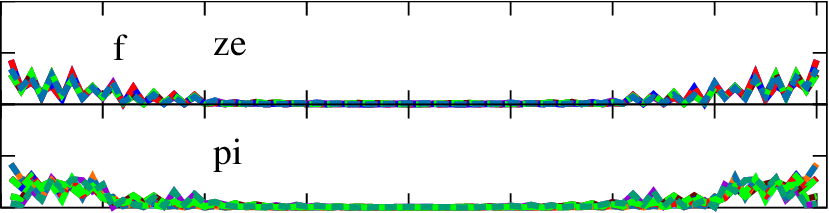}
  \end{minipage}
   \vskip -0.1 cm
   \hskip -0.31 cm
  \begin{minipage}{0.23\textwidth}\hskip 0.0cm
  \includegraphics[width=130pt,angle=0]{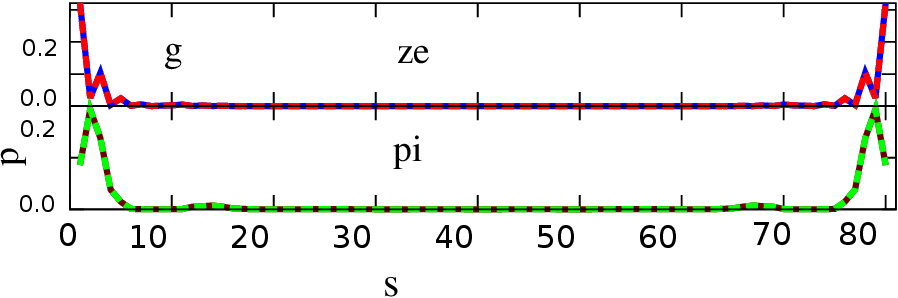}
  \end{minipage}\hskip 0.5cm
  \begin{minipage}{0.23\textwidth}
  \vskip 0.1cm
  \includegraphics[width=120pt,angle=0]{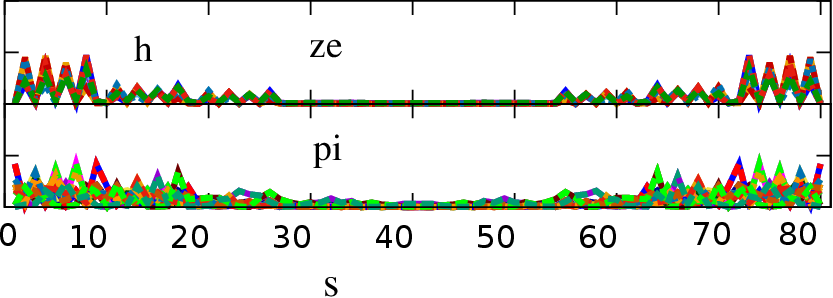}
  \end{minipage}
  \caption{Quasienergy spectra for open systems when $n=4$ are shown in (a) and (b)
    with respect to $\theta$. Variation of winding numbers with respect to
    $\theta$  is shown in (c) and (d). Edge states are indicated by the higher value of
  $I_{pr}$. }
\label{edge-states-floquet-fixed-q-m-4}
\end{figure}

Similarly, system exhibits phases with $w_{10}=1$ and $w_{1\pi}=-1,-2$, when
$t_0=0.9, \alpha=1,\beta=0.5,\gamma=0.7,\delta=0.01,t_p=0.5T$,
for PII.
Quasienergy band diagram and variation of winding numbers
with respect to $\theta$ 
are shown in Fig. \ref{edge-states-floquet-m-1} (b) and
(f). Likewise, the same diagrams for PIII have been
shown in Fig. \ref{edge-states-floquet-m-1} (c) and
(g), where topological phases with winding numbers
$w_{10}=1$, and $w_{1\pi}=-1,-2$ appear when
$t_0=0.8, \alpha=1,\beta=0.3,\gamma=0.5,\delta=0.01,t_p=0.5T$.
For PIV, system exhibits phases with winding numbers
$w_{10}=1$, and $w_{1\pi}=1,3$, when
$t_0=0.9, \alpha=0.02,\beta=0.4,\gamma=0.1,\delta=1,t_p=0.5T$.
The respective diagrams are shown in
Fig. \ref{edge-states-floquet-m-1} (d) and (h).

It is shown that the Floquet system with $n=1$ hosts
topological phases with winding numbers of extreme values 3 and $-2$.
Topological regions defined by nonzero values of
$w_{10}$ and $w_{1\pi}$ are accompanied by the presence of edge states
at energies $ET/\pi=0$ and $ET/\pi=1$, respectively
in addition to the respective band gap. 
Localization of the edge states has been determined by
estimating the inverse participating ratio,
$I_{pr}=\sum_j|\psi_j|^4$, where $\psi_j$ is the
probability amplitude of a particular normalized
eigenstate $\psi$ at the lattice site $j$, for the
eigenenergy $E$ and $\sum_j|\psi_j|^2=1$.
For a perfect localized state
$I_{pr}=1$, while for an extended state $I_{pr}\rightarrow 0$.
$I_{pr}$ of every quasienergy state has been enumerated
as shown in the colorbar of the Fig. \ref{edge-states-floquet-m-1}
(a)-(d). According to the colorbar it shows that value of 
$I_{pr}$ for the edge states ($ET/\pi=0,1$)
is higher than that for the other states ($ET/\pi\ne 0,1$).
It means edge sates are highly localized.

In order to study the localization of edge states
geometrically over the
1D lattice value of $|\psi_j|^4$ have been evaluated
for all of the edge states of 
$ET/\pi=0$ (0-energy) and $ET/\pi=1$ ($\pi$-energy),
for definite values of $\theta$ in the topological regions.
For this purpose, quasienergies obtained by
diagonalizing the Floquet matrix (Eq. \ref{H-nF})
without imposing PBC have been plotted within first Floquet-zone
as shown in Fig. \ref{edge-states-floquet-fixed-q-m-1} (a)-(d). 
Quasienergies are plotted in their ascending order with respect to a
number, where the total number of energies is exactly equal to the total 
number of sites, $N$ as noted in Eq. \ref{H-nj}.
In this case $N=80$. 
Fig. \ref{edge-states-floquet-fixed-q-m-1} (a) has been shown
for PI when 
$t_0=3/4,\, \alpha=1,\,\beta=0.8,\,\gamma=0.7,\,\delta=0,\,t_p=0.3T,\,\theta=0.4\pi$. In this case $w_{10}=1$ and $w_{1\pi}=-1$, which corresponds to
a pair of edge states each for 0 and $\pi$-energies according to
bulk-boundary correspondence rule\cite{Asboth1,Asboth2}.
Those edge states are shown in
Fig. \ref{edge-states-floquet-fixed-q-m-1} (e), where their localization
are found near the two boundaries.
\subsection{Floquet topological phases for $n=2$:}
In the same way, quasienergy band diagrams and variations
of $w_{20}$ and $w_{2\pi}$ have been shown in Fig. \ref{edge-states-floquet-m-2}
(a)-(d), for the sets PI, PII, PIII and PIV, respectively, when $n=2$.
Fig. \ref{edge-states-floquet-m-2} (a) and (e) 
has been drawn for PI when the parameters are 
$t_0=0.9,\, \alpha=1,\,\beta=0.2,\,\gamma=0.0,\,\delta=0.65,\,t_p=0.3T$.
The system exhibits the topological phases with
$w_{20}=1,-2$ and $w_{2\pi}=1,2,-1$, for PI.
As shown in Fig. \ref{edge-states-floquet-m-2} (b) and (f),
band diagram and winding numbers for the
topological phases are shown for PII, when
$t_0=0.9,\, \alpha=1,\,\beta=0.525,\,\gamma=0.7,\,\delta=0.01,\,t_p=0.3T$.
In this case, topological phases are characterized by
$w_{20}=3$ and $w_{2\pi}=-1,-3$.
Similarly, Fig. \ref{edge-states-floquet-m-2} (c) and (g),
demonstrate the band diagram and the winding numbers of the
topological phases for the set PIII, when
$t_0=0.9, \,\alpha=1,\,\beta=0.3,\,\gamma=0.525,\,\delta=0.01,\,t_p=0.3T$.
Winding numbers in this case are
$w_{20}=1$ and $w_{2\pi}=-1,-3$.
Likewise, topological phases with $w_{20}=-1,-2$
and $w_{2\pi}=1,3,5$ appear for the set PIV when
$t_0=0.9,\,\alpha=0.015,\,\beta=0.3725,\,\gamma=0.1,\,\delta=1.0,\,t_p=0.3T$, 
as shown in Fig. \ref{edge-states-floquet-m-2} (d) and (h).
Anyway, for $n=2$ topological phases with extreme values of
winding numbers are 5 and $-3$.

Quasienergy bands for fixed values of $\theta$ for $n=2$
have been shown in Fig. \ref{edge-states-floquet-fixed-q-m-2} (a)-(d),
for four different cases, PI-PIV. 
In Fig. \ref{edge-states-floquet-fixed-q-m-2} (a) and (e),
quasienergies and site-wise
probability densities of four 0-energy and six $\pi$-energy
for the set PI are shown when
$t_0=0.9,\, \alpha=1,\,\beta=0.2,\,\gamma=0.0,\,\delta=0.65,\,t_p=0.3T,\,
\theta=\pi$. Those edge states correspond to
$w_{20}=-2$ and $w_{2\pi}=3$. 
For the set PII, quasienergy band and probability
densities of six 0- and $\pi$-energy edge states for 
$t_0=0.9,\, \alpha=1,\,\beta=0.525,\,\gamma=0.7,\,\delta=0.01,\,t_p=0.3T,
\,\theta=0$ are shown
in Fig. \ref{edge-states-floquet-fixed-q-m-2} (b) and (f).
It corresponds to the topological phase
with $w_{20}=3$ and $w_{2\pi}=-3$.
Similarly, in Fig. \ref{edge-states-floquet-fixed-q-m-2} (c) and (g),
quasienergy band and probability
densities of edge states for 
$t_0=0.9, \,\alpha=1,\,\beta=0.3,\,\gamma=0.525,\,\delta=0.01,\,t_p=0.3T,
\,\theta=\pi/2$ are shown for the set PIII.
In this case two edge states each for 0- and $\pi$-energy appear
which correspond to the topological phase with $w_{20}=1$ and $w_{2\pi}=-1$.
Finally, when
$t_0=0.9,\,\alpha=0.015,\,\beta=0.3725,\,\gamma=0.1,\,\delta=1.0,\,t_p=0.3T,\,
\theta=0.325\pi$,
quasienergy band and probability
densities of four 0-energy and ten $\pi$-energy 
edge states have been shown in Fig. \ref{edge-states-floquet-fixed-q-m-2} (d) and (h), respectively, for the set PIV. Here,
the topological state is characterized by
$w_{20}=-2$ and $w_{2\pi}=5$.
\subsection{Floquet topological phases for $n=3$:}
Quasienergy band diagrams with respect to $\theta$ 
have been shown in Fig. \ref{edge-states-floquet-m-3}
(a)-(d), for the sets PI, PII, PIII and PIV, respectively, when $n=3$.
Values of $w_{30}$ and $w_{3\pi}$ for different topological phases
have been noted in Fig. \ref{edge-states-floquet-m-3}
(e)-(h), for the respective cases.
For PI, quasienergy band diagram and the variation of $w_{30}$ and $w_{3\pi}$ 
with respect to $\theta$ have been shown in Fig. \ref{edge-states-floquet-m-3}
(a) and (e) when 
$t_0=0.9,\,\alpha=0.3,\,\beta=0.05,\,\gamma=0.015,\,\delta=0.9,\,t_p=0.3T$.
In this case, topological phases characterized by $w_{30}=-3,7$ and
$w_{3\pi}=1,3,5,7$ have been noted. Similarly 
in Fig. \ref{edge-states-floquet-m-3} (b) and (f)
band diagram and the variation of $w_{30}$ and $w_{3\pi}$
are plotted when 
$t_0=0.9,\,\alpha=1,\,\beta=0.5,\,\gamma=0.6725,\,\delta=0.01,\,t_p=0.3T$,
for PII. Topological phases with $w_{30}=1,2,4$ and
$w_{3\pi}=-1,-3,-4$ have been found in this case.
For the set PIII, the diagrams in
Fig. \ref{edge-states-floquet-m-3} (c) and (g) are drawn when
$t_0=0.9,\,\alpha=0.76,\,\beta=0.325,\,\gamma=0.5,\,\delta=0.01,\,t_p=0.3T$.
For this set topological phases with $w_{30}=1,2$ and
$w_{3\pi}=-3,-4$ have been emerged. Finally for the set PIV 
topological phases with $w_{30}=-3$ and $w_{3\pi}=1,3,5,7$
have been noted as shown in
Fig. \ref{edge-states-floquet-m-3} (h). The corresponding
band diagram is shown in 
Fig. \ref{edge-states-floquet-m-3} (d) when
$t_0=0.9,\,\alpha=0.1,\,\beta=0.34,\,\gamma=0.01,\,\delta=0.8,\,t_p=0.3T$.
The topological phases with extreme values of
$-4$ and 7 are found in $n=3$. So, topological phases with
higher winding numbers are found for $n=3$.

Quasienergy diagrams for fixed $\theta$ when $n=3$
are shown in Fig. \ref{edge-states-floquet-fixed-q-m-3} (a)-(d),
for four cases, PI-PIV. 
Fig. \ref{edge-states-floquet-fixed-q-m-3} (a) and (e) are drawn
when 
$t_0=0.9,\,\alpha=0.3,\,\beta=0.05,\,\gamma=0.015,\,\delta=0.9,\,t_p=0.3T,\,
\theta=\pi$ for the set PI. In this case system exhibits
the phase with $w_{30}=-3$ and $w_{3\pi}=7$, where band diagram and
the distribution of probability densities for all the edge states are shown in
Fig. \ref{edge-states-floquet-fixed-q-m-3} (a) and (e), respectively. 
For the set PII, similar diagrams are drawn for 
$t_0=0.9,\,\alpha=1,\,\beta=0.5,\,\gamma=0.6725,\,\delta=0.01,\,t_p=0.3T,\,
\theta=0$, as shown in Fig. \ref{edge-states-floquet-fixed-q-m-3}
(b) and (f). For these parameters, system exhibits the
phase with $w_{30}=4$ and $w_{3\pi}=-4$. Likewise, for the set PIII
band diagram and probability densities of all the edge states are shown in
Fig. \ref{edge-states-floquet-fixed-q-m-3} (c) and (g), respectively,
when $t_0=0.9,\,\alpha=0.76,\,\beta=0.325,\,\gamma=0.5,\,\delta=0.01,
\,t_p=0.3T,\,\theta=0.4\pi$. At this point system exhibits the phase
with $w_{30}=2$ and $w_{3\pi}=-3$. System for the set PIV
demonstrates the topological phase with $w_{30}=-3$ and $w_{3\pi}=7$ for 
$t_0=0.9,\,\alpha=0.1,\,\beta=0.34,\,\gamma=0.01,\,\delta=0.8,\,t_p=0.3T,\,
\theta=0$, whose band diagram and
probability densities of all the edge states are shown
in Fig. \ref{edge-states-floquet-fixed-q-m-3} (d) and (h), respectively. 
\subsection{Floquet topological phases for $n=4$:}
Obviously phases with much higher values of winding numbers
have been found for the case $n=4$ than $n=1,2,3$.
Series of topological phases have been demonstrated
in Fig. \ref{edge-states-floquet-m-4}
(a)-(d), for the sets PI, PII, PIII and PIV, respectively,
where quasienergy band diagrams with respect to $\theta$
are displayed. Variation of winding numbers, $w_{40}$ and $w_{4\pi}$
are shown in Fig. \ref{edge-states-floquet-m-4}
(e)-(h), for the respective sets. 
The Fig. \ref{edge-states-floquet-m-4}
(a) and (e), are drawn with the parameters
$t_0=0.8,\,\alpha=1,\,\beta=0.28,\,\gamma=0.01,\,\delta=0.585,\,t_p=0.3T$, 
for the set PI. It exhibits the phases with 
$w_{40}=1,-2,-4$ and $w_{4\pi}=-1,1,3,5$.
Similarly, Fig. \ref{edge-states-floquet-m-4}
(b) and (f), are drawn for the set PII with the parameters
$t_0=0.9,\,\alpha=1,\,\beta=0.5,\,\gamma=0.6225,\,\delta=0.01,\,t_p=0.3T$,
where the system is found to exhibit the phases with 
$w_{40}=3,-2$ and $w_{4\pi}=-1,-3,-5$. Likewise, system exhibits the
phases $w_{40}=1$ and $w_{4\pi}=-1,-3,-5$ for the set PIII, when 
$t_0=0.9,\,\alpha=1,\,\beta=0.225,\,\gamma=0.2,\,\delta=0.01,\,t_p=0.3T$.
Band diagrams and winding numbers with respect to $\theta$ are shown in
Fig. \ref{edge-states-floquet-m-4} (c) and (g), respectively, for this set.
Fig. \ref{edge-states-floquet-m-4} (d) and (h) demonstrate the
variation of band structures and winding numbers for the
topological phases with respect to $\theta$, respectively  for the set PIV. 
These diagrams are drawn for 
$t_0=0.9,\,\alpha=0.2,\,\beta=0.335,\,\gamma=0.01,\,\delta=0.8,\,t_p=0.3T$, 
when the system hosts the topological phases with  
$w_{40}=-2,-4$ and $w_{4\pi}=1,3,5,7,9$. Extreme values of winding numbers
in this case are $-5$ and 9, which is higher than the previous
cases of $n=1,2,3$. 

System exhibits a particular topological phase with
definite value of winding numbers for a fixed value of $\theta$.
For this purpose, Fig. \ref{edge-states-floquet-fixed-q-m-4} (a)-(d)
are drawn for the sets PI-PIV, where band diagrams are shown  when $n=4$.
Probability densities per site for every edge state are shown
in  Fig. \ref{edge-states-floquet-fixed-q-m-4} (e)-(h) for the
respective sets. For example, Fig. \ref{edge-states-floquet-fixed-q-m-4} (a)
and (e) are drawn when 
$t_0=0.8,\,\alpha=1,\,\beta=0.28,\,\gamma=0.01,\,\delta=0.585,\,t_p=0.3T,\,
\theta=\pi$, for the set PI, where the system exhibits the phase with 
$w_{40}=-4$ and $w_{4\pi}=5$. Values of the parameters are fixed at
$t_0=0.9,\,\alpha=1,\,\beta=0.5,\,\gamma=0.6225,\,\delta=0.01,\,t_p=0.3T,\,
\theta=0$, while drawing the Fig. \ref{edge-states-floquet-fixed-q-m-4} (b)
and (f) for the set PII when the system hosts the phase with
$w_{40}=3$ and $w_{4\pi}=-5$. 
Fig. \ref{edge-states-floquet-fixed-q-m-4} (c) and (g) are drawn
for the set PIII when 
$t_0=0.9,\,\alpha=1,\,\beta=0.225,\,\gamma=0.2,\,\delta=0.01,\,t_p=0.3T,\,
\theta=\pi/2$, where the topological phase with
$w_{40}=1$ and $w_{4\pi}=-1$ are found. Finally, quasienergy band and
site-wise probability densities for the set PIV are drawn, respectively in the
Fig. \ref{edge-states-floquet-fixed-q-m-4} (d) and (h), when 
$t_0=0.9,\,\alpha=0.2,\,\beta=0.335,\,\gamma=0.01,\,\delta=0.8,\,t_p=0.3T,\,
\theta=0$, where topological phase with
$w_{40}=-4$ and $w_{4\pi}=9$ are observed.

As discussed above topological phase diagrams for four different sets
PI-PIV as well as four different values $n=1,2,3, 4$ have been drawn
for a variety of parameters. No common parameter set is found
for these cases where topologically nontrivial phases
are found for the entire range $0 \le \theta \le 2\pi$
with distinct transition points.
For every set phases of higher values of ($w_{n0},w_{n\pi}$) emerge with the
increase of $n$. 
For arbitrary value of $n$,
the extreme values of winding numbers for nontrivial
topology are $w_{n0}=-(n+1)$ and $w_{n\pi}=2n+1$,
which clearly indicates the increase of
the value of winding number linearly with $n$.
Which means the staggered eSSH models
introduced here are capable to host Floquet topological phase with
higher winding numbers when the extent of FN hopping are larger.
Comparing the results of all the sets PI-PIV, it is evident that
phases with higher winding numbers emerged for the set PIV. 
In every case topological state with definite values of
winding numbers $w_{n0}$ and $w_{n\pi}$ are found to 
associate with $2w_{n0}$ and $2w_{n\pi}$
numbers of edge states according to bulk-boundary correspondence rule
\cite{Asboth1,Asboth2}.
Localization of the edge states has been quantified by the
 values of $I_{pr}$. Distribution of probability density
$|\psi_j|^2$ over the sites $j$ of the edge states indicates their
 higher amplitudes near the boundaries, which is also supported by the
 higher values of $I_{pr}$.
\section{Discussion}
\label{Discussion}
In this study a series of static and dynamic eSSH models with varying
long range hopping terms are considered and their topological phase diagrams
are obtained. Topological characterization has been accomplished
in terms of winding number ($\nu$) for these chiral systems.
In these models effect of a pair of FN terms on the topological 
properties is taken into consideration in every time.
For the static models phases with higher winding numbers
are found with the increase of extent of the FN hopping, which is
generalized for the arbitrary value of $n$. Extensive phase diagrams
are drawn for four different eSSH models where boundaries of
the different phases have been identified exactly.
 The existence of topological phases of these eSSH
models can be verified
in various ways, like constructing optical lattices, 
photonic systems of coupled waveguides, and etc,
those are found successful to
demonstrate the zero energy edge states for the SSH model before
\cite{Atala,Rechtsman,Li3}. 

In the same way, topological phases have been 
obtained in the corresponding dynamic models
where phases with higher winding number is found with
the increase of extent of the FN hopping terms. 
For this purpose four different dynamic eSSH models with
quenching protocol is considered where extent of FN terms are different.
A definite Floquet system is composed of a pair of
static eSSH models with FN terms of fixed value of $n$. 
So, four different extents of FN terms are considered. Topological phases
are characterized by a pair of winding numbers, ($w_{n0},w_{n\pi}$).
For every Floquet system, 
four different types of hopping terms of the static eSSH models 
are formulated in terms of an angular parameter, $\theta \in (-\pi,+\pi)$,
which are denoted by the sets PI, PII, PIII and PIV, respectively.
Staggeredness of these sets is different. 
As a result, 16 different topological phase diagrams have been drawn
with respect to $\theta$ for four different values of $n=1,2,3,4$.
Topological edge states for `0' and `$\pi$' quasienergies are found
to emerge according to the bulk-boundary
correspondence rule which is shown for each case.
Phases of higher values of ($w_{n0},w_{n\pi}$) emerge with the
increase of $n$, and this is true for every set. Finally, those results are generalized for the
arbitrary values of $n$.
In this study, Floquet system is considered for a fixed value of
$n$. However, if one would construct Floquet systems out of eSSH models of
mixed values of $n$, different types of topological phase
diagrams might emerge. 
Interplay of topological and quantum phase transitions has been
noted before in various spin-1/2 systems
\cite{Kitaev,BKC3,Franchini,Rakesh2,Rakesh3,Susobhan}.
In the same way Floquet topological and dynamical phase transitions
and their interplay for those
two-band models can be studied. Effect of disorder on these
topological phases using Lindblad formalism will be studied in future.

\section*{Appendix-A}
Upon Fourier transformation of the Hamiltonian in Eq. \ref{essh}
for the two sublattice system has been converted to
$2\times 2$ matrix representation of that as shown in Eq. \ref{Hnk}.
The explicit form of the $H_n( k)$ is given by
\bea H_n( k)&=&\left(
\begin{array}{cc}\!\!\!\!\!\!\!\!\!\!\!\!\!\!\!\!\!\!\!\!\!\!\!\!\!\!\!\!\!\!\!\!\!\!\!\!\!\!\!\!\!0&\!\!\!\!\!\!\!\!\!\!\!\!\!\!\!\!\!\!\!\!\!\!\!\!\!\!\!\!\!\!\!\!\!\!\!\!\!\!\!\!\!\!\!\!\!\!\!\!\!\!\!\!\!t_1\!+\!t_2e^{-ik}\!+\!t_3e^{ink}\!+\!t_4e^{-i(n+1)k}\\[0.4em]
 t_1\!+\!t_2e^{ik}\!+\!t_3e^{-ink}\!+\!t_4e^{i(n+1)k} &\!\!\!\!\!\!\!\!\!0\end{array}
 \right)\!,\nonumber\\[0.5em]
&\!\!\!\!\!\!\!\!\!\!\!=&\!\!\!\!\!\!\!\!\left(
 \begin{array}{c}0\;\;\;\;t_1\!+\!t_2\cos{k}\!+\!t_3\cos{(nk)}\!+\!t_4\cos{((n\!+\!1)k)}\\[0.2em]
 \;\;\;\;-i(t_2\sin{k}\!-\!t_3\sin{(nk)}\!+\!t_4\sin{((n\!+\!1)k)})\\[0.7em]
 t_1\!+\!t_2\cos{k}\!+\!t_3\cos{(nk)}\!+\!t_4\cos{((n\!+\!1)k)}\;\;\;\;0\\[0.2em]
\!\!\!\!\!\!\!\!+i(t_2\sin{k}\!-\!t_3\sin{(nk)}\!+\!t_4\sin{((n\!+\!1)k)})
\end{array}
 \right)\!,\nonumber\\[0.5em]
 &\!\!\!\!\!\!\!\!\!\!\!=&\!\!\!\! (t_1\!+\!t_2\cos{k}\!+\!t_3\cos{(nk)}\!+\!t_4\cos{((n\!+\!1)k)})\, \sigma_x\nonumber\\[0.2em]
 &\!\!\!\!\!\!\!\!\!\!\!&\!\!\!\!+\,
 (t_2\sin{k}\!-\!t_3\sin{(nk)}\!+\!t_4\sin{((n\!+\!1)k)})\, \sigma_y,
 \nonumber\\[0.5em]
  &\!\!\!\!\!\!\!\!\!\!\!=&\!\!\!\! h_x(k)\sigma_x+h_y( k)\sigma_y,
 \label{Hnkk}\eea
where  $h_x(k)=t_1+t_2\cos{(k)}+t_3\cos{(nk)}+t_4\cos{((n\!+\!1)k)},$
and $h_y( k)=t_2\sin{(k)}-t_3\sin{(nk)}+t_4\sin{((n\!+\!1)k)}$.
On the other hand
introducing  $g_n(k)=t_1+t_2e^{-ik}+t_3e^{ink}+t_4e^{-i(n+1)k}$,
$ H_n( k)$ can be expressed as 
\be H_n( k)=\left(
\begin{array}{cc}0&g_n( k)\\[0.4em]
 g^*_n( k) &0\end{array}
 \right).\ee
\section*{Appendix-B}
In this appendix, derivation of the Eq. \ref{Np-Nz}
for the enumeration of winding number in terms of
number of poles ($\mathcal N_{\rm p}$), and number of zeros  ($\mathcal N_{\rm z}$) of
the complex function $g_n(z)$
has been explained.

The Hamiltonian in the $k$-space, $H_n(k)$
(Eq. \ref{Hnk}) can always be expressed as 
\be  H_n(k)=h_x(k)\sigma_x+h_y( k)\sigma_y,\nonumber
\ee
where,  $H_n(k)$ preserves the chiral symmetry due to the fact that
$ \sigma_z H_n(k) \sigma_z=-H({\rm k})$. 
Now introducing the vector 
${\boldsymbol h}(k)=h_x(k)\boldsymbol{\hat i}+h_y(k)\boldsymbol{\hat j}$, 
it may happen that 
tip of the vector $\boldsymbol h( k)$ has 
traceed out closed loops
in the $h_x\mbox{-}h_y$ plane 
if $k$ runs from $-\pi$ to $\pi$ on the BZ. Winding number, $\nu$,  
accounts the number of closed loops traced out 
by the tip of
vector $\boldsymbol h( k)$ around the origin of the
$h_x\mbox{-}h_y$ plane in the anticlockwise direction.
Mathematically $\nu$ is defined as
\be \nu=\frac{1}{2\pi}\oint_{\mathcal C} \left[\boldsymbol {\hat h}( k)\times
  \frac{d}{d k}\,\boldsymbol {\hat h}( k)\right]_z\!d{ k},
\label{nui}\ee
where $\boldsymbol {\hat h}( k)
=\boldsymbol h( k)/|\boldsymbol h( k)|$, and ${\mathcal C}$
is the closed contour. 
Upon simplification,
\be \left[
\boldsymbol{\hat h}\times
\frac{d\boldsymbol{\hat h}}{d k}\right]_z=
\frac{1}{h^2}\bigg(h_x\,\frac{dh_y}{dk}-h_y\,\frac{dh_x}{dk}
\bigg),\nonumber \ee
So, 
\be \nu=\frac{1}{2\pi}\oint_{\mathcal C}\,\frac{1}{h^2}
\bigg(h_x\,\frac{dh_y}{dk}-h_y\,\frac{dh_x}{dk}
\bigg)dk.\label{nu11}\ee
Now, introducing the complex function (Eq. \ref{gn}),
\[g_n(z)=t_1+\frac{t_2}{z}+t_3\,z^{n}+\frac{t_4}{z^{n+1}},\]
where $z=e^{ik}$, and 
$g_n(z)=h_x(k)+i h_y(k)$, or $|g_n(z)|=h(k)$, it can be shown that
\[\frac{1}{h^2}\bigg(h_x\,\frac{dh_y}{dk}-h_y\,\frac{dh_x}{dk}
\bigg)=i\left(\frac{d}{dk}\ln{g_n(z)}-
\frac{d}{dk}\ln{|g_n(z)|}\right).\]
Hence,
\bea \nu
&=&\frac{i}{2\pi}\oint_{\mathcal C}\left(\frac{d}{dk}\ln{g_n(z)}-
\frac{d}{dk}\ln{|g_n(k)|}\right)dk,\nonumber  \\[0.6em]
&=&\frac{ i}{2\pi}\oint_{\mathcal C}d\ln{g_n(z)}-\frac{ i}{2\pi}\oint_{\mathcal C}
  d\ln{|g_n(z)|},\nonumber  \\[0.4em]
&=&\frac{i}{2\pi}\oint_{\mathcal C}d\ln{g_n(z)},\nonumber \eea
  since the integration of the real integrand, $\ln{|g_n(z)|}$
  vanishes over the closed contour, $\mathcal C: \,|z|=1$.

  In order to evaluate the integral
  $\oint_{\mathcal C} d\ln{g_n(z)}$, and assuming the fact that 
  the function $g_n(z)$ has $m_z$ number of zeros of different orders
  within the contour ${\mathcal C}$, and  
for a particular zero at  $z=z_l$,  having order $m_l$,
  $g_n(z)$ can be expressed as
  \[g_n(z)=(z-z_l)^{m_l}f(z),\]
  where $f(z)$ is analytic and nonzero at $z_l$.
  Hence its derivative can be written as
  \[\frac{d}{dz}g_n(z)=m_l(z-z_l)^{m_l-1}f(z)+(z-z_l)^{m_l}\frac{d}{dz}f(z).\]
  Now dividing by $g_n(z)$, 
  \[\frac{1}{g_n(z)}\frac{d}{dz}g_n(z)=\frac{m_l}{z-z_l}+\frac{1}{f(z)}\frac{d}{dz}f(z).\]
  Hence, it reveals that the function $\frac{1}{g_n(z)}\frac{d}{dz}g_n(z)$
  has a simple pole 
  (or pole of order one) at $z_l$ with residue $m_l$, since
  the function $\frac{1}{f(z)}\frac{d}{dz}f(z)$ is analytic at $z_l$ by its definition.
  Now applying the Cauchy's residue theorem \cite{Com-Var}:
  \bea
  \oint_{\mathcal C}d\ln{g_n(z)}&=&\oint_{\mathcal C}\frac{1}{g_n(z)}\frac{d}{dz}g_n(z) dz\nonumber \\[0.4em]
  &=&\sum_{l=1}^{m_z} \oint_{\mathcal C} \frac{m_l}{z-z_l} dz\nonumber \\[0.4em]
  &=&2\pi i \sum_{l=1}^{m_z} m_l,\nonumber
  \eea
where the summation is implied over all the zeros of $g_n(z)$ within the contour ${\mathcal C}$.
  
  On the other hand, assuming the fact that
  the function $g_n(z)$ has $m_p$ number of poles of
variety of orders within the contour ${\mathcal C}$, and 
 for a particular pole of order $n_j$ at $z=z_j$,
  $g_n(z)$ can be expressed as
  \[g_n(z)=\frac{s(z)}{(z-z_j)^{n_j}},\]
  where $s(z)$ is analytic and nonzero at $z_j$.
   As a result its derivative can be written as
   \[\frac{d}{dz}g_n(z)=\frac{-n_j}{(z-z_j)^{n_j+1}}s(z)
   +\frac{1}{(z-z_j)^{n_j}}\frac{d}{dz}s(z).\]
  Now dividing by $g_n(z)$, 
  \[\frac{1}{g_n(z)}\frac{d}{dz}g_n(z)=\frac{-n_j}{z-z_j}+\frac{1}{s(z)}\frac{d}{dz}s(z).\]
  Hence, in this case, the function $\frac{1}{g_n(z)}\frac{d}{dz}g_n(z)$
  has a simple pole at $z_j$ with residue $-n_j$, since
  the function $\frac{1}{s(z)}\frac{d}{dz}s(z)$ is analytic at $z_j$ by its definition.
   Applying the Cauchy's residue theorem in this case: 
  \bea
  \oint_{\mathcal C}d\ln{g_n(z)}
  &=&\sum_{j=1}^{m_p} \oint_{\mathcal C} \frac{-n_j}{z-z_j} dz,\nonumber \\[0.4em]
  &=&-2\pi i \sum_{j=1}^{m_p} n_j,\nonumber
  \eea
  where the summation extends over all the poles of $g_n(z)$ within the contour ${\mathcal C}$.
  So, combing these two fact that the function,
   $g_n(z)$ has $l$ zeros with respective order $m_l$,
  along with $j$ poles with respective order $n_j$ within a closed contour
  ${\mathcal C}$, it can be shown that
   \bea
   \oint_{\mathcal C}d\ln{g_n(z)}&=&2\pi i \,\left(\sum_{l=1}^{m_z} m_l-\sum_{j=1}^{m_p} n_j\right),\nonumber\\[0.4em]
   &=&2\pi i \,\left({\mathcal N_{\rm z}}-{\mathcal N_{\rm p}}\right),
   \label{Arg-principle}
   \eea
   where ${\mathcal N_{\rm z}}=\sum_{l=1}^{m_z} m_l$, actually accounts the
   total number of zeros of order one and ${\mathcal N_{\rm p}}=\sum_{j=1}^{m_p} n_j$, 
   means total number of simple poles of the function $g_n(z)$ within contour
   ${\mathcal C}$. It is worthy at this point
   to note that $m_l$ ($n_j$) is known as the
   multiplicity of the $l$-th ($j$-th) zero (pole) at $z_l$ ($z_j$),
   and this result (Eq. \ref{Arg-principle}) is known as the argument principle \cite{Com-Var}. 
  Hence, the winding number can be expressed as:
 \bea  \nu
 &=&\frac{i}{2\pi}\oint_{\mathcal C}d\ln{g_n(z)},\nonumber \\[0.3em]
 &=&{\mathcal N_{\rm p}}-{\mathcal N_{\rm z}},
 \label{nu12}
 \eea
 where ${\mathcal N_{\rm p}}$ and ${\mathcal N_{\rm z}}$
 are non-negative integers.  
 It proves that winding number is equal to the
 difference of the number of poles and that of zeros of the function
 $g_n(z)$ lies within contour ${\mathcal C}$ with $|z|=1$.

 Alternately, Eq. \ref{nu11} can be expressed as: 
 \bea \nu&=&\frac{1}{2\pi}\oint_{\mathcal C}\,\frac{1}{h^2}
\bigg(h_x\,\frac{dh_y}{dk}-h_y\,\frac{dh_x}{dk}
\bigg)dk,\nonumber \\[0.5em]
&=&\frac{1}{2\pi}\oint_{\mathcal C}\,\frac{d \eta(k)}{dk}dk,
\label{nu13}
\eea
where $\eta(k)=\arctan{(h_y(k)/h_x(k))}$.
This expression has been used before in Eq. \ref{Nu11}.
Therefore, value of $\nu_{nj}$ can be derived using the
formula, $\nu_{nj}={\mathcal N_{{\rm p}nj}}-{\mathcal N_{{\rm z}nj}}$,
where ${\mathcal N_{{\rm p}nj}}$ and ${\mathcal N_{{\rm z}nj}}$ are the
number of poles and zeros of the complex function,
$g_{nj}(z)= g_{njx}+i g_{njy}$.
So, this extensive analysis clearly demonstrates that the
value of $\nu$ which is defined primarily in Eq. \ref{nui},
can be otherwise enumerated using one of the formulae among 
the Eq. \ref{nu11},  \ref{nu12},
and \ref{nu13} \cite{Maffei}.
 
  \section{ACKNOWLEDGMENTS}
  RC acknowledges the UGC fellowship 211610042893. 
  \section{Data availability statement}
  All data that support the findings of this study are
  included within the article.
   \section{Conflict of interest}
  Authors declare that they have no conflict of interest.
  
\end{document}